\documentclass[usenatbib]{mn2e}
\usepackage{natbibmnfix, graphicx, times}
 \usepackage{graphicx}
 \usepackage{subfigure}
\usepackage{color}

\usepackage{amssymb,amsmath}
\usepackage{bm}

\newcommand{\arepo}{{\footnotesize AREPO}}
\newcommand{\apj}{ApJ}

\newcommand{\mnras}{MNRAS}

\title[Viscous flow on a moving mesh]{Multi-dimensional, compressible viscous flow on a moving Voronoi mesh}
 \author[Mu\~noz et al.]{D. J. Mu\~noz$^{1}$\thanks{E-mail:dmunoz@cfa.harvard.edu}
 V. Springel$^{2,3}$, R. Marcus$^{1}$, M. Vogelsberger$^{1}$ and L. Hernquist$^{1}$\\
$^{1}$ Harvard Smithsonian Center for Astrophysics, 60 Garden Street, Cambridge, MA 02138 \\
$^{2}$ Heidelberg Institute for Theoretical Studies, Schloss-Wolfsbrunnenweg 35, 69118 Heidelberg, Germany \\
$^{3}$ Zentrum f\"{u}r Astronomie der Universit\"{a}t Heidelberg, ARI, M\"onchhofstr. 12-14, 69120 Heidelberg, Germany}

\begin{document}

\pagerange{\pageref{firstpage}--\pageref{lastpage}} \pubyear{2012}

\maketitle


\begin{abstract}
Numerous formulations of finite volume schemes for the Euler and
Navier-Stokes equations exist, but in the majority of cases they have
been developed for structured and stationary meshes.  In many
applications, more flexible mesh geometries that can dynamically
adjust to the problem at hand and move with the flow in a (quasi)
Lagrangian fashion would, however, be highly desirable, as this can
allow a significant reduction of advection errors and an accurate
realization of curved and moving boundary conditions. Here we describe
a novel formulation of viscous continuum hydrodynamics that solves the
equations of motion on a Voronoi mesh created by a set of
mesh-generating points.  The points can move in an arbitrary manner,
but the most natural motion is that given by the fluid velocity
itself, such that the mesh dynamically adjusts to the flow. Owing to
the mathematical properties of the Voronoi tessellation, pathological
mesh-twisting effects are avoided.  Our implementation considers the
full Navier-Stokes equations and has been realized in the \arepo\ code
both in 2D and 3D. We propose a new approach to compute accurate
viscous fluxes for a dynamic Voronoi mesh, and use this to formulate a
finite volume solver of the Navier-Stokes equations.  Through a number
of test problems, including circular Couette flow and flow past a
cylindrical obstacle, we show that our new scheme combines good
accuracy with geometric flexibility, and hence promises to be
competitive with other highly refined Eulerian methods. This will in
particular allow astrophysical applications of the  \arepo\ code where
physical viscosity is important, such as in the hot plasma in galaxy
clusters, or for viscous accretion disk models.
\end{abstract}


\section{Introduction} 

The last two decades have seen remarkable advances in the numerical
solution of the compressible Navier-Stokes (NS) equations, which lies
at the heart of computational fluid dynamics (CFD) and computational
aeroacoustics, but also as numerous applications in astrophysics.  In
particular, important progress has been made in approaches based on
the finite volume method (FVM), both using structured as well as
unstructured grids \citep[see][for a review]{mav97}. Other popular
techniques include finite element methods (FEM), discontinuous
Galerkin schemes, and even mesh-free approaches such as smoothed
particle hydrodynamics \citep{sij06}.

When unstructured grids have been employed, they were most most often
in the form of triangular grids in two dimensions, or tetrahedral
grids in three dimensions.  Indeed, finite-volume implementations of
the two-dimensional NS equations on triangular meshes date
back to work by \citet{mav90}, \citet{fri94} and \citet{coi96}.  Much
recent work has also focused on developing optimum mesh-generating
algorithms that require minimal human input and yield efficient
representations of geometrically complex simulation domains.  However,
little work has been done on dynamically evolving meshes, such as
those we shall consider here.
 
Because unstructured meshes have been demonstrated to be accurate and
efficient for both steady-state and transient compressible inviscid
flows \citep{bar92,ven96}, they are now used regularly in engineering
applications. Moreover, the geometric flexibility of unstructured
grids allows the use of simple coordinate systems (in the laboratory
frame) without the need to work with complex coordinate
transformations to describe curved surfaces
\citep[e.g. see][]{tor09}. Indeed, hard boundaries can be tailored by
carefully positioning a few cell faces or mesh generating points along
the surface, and creating the triangulation through Delaunay
tessellation.  As a result, most NS applications on unstructured
meshes for industrial design make use of triangular grids, typically
based on the finite element method, although finite volume schemes
have also been considered. Detailed reviews and stability analysis of
explicit FVM for the NS equations on Cartesian and Delaunay meshes can
be found, e.g, in the doctoral theses of \citet{coi94} and
\citet{mun09}.

In this work, we present a numerical scheme that solves the NS
equations on a general unstructured moving mesh that is constructed as
the Voronoi tessellation of a distributed set of points that move with
the local velocity field. Despite being, in the general sense, an
``unstructured" mesh, the Voronoi diagram has a mathematically
well-defined structure that makes the resulting schemes comparatively
simple and robust \citep[e.g.][]{mis98}.  In fact, this type of mesh
is commonly adapted for the construction of finite volume methods for
elliptic problems and has been in use in numerical studies of solid
state physics \citep{suk98, suk03} such as simulations of fractures
and cracks \citep{suk09}, as well as numerical simulations of oil
reservoirs. Some studies \citep{chr09} have also examined how
reconstructions designed for unstructured triangulations can be
extended to {\it static} Voronoi meshes.

However, Voronoi meshes have infrequently been applied to hyperbolic
conservation laws such as the Euler equations, let alone moving
Voronoi meshes. To our knowledge, the earliest attempts to use
  dynamically adaptive Voronoi tessellations for the NS
  equations data back to \citet{bor87}, although for very simplified,
  incompressible, two-dimensional problems. Around the same time,
  \citet{duk89} developed the General Topology Godunov Method. This
  method -- based on a mesh that is not quite a Voronoi tessellation,
  but similar in spirit -- was introduced as an alternative to the
  Lagrangian particle methods \citep[see, for example][]{bra88} which
  gained increasing popularity in computational plasma physics and
  astrophysics in the following years.

Recently, a complete three-dimensional implementation of the
Euler equations on a moving Voronoi mesh has been described and
implemented in the \arepo\ code by \citet{spr10} \citep[see
also][]{duf11}. The work we present here is an extension of the \arepo\ scheme to the
NS equations, which we have realized in this code as an
optional module. \arepo\ can be classified as an arbitrary
Lagrangian/Eulerian \citep[ALE;][]{hir74} code, in the sense that the
mesh can be moved with the velocity of the flow so that
quasi-Lagrangian behavior results and the mass flux between cells is
minimized (although it is not strictly zero, in general).  On the
other hand, the mesh may also be kept stationary if desired,
effectively yielding an Eulerian formulation.  We note that because
the mesh-generating points may also be arranged on a regular lattice
and arbitrarily refined with time, the \arepo\ code naturally includes
ordinary Eulerian techniques on a Cartesian grid and adaptive mesh
refinement (AMR) algorithms as special cases.

Besides the work of \citet{duf11}, the new Voronoi-ALE
method of \citet{nor10}, which includes viscous terms, 
is the approach most closely related to that presented here,
although it is restricted to the incompressible NS equations.  Also, \citet{ata09} have
applied a Voronoi-based finite volume scheme to the two-dimensional
inviscid shallow water equations, in terms of an algorithm they
referred to as the `natural volume' method.

Although primarily designed for astrophysical fluid dynamics where
self-gravity is an important ingredient \citep[see for example][]{vog11},
the moving Voronoi mesh
approach of \arepo\ offers a number of features than can be
advantageous for more general problems in fluid dynamics.  First, the
moving mesh geometry is adaptive in a continuous manner and can
naturally respond to the local flow, increasing the resolution
automatically and smoothly in regions where the flow converges.  (In
contrast, AMR codes refine the grid discontinuously in time, which can
introduce errors that are potentially difficult to assess.)
Importantly, this Lagrangian character of the dynamics yields reduced
advection errors and a very low numerical diffusivity of the scheme.
Second, the moving mesh formulation retains the Galilean-invariance of
the fluid dynamics at the discretized level of the equations
\citep{spr10}.  In other words, the truncation error of the scheme
does not depend on the bulk velocity of the system, unlike for
traditional Eulerian and AMR codes, and the quality of the solution
does not degrade when high-speed flows are present.  While
conventional fixed-mesh Eulerian codes may, in principle, be able to
suppress additional errors from large bulk velocities by using a
sufficiently fine mesh \citep[see][for a study of Galilean invariance
  in grid codes]{rob10}, this strategy can become computationally
prohibitive, and it also depends on the magnitude of the bulk velocity
involved.  It is therefore desirable to construct efficient methods
that yield manifestly Galilean-invariant solutions (modulo floating
point round-off errors). Third, the moving mesh approach allows much
larger timesteps in the case of rapidly moving flows, because it can
avoid the $\Delta t < d/v$ stability constraint (where $d$ is the cell
size and $v$ the bulk velocity) that augments the Courant condition in
the Eulerian case.
  
From an astrophysical standpoint, compressible viscous flow remains a
viable approximation to more complex or computationally expensive
momentum transport mechanisms such as magneto-hydrodynamic turbulence
or anisotropic plasma viscosity.  Global simulations of cold accretion
disks around protostellar objects \citep[e.g. see][]{val06} still
include shear viscosity coefficients in the form of a Shakura-Sunyaev
eddy viscosity coefficient \citep{sha73}. 

An even clearer case for the need of a viscous treatement of
astrophysical gasdynamics is given by the interacluster medium of hot
galaxy clusters. Here the Spitzer-Braginskii viscosity \citep{bra65}
becomes quite significant, certainly in the unmagnetized case, which
has been studied both using grid \citep{rus04} and SPH \citep{sij06}
codes. In this regime, the commonly adopted assumption of inviscid
behaviour with an effectively infinite Reynolds number is in principle
incorrect and should in future simulation work be replaced with a full
accounting of the correct physical viscosity.

Additionally, physical viscosity can be implemented on turbulent
cascades with resolved inertial range \citep[see][for an application
of our viscosity approach]{bau11} in order to prescribe a
well-specified Reynolds number and a physically correct shape for the
dissipation range, unaffected by the details of the numerical
viscosity of the hydro scheme, which would otherwise induce the
dissipation of turbulence on the grid scale.  This can in particular
inform the ongoing debate whether artificial viscosity effects in SPH can
affect the turbulent cascade above the formal resolution length
\citep{bau11,pri12}.

This paper is organized as follows. In Section~2, we briefly review
the basic NS equations we want to solve, and the role and
meaning of the different viscosity coefficients.  In Section~3, we
then introduce in detail our discretization and time integration
schemes, emphasizing a description of the calculation of suitable
velocity gradient estimates at face centers, and of second-order derivatives of
the velocity field.  We then move on to discuss the performance of our
new approach for a number of test problems in Section~3. Finally, we
summarize our results and present our conclusions in Section~4.

\section{The Navier-Stokes Equations}
%
%
The compact form of the Euler equations, when written in terms of the vector of conserved quantities
$\mathbf{U}$ \citep{tor09} is
\begin{equation}\label{eq:euler}
\partial_t\mathbf{U}+\nabla\cdot\mathbf{F}_\mathrm{adv}\left(\mathbf{U}\right)=0 ,
\end{equation}
with
\begin{equation}\label{eq:conserved}
\mathbf{U}=\left(
\begin{array}{c}
\rho\\ \rho\mathbf{v}\\ \rho e
\end{array}
\right)=\left(\begin{array}{c}
\rho\\ \rho u \\ \rho v \\ \rho w \\ \rho e
\end{array}
\right),
\end{equation}
and where
\begin{equation}\label{eq:ad_flux}
\begin{array}{l}
\mathbf{F}_\mathrm{adv}(\mathbf{U})=\left(\cfrac{}{}
\rho\mathbf{v}\;, \;\rho\mathbf{v}^T\mathbf{v}+P\mathbf{I}\;, \;(\rho e+P)\mathbf{v}
\cfrac{}{}\right)\\
\\
=
\begin{pmatrix}
\begin{array}{c}
\rho u \\ \rho v \\ \rho w
\end{array}
&
\left|
\begin{array}{c}
3\times3 \\
\text{{\large momentum }} \\ \text{{\large flux tensor}}
\end{array}
\right.
&
\left|
\begin{array}{c}
 (\rho e +P)u \\ (\rho e +P)v \\ (\rho e +P)w
\end{array}
\right.
\end{pmatrix}
\end{array}
\end{equation}
is the mass-momentum-energy flux density tensor ($3\times5$). The
operator $\nabla\cdot(\;)$ in Eq.~(\ref{eq:euler}) is a tensor
divergence, i.e.~in tensor notation we have
$\left\{\nabla\cdot\mathbf{F}_\mathrm{adv}\right\}^a=\partial_b{F_\mathrm{adv}}^{\,ba}$. The
momentum components in the conservative form of
Equation~(\ref{eq:euler}) represent a transfer of momentum,
owing merely to the mechanical transport of different particles of
fluid from place to place and to the pressure forces acting on the
fluid \citep[e.g.][]{landau}. In Eq.~(\ref{eq:euler}) we have made the
advective character of the fluxes explicit by denoting them
$\mathbf{F}_{\rm adv}$.

The internal friction present in any real fluid causes an irreversible
transfer of momentum from points where the velocity is large to those
where it is small. The momentum flux density tensor is thus altered
from its ideal from in Eq.~(\ref{eq:ad_flux}), where it only contains an
inertial and an isotropic component (described by a symmetric stress tensor
due to the local pressure $P$), to a modified expression that accounts
for an irreversible viscous transfer of momentum
\begin{equation}
\rho\mathbf{v}^T\mathbf{v}+P\mathbf{I}\longrightarrow \rho\mathbf{v}^T\mathbf{v}+P\mathbf{I}-\mathbf{\Pi},
\end{equation}
where $P\mathbf{I}-\mathbf{\Pi}$ is the total stress tensor and
$\mathbf{\Pi}$ is called the viscous stress tensor. The latter
includes the effects of isotropic compression and expansion forces
(``bulk viscosity'') as well as shearing forces (``shear viscosity'').

Similarly, the energy component of Eq.~(\ref{eq:ad_flux}) is affected by
the inclusion of the viscous stress tensor.  Because of the
dissipative nature of viscosity, a conservative formulation of the
NS equations must include a contribution of $\mathbf{\Pi}$
to the energy budget, i.e.~the work per unit area per unit time,
\begin{equation}
\left(\rho e+P\right)\mathbf{v}\longrightarrow \left(\rho e+P\right)\mathbf{v}-\mathbf{\Pi}\mathbf{v}
\end{equation}
needs to explicitly account for the work done by viscous forces.

A general parametrization of the viscous stress tensor $\mathbf{\Pi}$ is given by 
\begin{equation}\label{eq:viscous_tensorA}
\mathbf{\Pi}=\eta\left\{\left[\nabla \mathbf{v}+\left(\nabla\mathbf{v}\right)^T\right]
-\frac{2}{3}\mathbf{I}\left(\nabla\cdot\mathbf{v}\right)\right\}+\zeta\mathbf{I}\left(\nabla\cdot\mathbf{v}\right).
\end{equation}
Often, the viscous stress tensor is decomposed into a traceless part
and a diagonal part, such that the first corresponds to
constant-volume shear deformations (often called the {\it
  rate-of-deformation} tensor) and the second  to isotropic
expansions/contractions. Accordingly, $\eta$ in Eq.~(\ref{eq:viscous_tensorA}) is
commonly referred to as the {\it shear viscosity} and $\zeta$ as the
{\it bulk viscosity}.  The degree of resistance to uniform
contractions/expansions is  intrinsic to the
molecular/chemical properties of the fluid in question, and can be understood
through
kinetic theory. In this picture,  bulk viscosity arises
because kinetic energy of molecules is transferred to internal
degrees of freedom. 
Ideal monoatomic
gases (modeled as hard spheres interacting only through
elastic collisions)  have no internal degrees of freedom,
and are thus expected to have vanishing bulk viscosity.
At one time Stokes suggested that this might in general be true
(the so-called {\it Stokes'
  hypothesis} of $\zeta=0$) but
later wrote that he never put much faith in this relationship
\citep{gra07}. Indeed, when deviations from the ideal gas
equation of state are included in a hard-sphere, Chapman-Enskog
approach to kinetic theory, a non-zero value for the bulk viscosity is
obtained. In an extension of the hard sphere fluid model,
the Longuet-Higgins-Pople relation $\zeta=(5/3)\eta$ 
results \citep{mar02}, motivating the hypothesis that both viscosities are always related in a
linear fashion \citep[but see][]{mei05}. In general, we consider
$\eta$ and $\zeta$ as essentially arbitrary input properties to our
simulations, which may also depend on local physical parameters such
as temperature or density. Although the effects of physical bulk viscosity
are not harder to implement numerically than those of shear viscosity,
the physical origin of bulk viscosity is often less clear.
Also, we note that many numerical solvers for viscous flow focus on the
incompressible regime ($\nabla\cdot\mathbf{v}=0$), where the existence
of a physical bulk viscosity is in any case not of importance. However, for
compressible flow, the value of $\zeta$ may still become important in
certain situations. 

When the effects of viscosity are included, the formerly homogeneous differential equations of the Euler 
form (Eq.~\ref{eq:euler}) become
\begin{equation}\label{eq:euler_wsource}
\partial_t\mathbf{U}+\nabla\cdot\mathbf{F}_\mathrm{adv}\left(\mathbf{U}\right)=\mathbf{S}(\mathbf{U})
\end{equation}
where $\mathbf{S}(\mathbf{U})$ is a viscous source term given by
\begin{equation}\label{eq:source}
\mathbf{S}(\mathbf{U})\equiv\nabla\cdot \left(\cfrac{}{}
\mathbf{0}\;, \;\mathbf{\Pi}\;, \;\mathbf{\Pi}\mathbf{v}
\cfrac{}{}\right).
\end{equation}

The solution of the Euler equations with source terms is often handled
by operator-splitting methods \citep[e.g.][]{tor09,lev02}. That is,
the numerical scheme alternates between an advective step that solves
the homogeneous part, and a source-term step. Thus, the solution of
Eq.~(\ref{eq:euler_wsource}) is split into a two stage problem:
\begin{eqnarray}
\left.\begin{array}{ll}
\mathrm{PDE}:&\partial_t\mathbf{U}+\nabla\cdot\mathbf{F}_\mathrm{adv}\left(\mathbf{U}\right)=0\\
\mathrm{IC}:&\mathbf{U}(\mathbf{x},t)=\mathbf{U}^n
\end{array}\right\}\Rightarrow \widetilde{\mathbf{U}}^{n+1}&&\\ 
\left.\begin{array}{ll}
\mathrm{ODE}:&\frac{d}{dt}\mathbf{U}=\mathbf{S}(\mathbf{U})\\
\mathrm{IC}:&\widetilde{\mathbf{U}}^{n+1}
\end{array}\right\}\Rightarrow\mathbf{U}^{n+1}&&.
\end{eqnarray}
Typically, the source terms are more easily written in the primitive
variable formulation of the Euler equations. A common choice of the primitive-variable vector 
is $\mathbf{W}=(\rho,\mathbf{v},P)^T=(\rho,v_x,v_y,v_z,P)^T$, which we also adopt here. For sources
corresponding to the NS viscous terms (Eq.~\ref{eq:source}), only the $\mathbf{v}$
component of $\mathbf{W}$ is affected, thus simplifying the solution method of the source-term step.
The three-dimensional Euler equations  can be written in the primitive
variable form as \citep{tor09} 
\begin{equation}\label{eq:primitive3}
\partial_t\mathbf{W}+\mathbf{A}_1(\mathbf{W})\,\partial_x\mathbf{W}
+\mathbf{A}_2(\mathbf{W})\,\partial_y\mathbf{W}
+\mathbf{A}_3(\mathbf{W})\,\partial_z\mathbf{W}=\mathbf{0}.
\end{equation} 
 For this choice of variables, the coefficient matrices are given by \citep{tor09}
\begin{gather}
\mathbf{A}_1(\mathbf{W})=\begin{pmatrix}
v_x&\rho&0&0&0\\
0&v_x&0&0&1/\rho\\
0&0&v_x&0&0\\
0&0&0&v_x&0\\
0&\gamma P&0&0&v_x
\end{pmatrix},\;\;\;\;
\\
\mathbf{A}_2(\mathbf{W})=\begin{pmatrix}
v_y&0&\rho&0&0\\
0&v_y&0&0&0\\
0&0&v_y&0&1/\rho\\
0&0&0&v_y&0\\
0&0&\gamma P&0&v_y
\end{pmatrix},\;\;\;\;
\\
\mathbf{A}_3(\mathbf{W})=\begin{pmatrix}
v_z&0&0&\rho&0\\
0&v_z&0&0&0\\
0&0&v_z&0&0\\
0&0&0&v_z&1/\rho\\
0&0&0&\gamma P&v_z
\end{pmatrix},\;\;\;\;
\end{gather}
which is exactly equivalent to the familiar equations
\begin{subequations}
\begin{align}
\frac{\partial\rho}{\partial t}+\frac{\left(\partial{\rho v_i}\right)}{\partial x_i}=0~~\\
\frac{\partial v_i}{\partial t}+v_i\frac{\partial v_i}{\partial x_i}+\frac{1}{\rho}\frac{\partial P}{\partial x_i}=0~~\\
\frac{\partial P}{\partial t}+\gamma P \frac{\partial v_i}{\partial x_i}+v_i\frac{\partial P}{\partial x_i}=0~~.
\end{align}
\end{subequations}
In this formulation, the viscous terms of the NS equations, which affect only the velocity, are
\citep[e.g.][]{landau}  
\begin{equation}\label{eq:viscous_source}
\mathbf{S}(\mathbf{W})=\frac{1}{\rho}\begin{pmatrix}
0 \\ \eta \Delta \mathbf{v} +\left(\zeta+\frac{1}{3}\eta\right)\nabla\left(\nabla\cdot\mathbf{v}\right)\\ 0 
\end{pmatrix}.
\end{equation}

An alternative to expressing the viscosity effects as source terms is to absorb them directly
into the flux divergence, 
\begin{equation}\label{eq:navier}
\partial_t\mathbf{U}+\nabla\cdot\left[\cfrac{}{}\mathbf{F}_\mathrm{adv}\left(\mathbf{U}\right)
-\mathbf{F}_\mathrm{diff}\left(\mathbf{U}\right)\right]=0,
\end{equation}
which highlights the still conservative character of the NS equations.
Here diffusive fluxes, defined by
\begin{equation}\label{eq:viscous_fluxes}
\mathbf{F}_\mathrm{diff}(\mathbf{U})= \left(\cfrac{}{}
\mathbf{0}\;, \;\mathbf{\Pi}\;, \;\mathbf{\Pi}\mathbf{v}
\cfrac{}{}\right),
\end{equation}
are responsible for the effects of viscosity.  An implementation of
the diffusive fluxes in this conservation-law form is clearly the
preferred choice for FVM schemes, which are specifically designed for
solving the integral form of these conservation laws. In fact, in this
case they {\em exactly} conserve all the involved quantities to
machine precision. We will therefore focus on this method in our
study. The central aspect will be the numerical scheme used
for estimating the velocity gradients at the cell interfaces, and
hence the discretization of the diffusive fluxes. In the next section,
we describe our approach for this in detail.

\section{A Finite Volume Scheme with Viscous Fluxes on a Voronoi Mesh}\label{sec:finite_volume}
\subsection{Basic MUSCL-Hancock Finite Volume Scheme: Overview }\label{sec:muscl_overview}
Finite volume methods enforce the integral form of the conservation
laws on discrete meshes. This approach is manifestly conservative,
since fluxes of quantities that leave a cell simply enter the neighboring cell. 
The NS equations in finite-volume form are
\begin{eqnarray}
\frac{{\rm d}\mathbf{Q}_{i}}{{\rm d}t}=-\sum_jA_{ij}\mathbf{F}_{ij}~~,
&\text{     with      }&
\mathbf{Q}_{i}=\int_{V_i}\mathbf{U}_{i}{\rm d}V~~,
\end{eqnarray}
where, in general, the intercell fluxes contain both advective and diffusive contributions,
\begin{equation}
\mathbf{F}_{ij}=\mathbf{F}_{\mathrm{adv},ij}-\mathbf{F}_{\mathrm{diff},ij}.
\end{equation}

The scheme used by \arepo\ is the finite volume MUSCL-Hancock
approach, consisting of a MUSCL (Monotone Upstream-centered Schemes
for Conservation Laws) linear reconstruction stage, and a Hancock
two-stage time integration
\begin{equation}\label{eq:finite_vol}
\mathbf{Q}_i^{n+1}=\mathbf{Q}_i^{n}-\Delta t\sum_j A_{ij}\hat{\mathbf{F}}_{ij}^{n+1/2},
\end{equation}
where the numerical fluxes $\hat{\mathbf{F}}_{ij}^{n+1/2}$ represent
appropriately time-averaged approximations to the true flux
$\mathbf{F}_{ij}$ across the interface shared by cells $i$ and $j$.
The time label $n+1/2$ in Eq.~(\ref{eq:finite_vol}) indicates that an
intermediate-stage (a half time-step evolution) has been performed to
obtain the numerical estimate of $\mathbf{F}_{ij}$, meaning that the
time-stepping in Eq.~(\ref{eq:finite_vol}) uses time-centered fluxes,
giving it second-order accuracy. The Hancock part of the scheme is a
two-step approach (the familiar predictor-corrector algorithm) in
which the correction half-step is obtained from the solution of the 1-D
Riemann problem across each face of the control volume.
The general finite volume MUSCL-Hancock scheme has hence the following three steps \citep{tor09}:

\paragraph*{(I) Gradient Estimation, Linear Data Reconstruction and Boundary Value Extrapolation}

Once a local gradient estimate for the conserved quantities
$\mathbf{U}_i=(\rho,\rho\mathbf{v},\rho e)_i$ of cell $i$ is
available, linear data reconstruction takes the form
\begin{equation}
\begin{split}
\mathbf{U}_{ij}^L&=\mathbf{U}_{i}+\nabla\mathbf{U}_i^{n}(\mathbf{f}_{ij}-\mathbf{s}_i) \\
\mathbf{U}_{ij}^R&=\mathbf{U}_{j}+\nabla\mathbf{U}_j^{n}(\mathbf{f}_{ji}-\mathbf{s}_j)
\end{split}
\end{equation}
where we denote by $\mathbf{U}_{ij}^L$ the estimated vector of
conserved variables at the centroid of the $ij$-interface, obtained by
linearly extrapolating the cell-centered values $\mathbf{U}_{i}$ of
the $i$-th cell (on the ``left'' side) from $\mathbf{s}_i$, the cell's
center position, to $\mathbf{f}_{ij}$. Similarly, $\mathbf{U}_{ij}^R$
corresponds to the estimates of the face-centroid values obtained by
linear extrapolation of the cell-centered values of the $j$-th cell
(the ``right'' side), whose center position is $\mathbf{s}_j$.  In
both cases, $\mathbf{f}_{ij}=\mathbf{f}_{ji}$ is the position vector
of the face centroid between the cells. The Jacobian
$\nabla\mathbf{U}_i^{n}$ is explicitly labeled with superscript $n$ to
point out that it corresponds to the estimate of spatial derivatives
at the beginning of the time-step.

\paragraph*{(II) Evolution of Boundary Extrapolated Values}
This is, strictly speaking, the ``predictor" half time-step. The conserved
variables are evolved for $\Delta t/2$ with flux estimates obtained
from the values at the beginning of the time-step:
\begin{equation}
\begin{split}
\widehat{\mathbf{U}}_{ij}^L&=\mathbf{U}_{ij}^L-\frac{\Delta t}{2}\frac{1}{V_i}
\sum_j A_{ij}\mathbf{F}_{ij}^{n}\\
\widehat{\mathbf{U}}_{ij}^R&=\mathbf{U}_{ij}^R-\frac{\Delta t}{2}\frac{1}{V_j}
\sum_j A_{ij}\mathbf{F}_{ji}^{n}
\end{split}
\end{equation}

\paragraph*{(III) Solution of 1-D Riemann Problems and Computation of Godunov Fluxes}
This corresponds to the ``corrector'' half time-step in the two-stage
Hancock approach. Once the values to the right and left of the
interface at time $\Delta t /2$ are known, the discontinuity is
treated as a one-dimensional Riemann problem. An exact or approximate
Riemann solver is used to return values of $\rho$, $\rho\mathbf{v}$
and $\rho e$ at the interface, at a time corresponding to
$n+1/2$. From these values, the advective fluxes can be directly
computed (Eq.~\ref{eq:ad_flux}). These are time-centered fluxes
${\mathbf{F}}_{ij}^{n+1/2}$ used to update the system from the
beginning of the time-step to its end,
\begin{equation}
{\mathbf{U}}_{ij}^{n+1}=\mathbf{U}_{ij}^n-\Delta t\frac{1}{V_i}\sum_j A_{ij}{\mathbf{F}}_{ij}^{n+1/2}.
\end{equation}

\begin{figure*}
\includegraphics[width=0.95\textwidth]{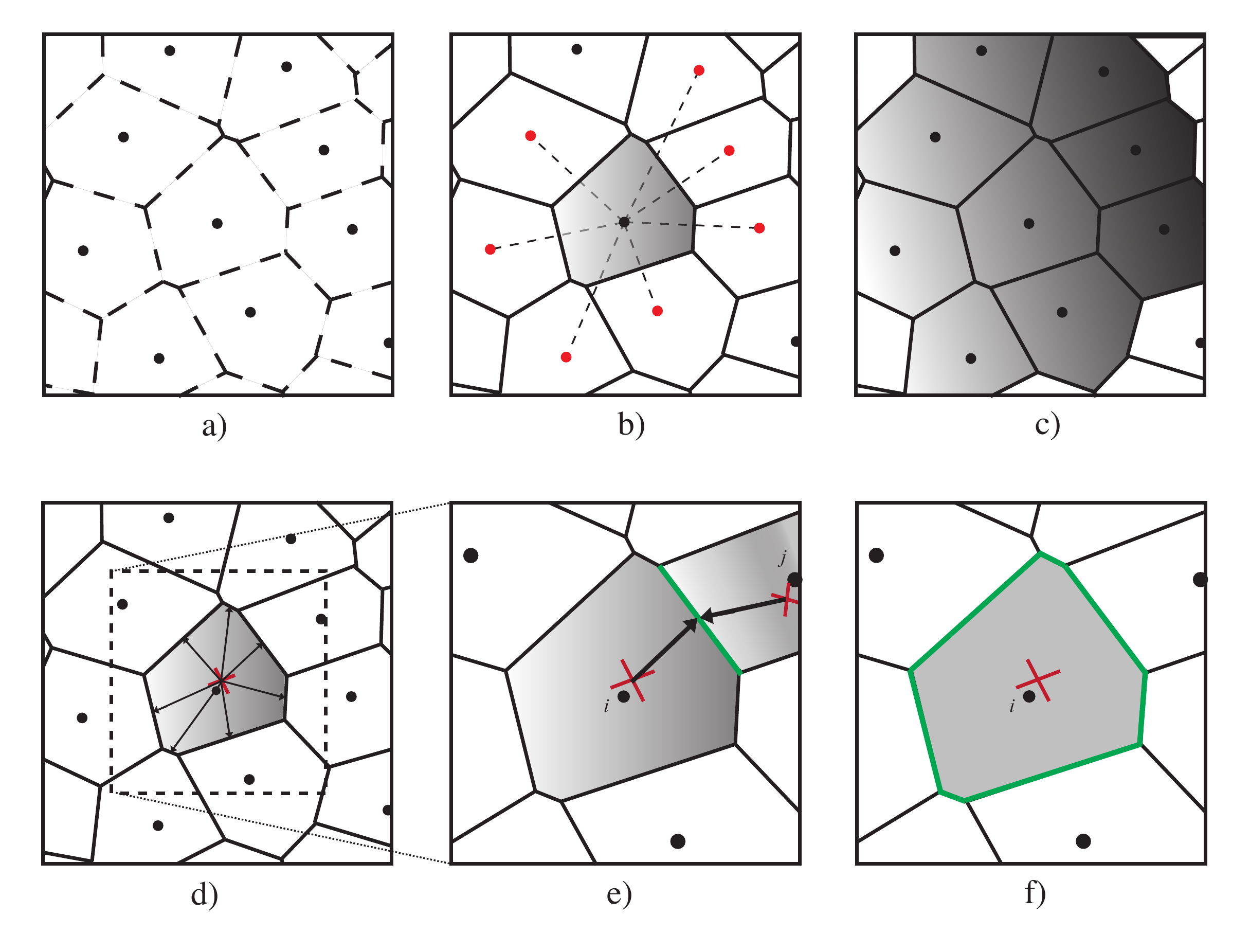}
\vspace*{-10pt}
\caption{Schematic representation of the mesh geometry and the
MUSCL-Hancock integration scheme implemented in {\small AREPO}:
a) TheVoronoi mesh is uniquely determined by the location of the
mesh-generating points. b) A gradient estimate for all primitive variables is
obtained from the immediate neighbors of a given cell. c)
The gradient-estimation process is repeated for each cell in the
domain and thus a piece-wise linear reconstruction is obtained for
each primitive variable.  d) The primitive variables are
extrapolated toward each interface and evolved for half a time-step.
e) For each face, a pair of extrapolated quantities for two
neighboring cells $i$ and $j$ forms a local Riemann problem. f) The
Riemann problem is solved for each face of a cell, yielding 
time-centered Godunov fluxes for the entire boundary of
the control volume $V_i$ of cell $i$.  These fluxes are used for updating the
conserved quantities of the cell through  Eq.~(\ref{eq:finite_vol}).\label{fig:arepo_scheme}}
\end{figure*}

\begin{figure*}
\centering
\includegraphics[width=16cm]{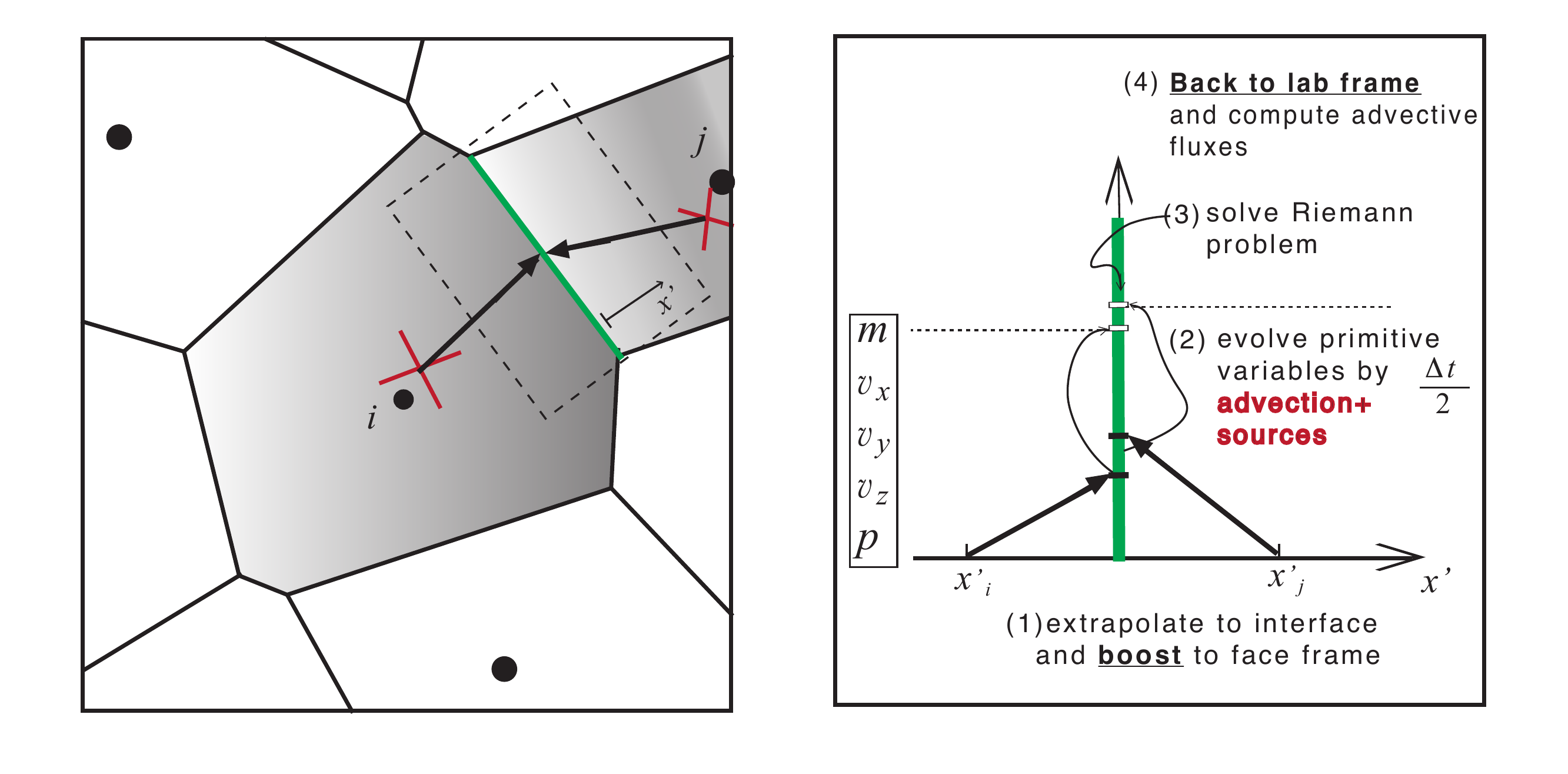}
\caption{Detailed description of the flux calculation with a Riemann
  solver in step e) of Fig.~\ref{fig:arepo_scheme}. For the case of a
  moving mesh, the standard MUSCL-Hancock method needs to be augmented
  with Galilean-boosts, as described by \citet{spr10}: (1) The
  extrapolation towards each interface is followed by a Galilean boost
  of the velocities to the rest frame of the face, and by a rotation
  of the coordinate axes. Each face is then treated as a
  one-dimensional discontinuity. Thus, the axes are oriented in the
  rotated frame such that the $x'$-axis coincides with the normal to
  the face (left panel). (2) The primitive variables in the moving
  frame are evolved for half a time-step, including source terms
  if present (e.g. gravity or viscosity). (3) A one dimensional
  Riemann problem is solved at the interface. (4) The velocities are
  translated back to the lab frame and the advective fluxes are
  computed.}
\label{fig:arepo_scheme_riemann}
\end{figure*}

Figures~\ref{fig:arepo_scheme} and \ref{fig:arepo_scheme_riemann}
illustrate the mesh geometry and the basic steps of this inviscid
numerical scheme implemented in \arepo. One additional point we have
not explicitly discussed here for simplicity is the treatment of
the mesh motion, as indicated in Fig.~\ref{fig:arepo_scheme_riemann}.
This is incorporated into the scheme by evaluating all fluxes in the
rest frame of the corresponding face, as described by
\citet{spr10}. This requires appropriate boosts of the fluid states
and the fluxes from the lab frame to the rest frame of each face, and
back. For a Voronoi mesh, the face velocities are fully specified by
the velocities of all the mesh generating points. The latter can be
chosen freely in principle, but if they are set equal to the fluid
velocities of the corresponding cells, a Lagrangian behavior and a
manifestly Galilean-invariant discretization scheme is obtained in which the 
truncation error does not depend on the bulk velocity of the system.

\subsection{A MUSCL-Hancock Finite-Volume Scheme with Viscous Terms}

A cell-centered, finite-volume solution of the NS equation can be written as
\begin{equation}\label{eq:finite_vol2}
\mathbf{Q}_i^{n+1}=\mathbf{Q}_i^{n}-\Delta t\sum_j A_{ij}\hat{\mathbf{F}}_{\mathrm{adv},ij}^{n+1/2}
-\Delta t\sum_j A_{ij}\hat{\mathbf{F}}_{\mathrm{diff},ij}^{n+1/2},
\end{equation}
where we have retained the distinction between advective and viscous
fluxes. As in the case of the Euler equations, 
the numerical method essentially consists of the problem of
finding accurate time-centered numerical fluxes across each  of the
interfaces of a cell. How to do this in detail for the diffusive part
of the fluxes has been the focus of numerous efficiency and stability
analyses \citep[see][for a detailed description]{pui10}.

Eq.~(\ref{eq:finite_vol2}) uses time-centered fluxes, obtained here
with the two-step Hancock technique, as described
above. Thus, for estimating both
$\hat{\mathbf{F}}_{\mathrm{adv},ij}^{n+1/2}$ and
$\hat{\mathbf{F}}_{\mathrm{diff},ij}^{n+1/2}$ a half time-step
predictor stage is required. In the MUSCL-Hancock approach for
inviscid flow, this step is carried out by linear reconstruction from
each cell center to the interface, followed by solving a
one-dimensional Riemann problem at the interface where the
extrapolations meet. The traditional formulation of the Riemann
problem and its solution are exclusive to hyperbolic differential
equations and thus do not provide exact solutions for the
NS equations. Since a general solution for the viscous
Riemann problem does not exist, we will treat the viscous fluxes in
Eq.~(\ref{eq:finite_vol2}) as a correction to the solution of an
otherwise inviscid flow.

Our NS version of the MUSCL-Hancock scheme consists of the
following three different stages (in addition to those described in Section.~\ref{sec:muscl_overview}):
\begin{itemize}
\item[(A)] {Correct the MUSCL linear extrapolation of primitive variables by applying a viscous kick.}
\item[(B)] {Extrapolate the cell-centered gradients linearly and evolve them for half a time-step.}
\item[(C)] {Average the extrapolated velocity gradients at the interface and use them to estimate viscous fluxes.}
\end{itemize}

To extrapolate the gradients from their cell-centered values to the
interfaces, information about the higher-order derivatives of the
primitive variables is needed. If gradients are assumed to vary
linearly in space, an estimator for the Hessian matrix for each of the
five primitive variables is sufficient. Evidently, enough information
is contained in the cell-centered quantities to estimate both the
local gradient $\nabla\phi$ and the Hessian $\mathbf{H}^\phi$
corresponding to a given scalar quantity $\phi$. However, estimating
both of these simultaneously is significantly more difficult than
estimating them one after the other. Therefore, we will effectively
treat $\phi$ and $\nabla\phi$ as two independent fields that vary
linearly in space, and this variation needs to be estimated from the
mesh data through a suitably discretized differential operator.

As a simpler alternative to the gradient reconstruction approach, we
briefly describe how one can use the gradients already available from
the linear reconstruction step. In this approximation, a given
quantity varies only linearly within the control volume, such that
consistently evaluated gradients are piece-wise constant. This means
that each interface represents a discontinuity in the gradient field
$\nabla\phi$. Naively, one may think that the arithmetic average of
both gradients that meet at a face is a good estimate for the gradient
at the interface itself. However, on second thought, one realizes that
both cells do no necessarily have the same weight if cells of
different volume meet. Furthermore, the unweighted average of the two
cell-centered values really represents the value at the midpoint of
the two mesh-generating points, which, for a Voronoi mesh, can be
substantially offset from the mid-point of the face. We therefore
adopt the approach of \citet{loh07}, which consists in choosing one of
the two gradients that meet at the interface, based on prior knowledge
of the direction of the flow across the interface. Thus the
three-stage scheme introduced above could be alternatively replaced by
the simpler method:
\begin{itemize}
\setlength{\itemindent}{33pt}
\item[(A'- C')]At the cell interface where two different gradients meet, choose the upwind gradient.
\end{itemize}
In either method, once we have an estimate of both viscous and
advective fluxes, the time-step evolution of the conserved quantities
$\mathbf{Q}_i$ is carried out as in Eq.~(\ref{eq:finite_vol2}).
However, the approach (A-C) is preferable to the \citet{loh07} scheme
because it uses time-centered estimates for both
$\hat{\mathbf{F}}_{\mathrm{adv},ij}^{n+1/2}$ and
$\hat{\mathbf{F}}_{\mathrm{diff},ij}^{n+1/2}$, hence preserving the
order of accuracy of the original inviscid scheme.  We therefore now
provide a more detailed description of the individual steps in this
three-stage approach.

\subsubsection*{(A) Viscosity Kicks}\label{sec:viscous_kick}

Although Eq.~(\ref{eq:finite_vol2}) is written in an unsplit form, the
predictor step is indeed operator split, evolving the advective and
diffusive terms separately \citep[e.g.][]{coi96}.  While our method for
estimating the advective fluxes remains the MUSCL-Hancock scheme, the
technique for estimating the diffusive fluxes is essentially contained
in the estimation of the velocity gradients at each interface
\citep[see][for a series of tests on different interface gradient
estimates]{coi94,pui10}. Looking for better accuracy, 
we have chosen to couple these two otherwise independent procedures by
correcting/biasing the linear extrapolation of the velocity field
(stage $(I)$ in Section~\ref{sec:muscl_overview}) with a viscous
source term.

The benefit of carrying out a linear extrapolation to cell interfaces
in primitive variables is the simplicity of the Galilean
transformation needed to boost the quantities to the frame of a moving
interface. Since the Galilean boost does not affect the mass and
pressure of a given cell, only the local velocity field is
transformed. In addition, adding force source terms to the equations
of motion in primitive variable formulation is simpler, since these
only couple to the momentum equations. Thus, a ``viscous kick" can
be applied to the velocity field in the half time-step evolution stage:
\begin{equation}\label{eq:visc_kick}
\Delta\mathbf{v}_\mathrm{visc}=\frac{\Delta t}{2}
\left[\frac{\eta}{\rho} \nabla^2 \mathbf{v} +
\frac{\zeta+\frac{1}{3}\eta}{\rho}\nabla
\left(\nabla\cdot\mathbf{v}\right)\right].
\end{equation}
In this way, the subsequent linear extrapolation of primitive
variables will already include viscosity effects to first order in
time.

While working with numerical fluxes across interfaces requires
velocity gradients, the use of cell-centered source terms in
Eq.~(\ref{eq:visc_kick}) calls for second order derivatives of the
velocity field.  Thus, in addition to the cell-centered velocity
gradients $\nabla v_x$, $\nabla v_y$ and $\nabla v_z$, the
cell-centered Hessian matrices $\mathbf{H}^{v_x}$, $\mathbf{H}^{v_x}$
and $\mathbf{H}^{v_x}$ are now needed.  As we will see below, these
matrices will be of use in more than one occasion, justifying the computational
cost incurred to calculate them.

\subsubsection*{(B) Linear Extrapolation of Gradients}

The linear reconstruction implemented in our MUSCL-Hancock approach
essentially assumes that the gradient of a scalar quantity $\phi$ does
not vary significantly across the spatial scale of a cell. For smooth
flows, the gradients of two neighboring cells $\nabla\phi\Big|_i$ and
$\nabla\phi\Big|_j$ will not differ significantly. Furthermore, in the
presence of strong discontinuities, gradients on each side will be
slope-limited, and therefore will not differ by much
either. Hence, a first guess for the gradient at the interface
between two cells is just the average of the cell-centered estimates
at each side of the face
\begin{equation}\label{eq:grad_estimate0}
\widetilde{\nabla\phi}\Big|_{ij}=\frac{\langle\nabla\phi\rangle_i+\langle\nabla\phi\rangle_j}{2}.
\end{equation}
However, as we pointed out earlier, the gradient average above is
actually representative of the midpoint between the two cell centers
$\mathbf{r}_i$ and $\mathbf{r}_j$, which in general does not lie close
to the center of the face in a Voronoi mesh, and may in fact lie
within a third cell.  Unless gradients are assumed to vary within a
cell, it will not be possible to assign the estimate to the center of
the interface with any confidence.

\begin{figure*}
\centering
\includegraphics[width=17cm]{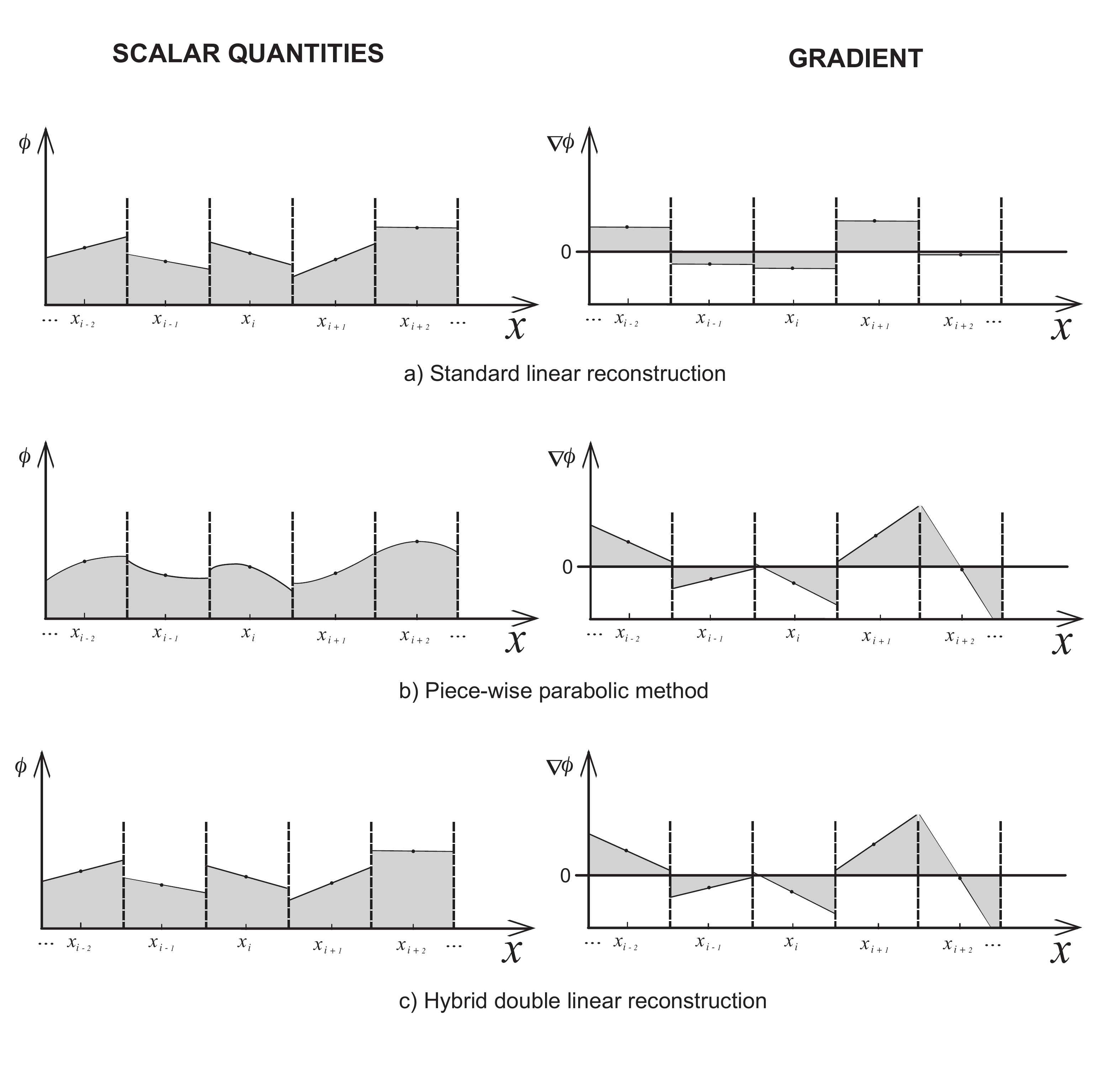}
  \vspace*{-15pt}
\caption{Schematic representation of the double linear reconstruction
proposed in this work compared to standard linear reconstruction and parabolic
reconstruction.\label{fig:gradient_reconstruction}}
\end{figure*}

Let us assume that the scalar field $\phi(\mathbf{r})$ is infinitely
differentiable and, consequently, so is its first derivative. Thus, we
can Taylor expand both quantities to arbitrary order around a mesh
generating point $\mathbf{r}_0$:
\begin{align}
\begin{split}
\phi(\mathbf{r})=&\phi(\mathbf{r}_0)+\nabla\phi\Big|_{\mathbf{r}_0}(\mathbf{r}-\mathbf{r}_0) \\
&+\frac{1}{2}(\mathbf{r}-\mathbf{r}_0)^T\mathbf{H}^{\phi}\Big|_{\mathbf{r}_0}(\mathbf{r}-\mathbf{r}_0)+\mathcal{O}(\mathbf{d}^3)
\end{split}\\
\begin{split}
 \nabla\phi(\mathbf{r})=&\nabla\phi\Big|_{\mathbf{r}_0}+\mathbf{H}^{\phi}\Big|_{\mathbf{r}_0}(\mathbf{r}-\mathbf{r}_0) \\&+
 \frac{1}{2}(\mathbf{r}-\mathbf{r}_0)^T\mathbf{D}^{\phi}\Big|_{\mathbf{r}_0}(\mathbf{r}-\mathbf{r}_0)+\mathcal{O}(\mathbf{d}^3)
\end{split}
\end{align}
where $\mathbf{H}^{\phi}$ is the Hessian matrix of the scalar quantity
$\phi$ and $\mathbf{D}^{\phi}$ is a $3\times3\times3$ tensor
containing the third-order derivatives of $\phi$
(i.e.~$D_{abc}=\partial^3\phi/\partial x_a\partial x_b\partial
x_c$). Truncating both Taylor expansions to first-order in
$\mathbf{d}=\mathbf{r}-\mathbf{r}_0$, we see that we can obtain linear
reconstructions for both the physical quantities and their gradients
provided that we have numerical estimates for both the gradients and
the Hessians at each mesh generating point.

We emphasize that a Taylor expansion is not equivalent to a
polynomial data reconstruction.  Indeed, it is desirable that
reconstruction schemes are manifestly conservative, in the sense that
the average of the reconstruction over the cell should be identical to
the value of $\phi$ at the geometric center of the cell. This property
of reconstruction schemes is sometimes referred to as $K$-exactness,
meaning that if a polynomial reconstruction is cell-averaged over the
mesh, the reconstruction procedure recovers the same polynomial.  This
condition is trivially satisfied for a linear reconstruction of the
form
$\phi(\mathbf{r})=\phi_i+\langle\nabla\phi\rangle_i(\mathbf{r}-\mathbf{s}_0)$. However,
higher-order reconstruction schemes require the use of zero-mean
polynomials, which, beyond first-order, differ from the Taylor series
\citep[e.g.][]{col84,coi96}.

The linear reconstruction of the scalar field $\phi$ and of the vector
field $\nabla\phi$, treated as if they were independent quantities,
effectively constitutes a hybrid method between standard linear
reconstruction and fully K-exact second-order reconstruction, as
illustrated in Figure~\ref{fig:gradient_reconstruction}. In this
approximation, second derivatives are considered negligible for the
spatial reconstruction of the primitive quantities, but they are still
included for a more accurate estimate of the gradients near the cell
interfaces. We also note, that in this way our numerical scheme
reduces to that originally in \arepo\ (which is second-order-accurate)
when the viscous fluxes are disabled.

Once an estimate for the Hessian matrix
$\mathbf{H}^{\phi}\Big|_{\mathbf{r}_0}$ is available (Section~\ref{sec:hessians}), a linear
extrapolation of the gradients from the cell centers to the interfaces
can be obtained from
\begin{equation}\label{eq:grad_estimate1}
\widetilde{\nabla\phi}\Big|_{ij}=\langle\nabla\phi\rangle_i+\langle\mathbf{H}^{\phi}\rangle
(\mathbf{f}_{ij}-\mathbf{r}_i),
\end{equation}
which is a better approximation than Eq.~(\ref{eq:grad_estimate0}).
However, the time evolution of the gradients during a single step
could be equally important as their spatial variation over the length
scale of a cell, hence we also need to evolve them for half a
time-step to obtain a time integration scheme that is consistent
with the second-order accurate two-stage MUSCL-Hancock approach. In
the latter, to extrapolate and evolve a scalar quantity $\phi$ we consider
\begin{equation}\label{eq:primitive_extrapolation}
\phi\Big|_{ij}=\phi_i+\nabla\phi\Big|_{\mathbf{r}_0}\left(\mathbf{f}_{ij}-\mathbf{s}_i\right)
-\frac{\Delta t}{2}\left\langle\frac{\partial\phi}{\partial t}\right\rangle_i
\end{equation}
where the time derivative of the quantity $\phi$ in the control volume
of the $i$-th cell can be obtained from the primitive variable formulation
of the Euler equations in tensor notation:
\begin{equation}\label{eq:euler_primitive}
\partial_t W_\alpha +A_{\alpha\beta b}(\mathbf{W})\partial_b W_\beta=0 .
\end{equation} 
Here sums over repeated indices
are implied.
Latin indices $a,b,c,d...$ take the values $1,2,3$ or $x,y,z$,
while Greek indices $\alpha,\beta,\gamma,...$ take the values
$1,2,3,4,5$ and are used to number the components of the primitive
quantity vector ($W_\alpha=\rho,v_x,v_y,v_z,P$ for $\alpha=1,2,3,4,5$, respectively). 
As with our previous notation, the indices $i,j$ and $k$ are reserved for
labeling the mesh generating points and their associated cells.

Eq.~(\ref{eq:euler_primitive}) is an advection equation for the
primitive variables.  Analogously, to ``advect" the gradients of the
primitive variables from the cell center to the interface, we can
ignore the viscous terms and derive an equation of motion for the
spatial derivatives by differentiating Eq.~(\ref{eq:euler_primitive}):
\begin{equation}
\partial_a\partial_tW_\alpha+\left(\partial_a A_{\alpha\beta b}\right)\partial_bW_\beta+A_{\alpha\beta b}\partial_a\partial_bW_\beta=0,
\end{equation} 
where we can identify the Jacobian matrix of the primitive variables
as $J_{\alpha a}\equiv\partial_aW_\alpha=W_{\alpha,a}$, and the
Hessian tensor ($5\times3\times3$) of the primitive variables as
$H_{\beta
  ba}\equiv\partial_b\partial_aW_\beta=W_{\beta,b,a}$. Therefore, the
time derivative of each component of the primitive variable Jacobian
matrix is
\begin{equation}\label{eq:euler_derivatives}
\partial_tJ_{\alpha a}=B_{\alpha\beta ba}J_{\beta b}-A_{\alpha\beta b} H_{\beta b a},
\end{equation} 
where we introduced the rank-$4$ tensor $B_{\alpha\beta ba}\equiv
\partial_a A_{\alpha\beta b}=A_{\alpha\beta b,a}$. Since
$A_{\alpha\beta b}$ is a function of the primitive variables
$W_\alpha$, the tensor $B_{\alpha\beta ba}$ can also be written as
(see Appendix)
\begin{equation}
B_{\alpha\beta ba}=\frac{\partial A_{\alpha\beta b}}{\partial
  W_\gamma}\partial_a W_\gamma ,
\end{equation}
and therefore its numerical estimate is given by the product of the
exact derivatives $\partial A_{\alpha\beta b}/\partial W_\gamma$
(evaluated with values of the primitive variables at the center of the
cell) and the (already available) numerical estimates for the gradients $\partial_a
W_\gamma=J_{\gamma a}$. The second term on the right hand side of
Eq.~(\ref{eq:euler_derivatives}) is the product of the known
coefficients $A_{\alpha\beta b} $ (evaluated at the center of the
cell) and the numerical estimates of the Hessian tensor $H_{\beta b
  a}$.

Finally, with a numerical estimate of $H_{\beta b a}$ at hand (see
Section~\ref{sec:hessians}), the extrapolated and half time-step
evolved gradients of the velocity are (in analogy to
Eq.~\ref{eq:primitive_extrapolation}):
\begin{equation}\label{eq:velocity_extrapolation}
\nabla v_x\Big|_{ij}=\langle \nabla v_x\rangle_i+\langle\mathbf{H}^{v_x}\rangle_i\left(\mathbf{f}_{ij}-\mathbf{s}_i\right)
+\frac{\Delta t}{2}\left\langle\frac{\partial \nabla v_x}{\partial t}\right\rangle_i,
\end{equation}
with analogous expressions for $\nabla v_y\,|_{ij}$ and $\nabla
v_z\,|_{ij}$. In Eq.~(\ref{eq:velocity_extrapolation}), the term
$\left\langle\partial \nabla v_x/\partial t\right\rangle_i$
is obtained from Eq.~(\ref{eq:euler_derivatives}) with $\alpha=2$
and $a=1,2,3$.

\begin{figure}
 \vspace*{20pt}
\includegraphics[width=9cm]{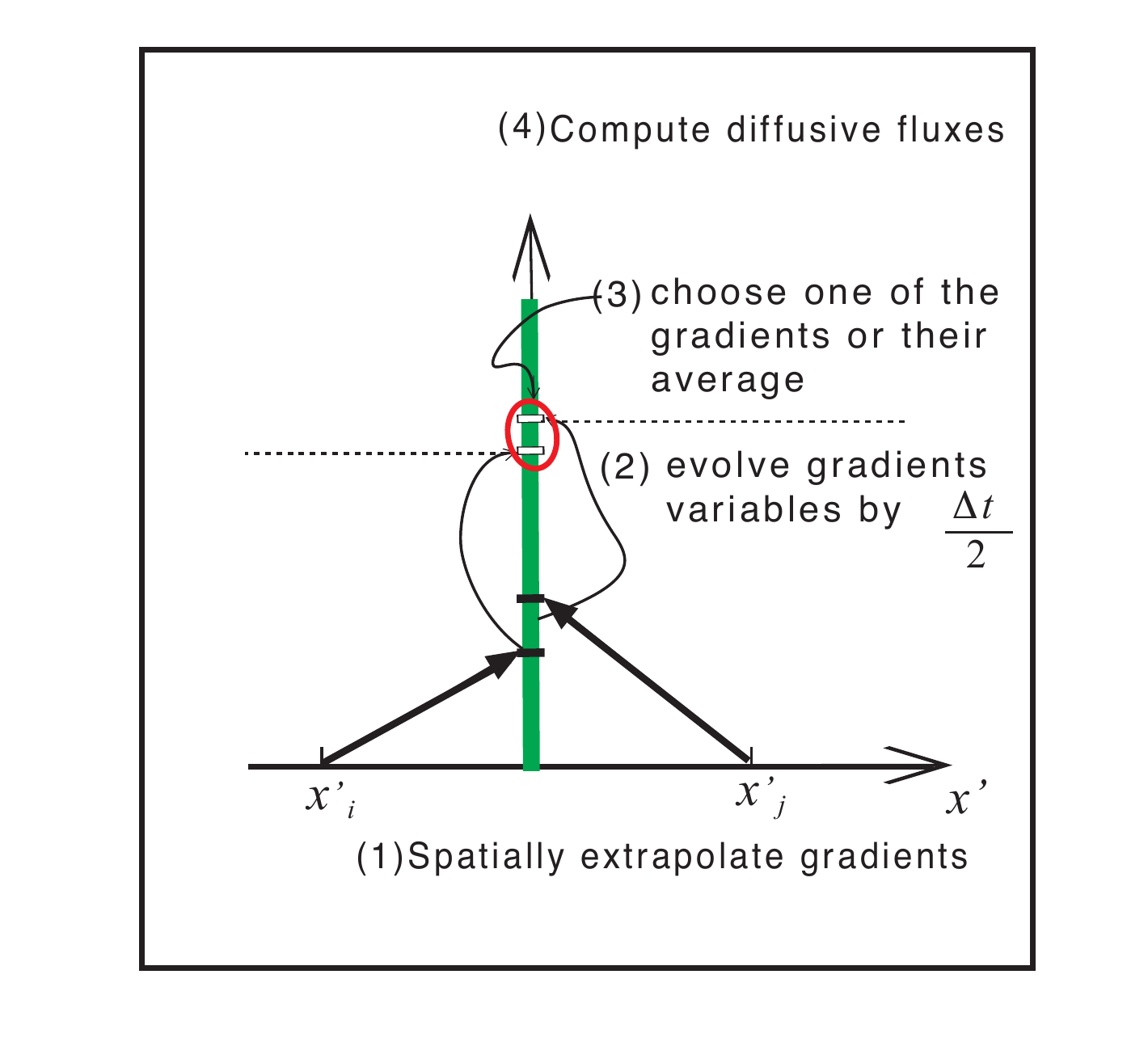}
\caption{Sketch illustrating the individual steps involved in the
  extrapolation and half time-step evolution of the gradients,
  analogous to the advective flux calculation shown in
  Fig.~\ref{fig:arepo_scheme_riemann}.  The different steps are: (1)
  spatial extrapolation of the gradients, followed by (2) a time
  advance by $\Delta t/2$ according to
  Eq.~(\ref{eq:primitive_extrapolation}), and (3) an approximate
  evaluation right at the interface. In step (4), the viscous fluxes
  are determined by evaluating Eq.~(\ref{eq:viscous_fluxes}) with the
  values of the primitive variables and the velocity gradients at the
  interface.\label{fig:gradient_extrapolation}}
   \vspace*{0pt}
\end{figure}

In Fig.~\ref{fig:gradient_extrapolation}, we show a sketch of the
different steps involved in obtaining time-centered diffusive fluxes.
We point out that taking the Hessian matrices of the velocity field to
be identically zero is $not$ equivalent to the alternative scheme
$(A')$.  The third term to the right hand side of
Eq.~(\ref{eq:velocity_extrapolation}) is still different from zero
even if $H_{\beta ba}=0$ (Eq.~\ref{eq:euler_derivatives}) since, in
general, $B_{\alpha\beta ba}J_{\beta b}\neq0$. By advecting the
gradients according to Eq.~(\ref{eq:euler_derivatives}) we gain
additional accuracy at no additional computational expense because the
terms $B_{\alpha\beta ba}J_{\beta b}$ are known exactly (see
Appendix), given the values of the primitive variables and their
respective gradients at the center of each cell.

\subsubsection*{(C) Viscous Flux Calculation}

An accurate estimate of the viscous fluxes between two cells requires
an accurate estimate of the velocity gradients at the interface. The
gradient extrapolation method described above produces in general two
different values of the velocity gradient that meet at the interface.
This defines a general Riemann problem for the differential equation
in Eq.~(\ref{eq:euler_derivatives}) which is no longer a homogeneous
hyperbolic differential equation. Therefore, attempting to solve this
new Riemann problem for the spatial derivatives of the scalar
quantities introduces a significant additional difficulty. For
simplicity, we will assume that the differences between two gradient
extrapolations meeting at an interface are small enough such that a
simple arithmetic mean can be used. This assumption, of course, is
valid only when the field of second derivatives is sufficiently smooth
(see Section~\ref{sec:hessians}).

The time and area averaged flux across the face $i$-$j$ that moves with speed $\mathbf{w}$ is defined as
\begin{equation}
\begin{split}
\hat{\mathbf{F}}_{ij}=\frac{1}{\Delta t}\frac{1}{A_{ij}}\int_{\Delta t}\int_{A_{ij}}
&\left[\frac{}{}\mathbf{F}_\mathrm{adv}(\mathbf{U})-\mathbf{U}\mathbf{w}^T \right.\\
&\left.-\mathbf{F}_\mathrm{diff}(\mathbf{W,\partial\mathbf{W}/\partial{\mathbf{r}}})
\frac{}{}\right]{\rm d}\mathbf{A}_{ij}\;{\rm d}t \\
\equiv\hat{\mathbf{F}}_{\mathrm{adv},ij}-\hat{\mathbf{F}}_{\mathrm{diff},ij}&~~.
\end{split}
\end{equation}
The numerical or Godunov estimate of these fluxes is chosen so that
the analytic expressions for $\mathbf{F}_\mathrm{adv}(\mathbf{U})$ and
$\mathbf{F}_\mathrm{diff}(\mathbf{W,\partial\mathbf{W}/\partial{\mathbf{r}}})$
are evaluated with numerical estimates of $\mathbf{U}$, $\mathbf{W}$
and $\partial\mathbf{W}/\partial\mathbf{r}$ at the centroid of the
interface. The
advective Godunov fluxes are
\begin{equation}\label{eq:advective_flux}
\hat{\mathbf{F}}_{\mathrm{adv},ij}=\left[\mathbf{F}_\mathrm{adv}(\mathbf{U}^\mathrm{lab}_\mathrm{Riem})-\mathbf{U}^\mathrm{lab}_\mathrm{Riem}\mathbf{w}^T\right]\hat{\mathbf{n}}_{ij},
\end{equation}
where $\mathbf{U}^\mathrm{lab}_\mathrm{Riem}$ is the conserved
variable vector at the centroid of the interface, as seen in the lab
frame, obtained from the solution of a 1-D Riemann problem across the
$i$-$j$ interface and along its normal. Multiplying by
$\hat{\mathbf{n}}_{ij}$ is equivalent to projecting the flux matrix
$\mathbf{F}_\mathrm{adv}$ (Eq.~\ref{eq:ad_flux}) along the normal of
each face. The Godunov fluxes $\hat{\mathbf{F}}_{\mathrm{adv},ij}$ and
$\hat{\mathbf{F}}_{\mathrm{diff},ij}$ are thus $5$-component
vectors. The diffusive Godunov flux vector is obtained from the
diffusive flux $5\times3$ matrix
 \begin{equation}\label{eq:viscous_flux_matrix}
 \begin{split}
 \mathbf{F}_\mathrm{diff}=
 \left[ \begin{matrix}\\
 0\\
\Pi_{xx}\\
\Pi_{xy}\\
\Pi_{xz}\\
 v_x \Pi_{xx}+v_y\Pi_{xy}+v_z \Pi_{xz} \\\\ \end{matrix}\right.
 &
 \begin{matrix}\\
 0\\
\Pi_{yx}\\
\Pi_{yy}\\
\Pi_{yz}\\
\;\;\;v_x \Pi_{yx}+v_y\Pi_{yy}+v_z \Pi_{yz} \\\\ \end{matrix}
\\ \\
& \left. \begin{matrix}\\
 0\\
\Pi_{zx}\\
\Pi_{zy}\\
\Pi_{zz}\\
v_x \Pi_{zx}+v_y\Pi_{zy}+v_z \Pi_{zz} \\\\ \end{matrix}\right]\\
\end{split}
 \end{equation}
 where $\Pi_{ab}$ are the components of the viscous stress tensor $\mathbf{\Pi}$, which depend
 on the local value of the velocity and the velocity gradients. These components
 are:
 \begin{equation}\label{eq:viscous_tensor}
 \begin{array}{ccl}
 \Pi_{xx}&=& \cfrac{4}{3}\eta \partial_xv_x-\cfrac{2}{3}\eta\left(\partial_yv_y+\partial_zv_z\right)+\zeta\nabla\cdot
 \mathbf{v}\\
 \Pi_{yy}&=&\cfrac{4}{3}\eta \partial_yv_y-\cfrac{2}{3}\eta(\partial_zv_z+\partial_xv_x)+\zeta\nabla\cdot
 \mathbf{v}\\
\Pi_{zz}&=&\cfrac{4}{3}\eta \partial_zv_z-\cfrac{2}{3}\eta(\partial_xv_x+\partial_yv_y)+\zeta\nabla\cdot
 \mathbf{v}\\
 \\
  \Pi_{xy}&=&\Pi_{yx}=\eta\left(\partial_yv_x+\partial_xv_y\right)\\
  \\
   \Pi_{yz}&=&\Pi_{zy}=\eta\left(\partial_zv_y+\partial_yv_z\right)\\
 \\
 \Pi_{zx}&=&\Pi_{xz}=\eta\left(\partial_xv_z+\partial_zv_x\right)\\
\end{array}
  \end{equation}
Just like with the advective fluxes, the flux tensor (Eq.~\ref{eq:viscous_flux_matrix}) must be projected
onto the normal $\hat{\mathbf{n}}_{ij}$ of each $ij$-interface to
obtain the 5-component vector
\begin{equation}\label{viscous_flux}
\hat{\mathbf{F}}_{\mathrm{diff},ij}=\mathbf{F}_\mathrm{diff}\left(\mathbf{W}^\mathrm{lab}_\mathrm{Riem},\left(\partial\mathbf{W}/\partial\mathbf{r}\right)^\mathrm{lab}_\mathrm{approx}\right)\hat{\mathbf{n}}_{ij},
\end{equation}
where $\mathbf{W}^\mathrm{lab}_\mathrm{Riem}$ is the primitive
variable vector at the centroid of the interface, as seen in the lab
frame (whose associated conserved variables are
$\mathbf{U}^\mathrm{lab}_\mathrm{Riem}$ in
Eq.~\ref{eq:advective_flux}). The spatial derivatives
$\left(\partial\mathbf{W}/\partial\mathbf{r}\right)^\mathrm{lab}_\mathrm{approx}$
correspond to our extrapolate-and-average scheme for linearly varying
gradients.  As with $\mathbf{W}^\mathrm{lab}_\mathrm{Riem}$, we are
interested in estimates of $\partial\mathbf{W}/\partial\mathbf{r}$ at
the centroid of the face.  For both these quantities, only the
velocity and its spatial derivatives are relevant when viscous fluxes
are calculated.

\subsection{Hessian Estimation}\label{sec:hessians}

In analogy to the gradient calculation for Voronoi meshes discussed by
\citet{spr10}, here we discuss the estimates of the cell-centered
Hessian matrices for each of the primitive variables $W_\alpha$. To
this end, let us consider a vector field $\mathbf{u}$ that varies
approximately linearly with distance as
$\mathbf{u}\approx\mathbf{u}_i+\mathbf{h}\left(\mathbf{r}-\mathbf{r}_i\right)$
near $\mathbf{r}_i$. Up to linear order, the first derivative of
$\mathbf{u}$ is simply $\mathbf{h}$. The volume-average of the spatial
derivatives of $\mathbf{u}$ in the vicinity of $\mathbf{r}_i$ is
\begin{equation}\label{eq:hess_estimate1}
\begin{split}
V_i\left\langle\frac{\partial\mathbf{u}}{\partial\mathbf{r}}\right\rangle_i=&\int_{V_i}\frac{\partial\mathbf{u}}{\partial\mathbf{r}}\,{\rm d}V\\
=&\int_{\partial V_i}\mathbf{u}\,{\rm d}\mathbf{A}\\
=&\sum_{j\neq i}\int_{A_{ij}}\left[\mathbf{u}_i+\mathbf{h}(\mathbf{r}-\mathbf{r}_i)\right]\frac{\mathbf{r}_j-\mathbf{r}_i}{r_{ij}}\,{\rm d}A,
\end{split}
\end{equation}
where we have assumed that the linear approximation is valid up to all
the neighboring mesh generating points $\mathbf{r}_j$. 
It is straightforward to verify that the average matrix
$\langle \partial\mathbf{u}/\partial\mathbf{r} \rangle_i$ can be
written as
\begin{equation}\label{eq:hess_estimate2}
\begin{split}
\left\langle\frac{\partial\mathbf{u}}{\partial\mathbf{r}}\right\rangle_i
=&\frac{1}{V_i}\sum_{j\neq i}A_{ij}\left(\frac{\mathbf{u}_i+\mathbf{u}_j}{2}\otimes\widehat{\mathbf{n}}_{ij}\right)\\
&- \frac{1}{V_i}\sum_{j\neq i}A_{ij}\left(\mathbf{h}\,\mathbf{c}_{ij}\otimes\frac{\mathbf{r}_{ij}}{r_{ij}}\right).
\end{split}
\end{equation}
Writing the vector product $\left(\mathbf{A}\,\mathbf{u}\right)\otimes
\mathbf{v}$ in tensor form (where $\mathbf{A}$ is a $n\times n$ square
matrix and $\mathbf{u}$ and $\mathbf{v}$ are vectors of dimension
$n$), it is easy to prove the identity
$A_{ac}u_cv_b=A_{ac}v_cu_b+\varepsilon_{bfc}\varepsilon_{fde}u_dv_e
A_{ac}$. Equivalently, going back to vector notation,
we have $\left(\mathbf{A}\,\mathbf{u}\right)\otimes
\mathbf{v}=\left(\mathbf{A}\,\mathbf{v}\right)\otimes
\mathbf{u}+\left(\mathbf{u}\times\mathbf{v}\right)\times \mathbf{A}$,
where, for simplicity, we used vector notation to denote a
``cross product'' between a vector and a matrix.

Therefore, the second term on the right hand side of
Eq.~(\ref{eq:hess_estimate2}) can be written as
\begin{equation}\label{eq:hess_estimate3}
\begin{split}
\sum_{j\neq
  i}A_{ij}\left(\mathbf{h}\,\mathbf{c}_{ij}\otimes\frac{\mathbf{r}_{ij}}{r_{ij}}\right)
=&\sum_{j\neq
  i}A_{ij}\left(\mathbf{h}\,\mathbf{r}_{ij}\otimes\frac{\mathbf{c}_{ij}}{r_{ij}}\right)\\
&+\sum_{j\neq
  i}\left(A_{ij}\mathbf{c}_{ij}\times\frac{\mathbf{r}_{ij}}{r_{ij}}\right)\times\mathbf{h}~~.
  \end{split}
\end{equation}
Here, the second term on the right hand side vanishes identically, because
\begin{equation}
\begin{split}
 \sum_{j\neq i}\left(A_{ij}\mathbf{c}_{ij}\times\frac{\mathbf{r}_{ij}}{r_{ij}}\right)\times\mathbf{h}
=&\left\{\int_{\partial V_i}\left(\mathbf{r}-\frac{\mathbf{r}_i+\mathbf{r}_j}{2}\right)\times {\rm d}\mathbf{A}\right\}\times\mathbf{h}\\
=&\left\{\int_{V_i}\nabla\times\left(\mathbf{r}-\frac{\mathbf{r}_i+\mathbf{r}_j}{2}\right){\rm d}V\right\}\times\mathbf{h}\\
=&~0~~.
\end{split}
\end{equation}
On the other hand, the first term on the right hand side of
Eq.~(\ref{eq:hess_estimate3}) can be rewritten by means of the
replacement
$\mathbf{h}\,\mathbf{r}_{ij}=-\mathbf{h}\,(\mathbf{r}_j-\mathbf{r}_i)=\mathbf{u}_i-\mathbf{u}_j$. Finally,
identifying the vector $\mathbf{u}_i$ with the gradient
$\langle\nabla\phi\rangle_i$ of a scalar quantity $\phi$, and the matrix $\langle
\partial\mathbf{u}/\partial\mathbf{r} \rangle_i$ with the
cell-averaged Hessian matrix $\langle \mathbf{H}^\phi \rangle_i$,
Eq.~(\ref{eq:hess_estimate3}) takes the form
\begin{equation}\label{eq:hessian_estimate}
\begin{split}
\left\langle\mathbf{H}^\phi \right\rangle_i =& \frac{1}{V_i}\sum_{j\neq i}A_{ij}\left\{
-\left(\frac{\langle\nabla\phi\rangle_i+\langle\nabla\phi\rangle_j}{2}\right)\otimes\frac{\mathbf{r}_{ij}}{r_{ij}}\right.\\
&\left.+\left(\frac{}{}\langle\nabla\phi\rangle_j-\langle\nabla\phi\rangle_i\right)\otimes\frac{\mathbf{c}_{ij}}{r_{ij}} \right\}.
\end{split}
\end{equation}
The most noteworthy characteristic of this expression is that it is
purely algebraic and explicit in nature.  That is, the Hessian matrix
of $\phi$ is simply a linear combination of the neighboring gradients
in which the coefficients are predetermined quantities that depend
only on the local mesh geometry. Each one of those neighboring gradients is, 
at the same time, a linear combination of its immediate neighbors' scalar quantities
 \citep[see Eq.~21 of][]{spr10}. Therefore, the Hessian estimate of 
Eq.~(\ref{eq:hessian_estimate}) is a weighted linear combination of scalars
from its immediate neighbors and from its neighbors' neighbors and, as such,
it  implicitly employs a larger stencil than the one used for the gradients.

\subsection{Slope-Limiting the Hessians}\label{sec:hessian_limiting}

It is well known that higher-order reconstruction schemes are prone to
produce spurious oscillations in the vicinity of steep gradients,
unless this is prevented by appropriate {\it slope limiter methods}
\citep{tor09}. These non-linear corrections in the data reconstruction
procedure prevent overshoots and allow for true discontinuities in the
solutions.  So far, we have discussed how to estimate second
derivatives from first derivatives, which in turn are already
reconstruction estimates obtained from the scalar physical
quantities. Potential irregularities in the second derivative fields
can lead to spurious oscillations and unphysical values of the viscous
stress tensor at the cell boundaries.  To alleviate this problem, we
enforce local monotonicity of each component of the gradients, which
is equivalent to smoothing out the Hessian estimates. In practice,
this is achieved by replacing the Hessian matrix by a `slope limited'
version
\begin{equation}
\overline{\left\langle\mathbf{H}^\phi \right\rangle_i}=\mathbf{A}_i\left\langle\mathbf{H}^\phi \right\rangle_i
\end{equation}
with
\begin{equation}
\mathbf{A}_i=\begin{pmatrix}\alpha_i^x &0 &0 \\ 0&\alpha_i^y&0\\ 0 & 0 & \alpha_i^z\end{pmatrix}
\;\;\;\;\;\;\;\;\;\;\;\;\;\;\;
\text{and  where}\;\;\;\alpha_i^a=\mathrm{min}\left(1,\psi^a_{ij}\right)
\end{equation}
are the slope limiters $0\leq\alpha_i^a\leq1$ for each direction
$x$, $y$ and $z$. This MINMOD-type slope-limiting method is readily applicable for
irregular meshes. The quantities
$\psi^a_{ij}$ are defined as
\begin{equation}
\psi^a_{ij}=\left\{\begin{array}{lcr} \left(\langle\partial_a\phi\rangle_i^\mathrm{max}-\langle\partial_a\phi\rangle_i\right)/\Delta \left(\partial_a\phi\right)_{ij} & &\Delta \left(\partial_a\phi\right)_{ij} >0\\
\left(\langle\partial_a\phi\rangle_i^\mathrm{min}-\langle\partial_a\phi\rangle_i\right)/\Delta \left(\partial_a\phi\right)_{ij} & &\Delta \left(\partial_a\phi\right)_{ij} <0\\
1&&\Delta \left(\partial_a\phi\right)_{ij} =0
\end{array}\right.
\end{equation}
where the $\Delta \left(\partial_a\phi\right)_{ij}$ are the components
of the vector
$\Delta\left(\nabla\phi\right)_{ij}=\langle\mathbf{H}^\phi\rangle_i(\mathbf{f}_{ij}-\mathbf{s}_i)$,
i.e.~the estimated change in the gradient $\nabla\phi$ between the
centroid $\mathbf{f}_{ij}$ of the cell and the center of cell $i$.
The quantities $\langle\partial_a\phi\rangle_i^\mathrm{max}$ and
$\langle\partial_a\phi\rangle_i^\mathrm{min}$ are the maximum and
minimum of the $a$-th component of the cell-centered gradient
estimates among all neighboring cells of cell $i$, including $i$
itself.

To our knowledge, the slope-limiting technique has not been
applied to the second derivatives before. However, its purpose is
equivalent to the ``flattening" procedure near shocks carried out by
\citet{col84} for parabolic reconstruction. In our approach, the
suppression of oscillations near shocks is exclusively handled by the
limitation of the gradients, since the reconstruction of hydrodynamic
quantities is only of linear order. Thus, the Hessian limitation
procedure serves the sole purpose of guaranteeing a smooth variation of
the gradients and avoiding spuriously large viscous fluxes. Future
improvements of the present method could however also employ these second
derivatives for higher-order reconstructions of the scalars.

\subsection{Time integration and time-step criterion}

Because of the more complex mathematical properties of the
NS equations compared with the Euler equations, obtaining a
rigorous analytic expression analogous to the CFL stability criterion
for the allowed time step size is not possible. However, \citet{mac75}
obtained an approximate semi-empirical stability criterion when
advective, viscous and heat diffusion terms are considered.
When there is no heat flux, the time-step criterion can be written
as \citep[e.g.][]{kun08}
\begin{equation}\label{eq:time-step2}
\Delta t\leq \frac{\sigma \Delta t_\mathrm{CFL}}
{1+2/\{\mathrm{Re}\}_i} ,
\end{equation}
where $\Delta t_\mathrm{CFL}$ is the standard CFL-criterion time-step
except for the Courant-Friedrichs-Levy coefficient, which is absorbed
into a ``safety factor" $\sigma$ (usually $\approx0.9$). In 
Eq.~(\ref{eq:time-step2}), the {\it cell Reynolds number} $\{\mathrm{Re}\}_i$
is
\begin{equation}
\{\mathrm{Re}\}_i=\frac{\rho|\mathbf{v}_i'|R_i}{\eta}, 
\end{equation}
where $\mathbf{v}_i'$ is the velocity of the gas relative to the motion of the grid  and
$R_i$ is the effective radius of the cell, calculated as $R_i = (3V_i/4\pi)^{1/3}$ from the volume of a cell
 (or as $R_i = (A_i/\pi)^{1/2}$ from the area in 2D).
Similar approaches to derive an appropriate NS time-step 
have also been described by \citet{mav90} and \citet{coi96}.


The numerical integration scheme we employ is time unsplit, that is,
advective and diffusive fluxes are applied simultaneously during each
hydrodynamic time-step and not sequentially
(Eq.~\ref{eq:finite_vol2}).  The prediction stage, on the other hand,
is operator-split, since the advective and diffusive terms are
computed almost independently of each other. This is in part due to
the nature of the standard one-dimensional Riemann problem, whose
solutions -- strictly speaking -- are only valid for the hyperbolic
Euler problem, but are not solutions to the full NS equations
with their additional parabolic terms. The validity of this approach
ultimately relies on the assumption that the viscous terms in the
NS equations are typically small perturbations to the Euler
equations.

\section{Numerical Test Results}

To test the performance of \arepo\ when our new treatment of viscous
fluxes is included, we have carried out a number of test simulations
for physical situations with known analytic or quantitative
solutions. Usually, the problems with known exact solutions are either
of self-similar type or have symmetries that make the non-linear term
proportional to $(\mathbf{v}\cdot\nabla)\mathbf{v}$ vanish
identically.  Owing to these limitations, numerical simulations of situations with
experimentally well-established behavior, such as flow past a circular
cylinder, have become common-place in testing the performance of
NS codes.  We will therefore also carry out such
qualitative benchmarks, besides looking at a few simple problems with
analytic solutions.

\begin{figure*}
\centering
\includegraphics[width=16cm]{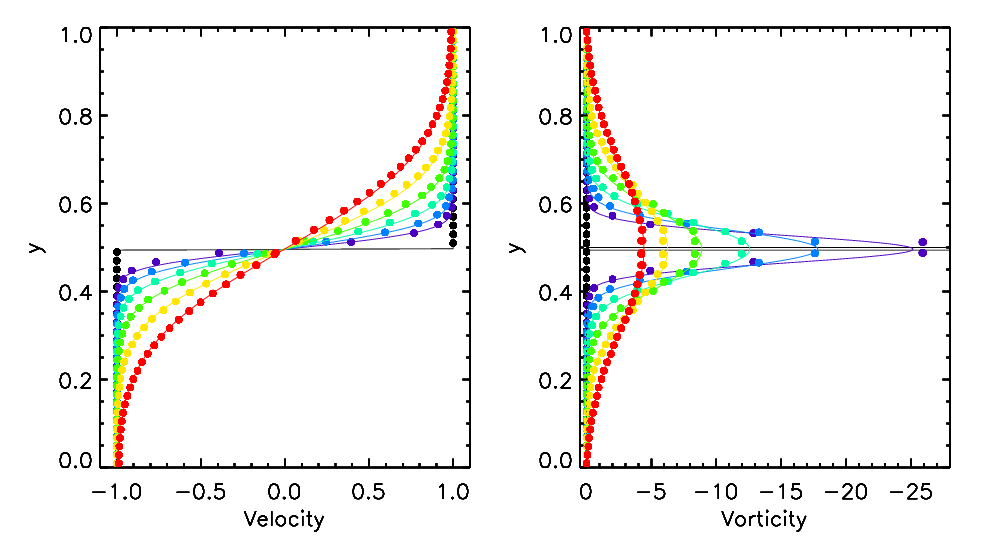}
 \vspace*{-10pt}
\caption{Diffusion of a vortex sheet. The two panels show the velocity
  $u$ along the $x$-axis ({\it left panel}), and the vorticity ({\it
    right panel}), at times $t=0$, 0.1, 0.2, 0.4, 0.8, $1.6$ and $3.2$
  (from black to red), for a dynamic viscosity coefficient
  $\mu=\nu\rho=0.005$. The solid lines are given by the analytic
  solution described by Eqs.~(\ref{eq:vortex_sheet}), while the solid
  circles are {\it all} 2500 cell-centered velocity and vorticity
  values of the initially Cartesian $50\times50$ mesh. Note that the
  simulation is started with a sharp discontinuity in velocity and
  thus the $\delta$-function vorticity field is initially
  unresolved. If the mesh would remain exactly Cartesian, the
  diffusion of vorticity would actually be suppressed in this
  case. Nevertheless, the small asymmetries introduced by the moving
  mesh trigger the diffusion regardless of the initially unresolved
  setup, and the time-dependent numerical result closely follows the
  expected exact solution.\label{fig:vortex_sheet}}
\vspace*{-5pt}
\end{figure*}

\begin{figure*}
\subfigure[$t=0.0$]{\includegraphics[width=7.5cm]{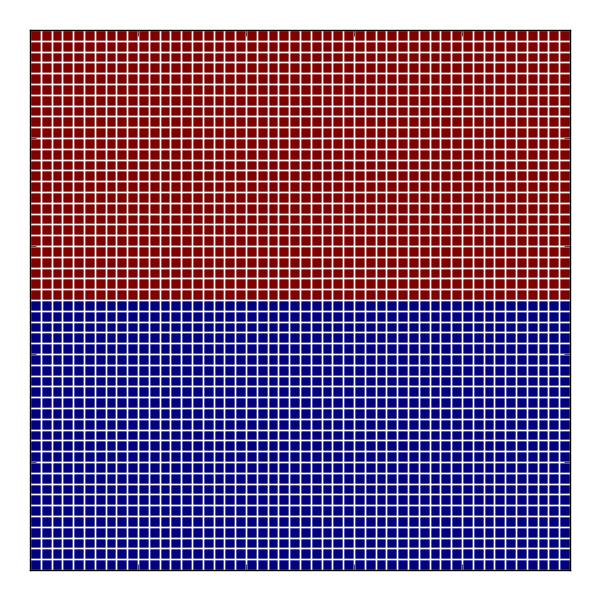}}
\subfigure[$t=0.06$]{\includegraphics[width=7.5cm]{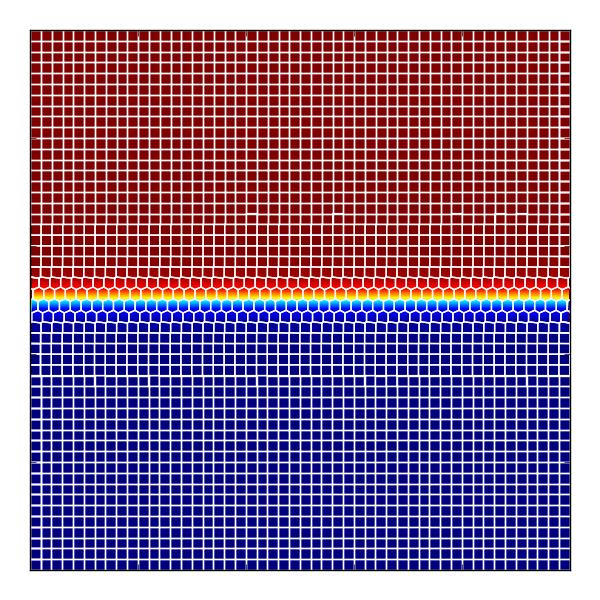}}
\subfigure[$t=0.6$]{\includegraphics[width=7.5cm]{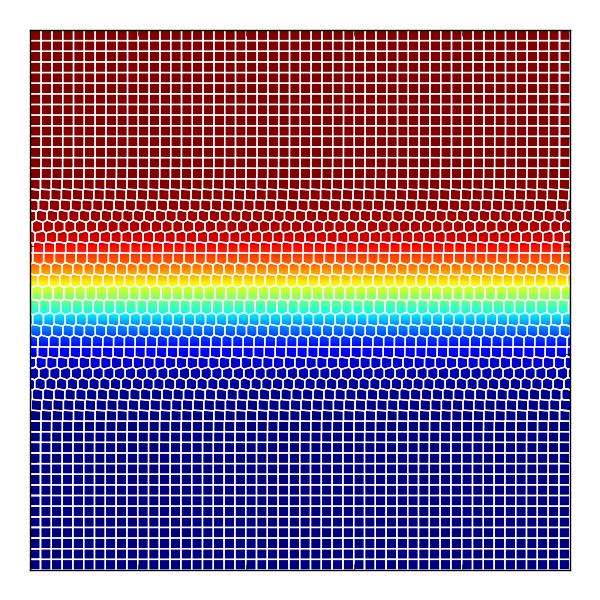}}
\subfigure[$t=1.8$]{\includegraphics[width=7.5cm]{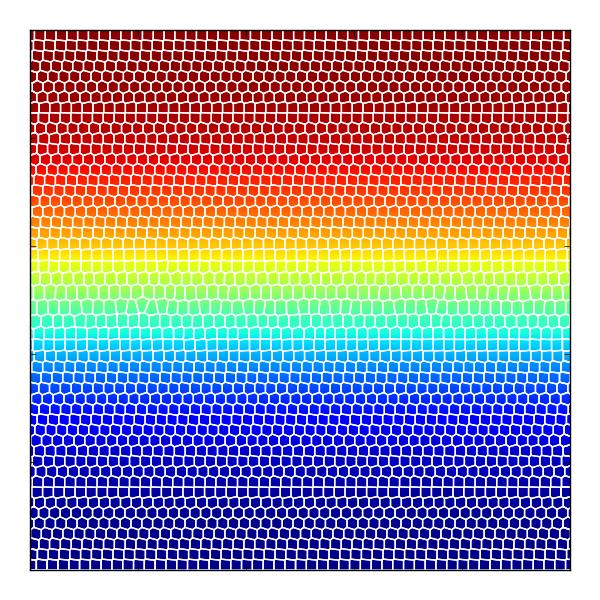}}
\caption{Time evolution of the mesh geometry and the velocity field
  for a diffusing vortex sheet test. As the vorticity spreads from the
  center of the domain to the upper and lower boundaries, the mesh
  adapts to the continuous change in velocity until its original
  Cartesian structure disappears entirely. The color table (from blue to red)
 corresponds to the range between $u=-1.0$ and $u=1.0$ in linear scale.}
\label{fig:vortex_sheet_mesh}
\end{figure*}

\subsection{Diffusion of a Vortex Sheet}
A simple problem of laminar flow in the presence of viscosity is given
by the vortex sheet diffusion test.  In this problem, the initial
velocity field at $t=0$ is given by $\mathbf{v}=(u,0,0)$ with $u=1$
for $y>0$ and $u=-1$ for $y<0$. Because of the symmetry of the
problem, the NS equations reduce to a 1D diffusion equation
\begin{equation}\label{eq:ns_1d}
\frac{\partial u}{\partial t}=\nu\frac{\partial^2 u}{\partial^2 y},
\end{equation}
with solution \citep[e.g.][]{kun08}
\begin{equation}\label{eq:vortex_sheet}
u=\mathrm{erf}\left[\frac{y}{2\sqrt{\nu t}}\right]
\;\;\;\;\;\;\;\;\;\;\omega=\frac{\partial u}{\partial y}=\frac{e^{-y^2/4\nu t}}{\sqrt{\pi\nu t}}.
\end{equation}

In Figure~\ref{fig:vortex_sheet}, we show the time evolution we obtain
for a two-dimensional simulation domain with initially uniform
pressure and density ($\rho=P=1$), and with a velocity field given by
$\mathbf{v}=({\rm sgn}(y),0,0)$. The mesh generating points were
distributed regularly at the initial time to produce a Cartesian
mesh. As the system evolves, the velocity and the vorticity fields as
a function of time and vertical coordinate $y$ follow the exact
solution remarkably well. It is worth pointing out that the initial
singularity in the vorticity field is unresolved numerically (and thus
appears as being uniformly zero throughout the domain), since the
system is started with an exact sharp discontinuity. Static, perfectly
aligned meshes with slope limitation techniques will typically
maintain this unresolved vorticity and thus no diffusion will proceed
unless some numerical perturbations are seeded that break the mesh
alignment of the initial state (a common way to overcome this
difficulty is to start the system according to
Eq.~(\ref{eq:vortex_sheet}) at $t>0$ such that there is initial
vorticity). However, the moving mesh of \arepo\ ``sees'' a non-zero
velocity gradient as soon as the upper and lower halves of the domain
become unaligned with respect to each other.  This happens because, as
soon as a cell shifts its position, the number of its neighbors that
have a drastically different velocity increases and so does the
``statistical weight" of the discontinuity.  At this point, the
slope-limiting technique, which had ignored the discontinuity in the
perfectly aligned mesh, now identifies the local variation as ``real"
and the vorticity field is ``detected".

Fig.~\ref{fig:vortex_sheet_mesh} shows the corresponding
two-dimensional velocity field of the diffusing vortex sheet test at
four different times, together with the geometry of the underlying
Voronoi mesh. The mesh geometry nicely shows how the cells transform
from a Cartesian configuration to an unstructured mesh, while the
velocity field evolves from a piece-wise constant state with a central
discontinuity to a smoothly varying shear flow due to the effects of
viscosity.

\subsection{Diffusion of a Gaussian Vortex}
The two-dimensional circular velocity distribution corresponding to an
irrotational vortex of circulation $\Gamma$ is
\begin{equation}
v_\theta=\frac{\Gamma}{2\pi R},
\end{equation}
where the vorticity
$\omega=|\nabla\times\mathbf{v}|=(1/R)\partial(R\,v_\theta)/\partial
R$ is zero everywhere except at the origin ($\omega=\delta(R)$, i.e. a
vortex line). In a viscous fluid, this velocity profile has to be
sustained by a point source of vorticity at the origin (e.g. an
infinitely thin rotating cylinder) otherwise the vortex line will
decay in a similar way as the vortex sheet in the previous example.
If the velocity at the origin is set impulsively to zero, the
subsequent evolution of the azimuthal velocity is given by
\begin{equation}
v_\theta(R,t)=\frac{\Gamma}{2\pi R}\left[1-e^{-R^2/4\nu t}\right]~~,
\end{equation}
while the vorticity $
\omega=\left[\nabla\times(v_\theta\hat{\bm\theta})\right]\cdot\hat{\mathbf{z}}$ 
evolves as
\begin{equation}
\omega=
-\frac{\Gamma}{4\pi \nu t }e^{-R^2/4\nu t}
\end{equation}
and the Laplacian of the velocity field is
\begin{equation}
|\nabla^2\mathbf{v}|
=\frac{\Gamma}{2\pi}\frac{R}{(2\nu t)^2}e^{-R^2/4\nu t} \hat{\bm\theta}~~.
\end{equation}
Because of its geometry, this problem is significantly more
challenging than the vortex sheet test considered above and cannot be
impulsively started at precisely $t=0$.  Besides the initial
singularity in the vorticity field, the velocity field is divergent as
we approach the origin. In addition, it is not possible to capture the
azimuthal velocity field when the distance from the origin is
comparable to the grid resolution. At the same time, the
azimuthal velocity field is challenging for the boundary conditions,
because the problem is self-similar in nature and therefore natural
boundaries do not exist. These problems did not exist for the
vortex-sheet problem, which is of one-dimensional
nature. Nevertheless, evolving the system from an initial time $t>0$
minimizes most of these complications. In addition, we extend the
computational domain far beyond the region of interest, such that
boundaries become essentially irrelevant during the timespan of the
numerical solution.

We setup a Cartesian mesh ($100\times100$) with an imposed initial velocity profile of
\begin{equation}
v_{\theta,0}=\frac{\Gamma}{2\pi R}\left[1-\exp\left(
-\frac{R^2}{4\nu t_0}
\right)\right]
\;\;\;\;\;\text{  with  }
\nu=\frac{\mu}{\rho} ,
\end{equation}
corresponding to a Gaussian vortex that we center in the middle of the
domain, which extends over the range $[0,40]\times[0,40]$, and thus
accommodates a radial range from $R=0$ to $R=20$.  The adopted physical
parameters are $t_0=10$, $\mu=0.08$, $\Gamma=1.0$, and the initial
density field is constant with $\rho=1$. The pressure field, however, is
not uniform because the fluid is not started from rest. We obtain the
correct pressure profile from the radial component of the equation of
motion:
\begin{displaymath}
-\frac{v_{\theta}^2}{R}=-\frac{1}{\rho}\frac{{\rm d}P}{{\rm d}R},
\end{displaymath}
and thus the initial pressure profile is
\begin{displaymath}
\begin{split}
P_\mathrm{init}=P_0-\frac{\Gamma^2\rho}{4\pi^2}\left\{\frac{1}{2R^2}e^{-R^2/(2\nu t_0)}
\left[e^{R^2/(4\nu t_0)}
-1\right]^2\right.\\
\left.+\frac{1}{4\nu t_0}\left(\mathrm{Ei}\left(-\frac{R^2}{2\nu t_0}\right)
-\mathrm{Ei}\left(-\frac{R^2}{4\nu t_0}\right)\right)\right\}~~,
\end{split}
\end{displaymath}
where $P_0$ is an integration constant.  The precise value of $P_0$
is irrelevant for the similarity solution presented here, because it
is obtained for incompressible flow. In our numerical experiments
(which are compressible), we set $P_0$ such that $P=1$ at $R=0$.

\begin{figure*}
\vspace*{-10pt}
\includegraphics[width=11.5cm]{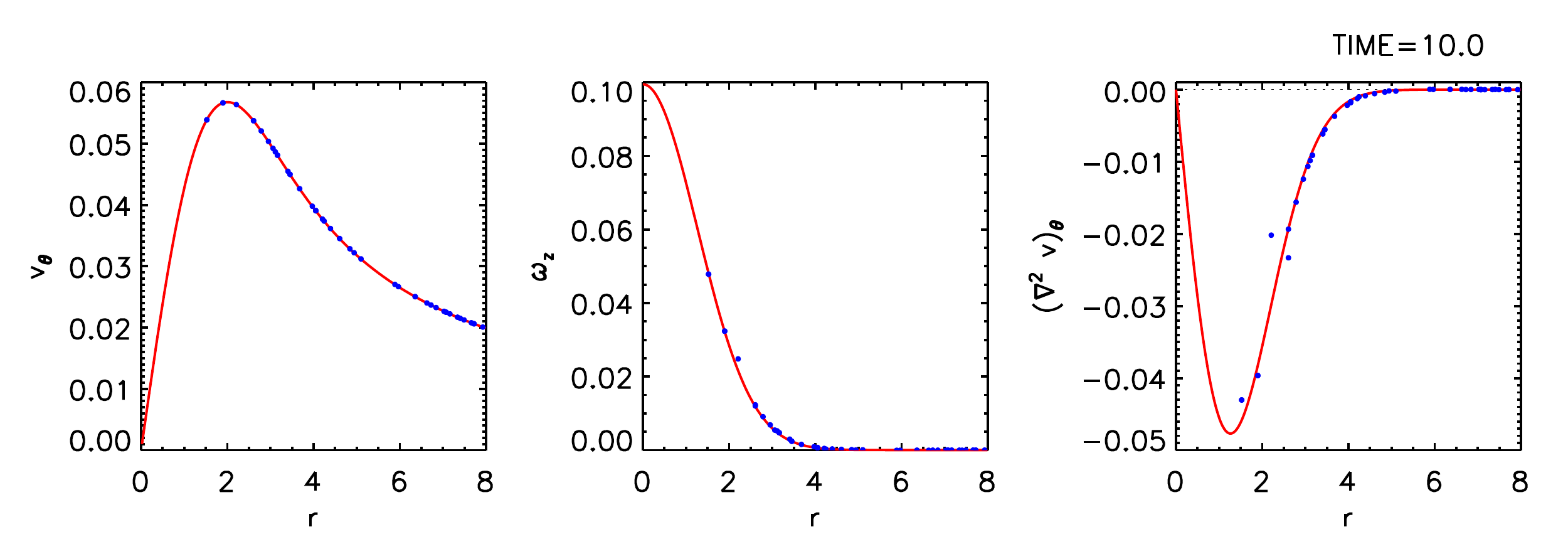}
\includegraphics[width=11.5cm]{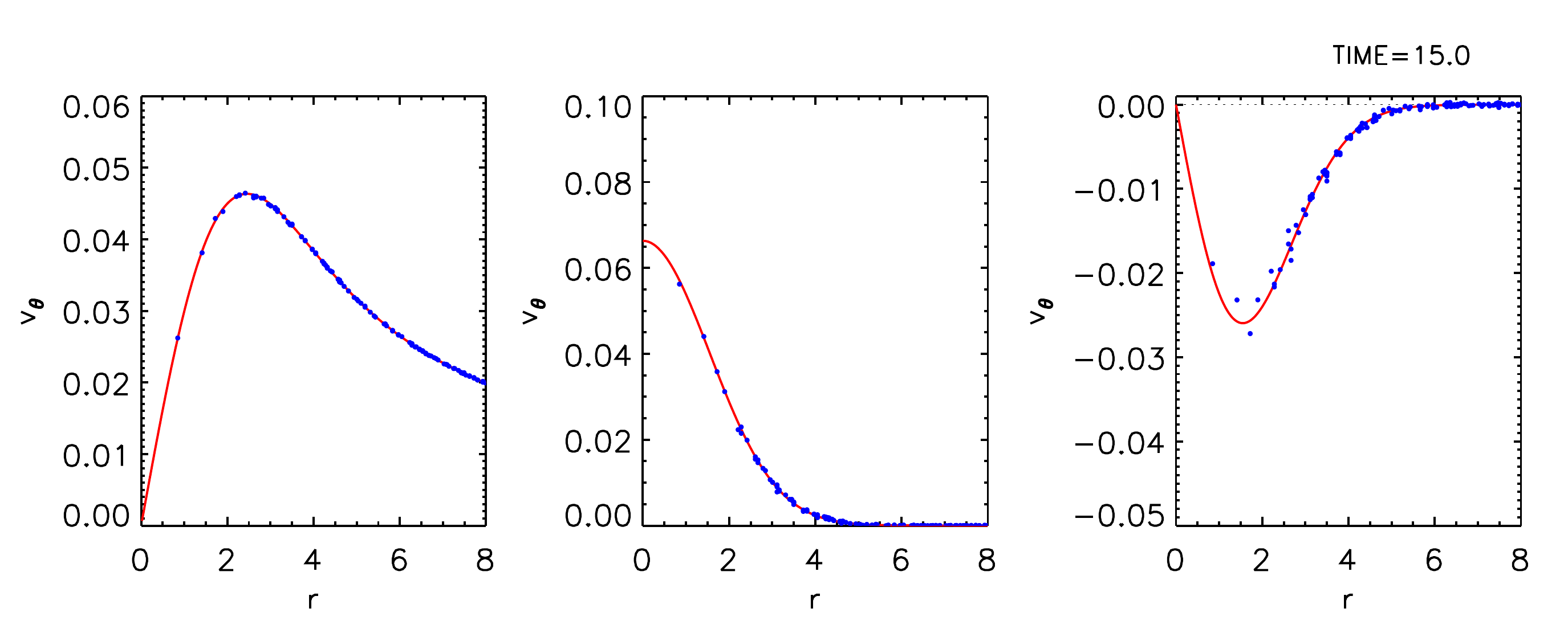}
\includegraphics[width=11.5cm]{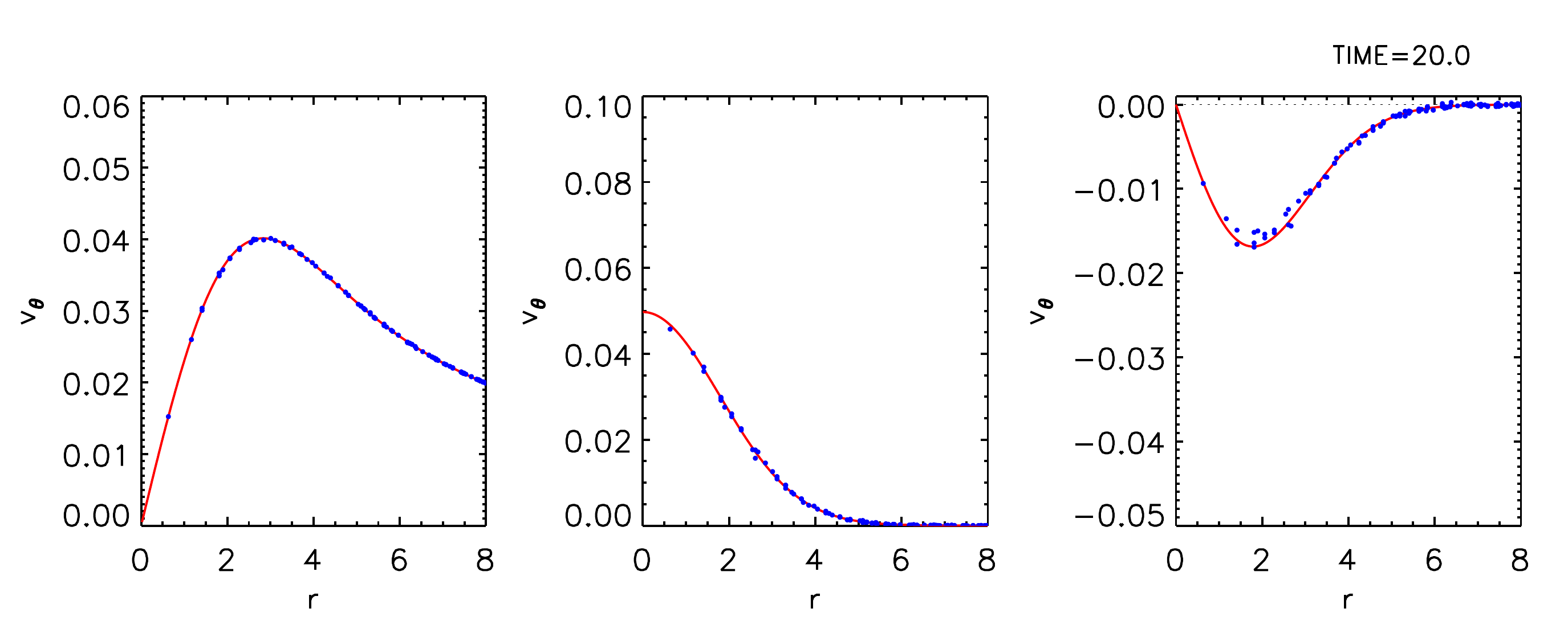}
\includegraphics[width=11.5cm]{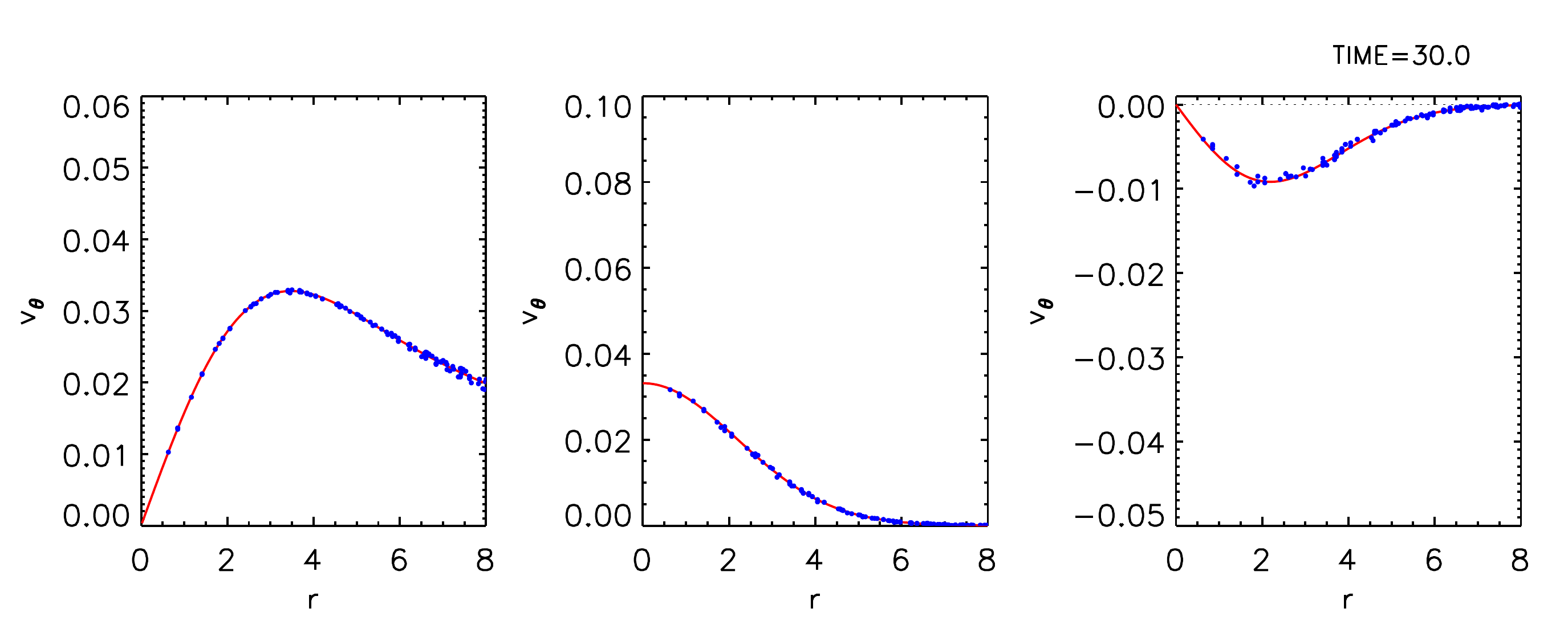}
 \vspace*{-10pt}
\caption{Time evolution of a diffusing Gaussian vortex. For each time (as
labeled), we show the azimuthal velocity profile $v_\theta(R)$, the
vorticity profile $\omega(R)$ and the Laplacian profile
$\nabla^2v_\theta$, as computed by {\small AREPO} (blue points; only a random $10\%$ of the total shown) 
and compare it to the corresponding analytic expressions (solid red lines).\label{fig:gaussian_vortex}}
 \vspace*{-12pt}
\end{figure*}

Fig.~\ref{fig:gaussian_vortex} shows the time evolution of the
velocity field, the vorticity field and the Laplacian field for a
Gaussian vortex started on an initially Cartesian mesh. We find 
not only that the velocity evolves as expected based on the similarity
solution, but the first and second derivatives also show excellent
agreement with the analytic expectations. These results validate both
the space- and time-accuracy of our viscous integration scheme, as well
as the accuracy with which the second derivatives are estimated.

\begin{figure*}
\centering \subfigure[Plane Poiseuille
  flow]{\label{fig:plane_poiseuille}\includegraphics[width=8.1cm]{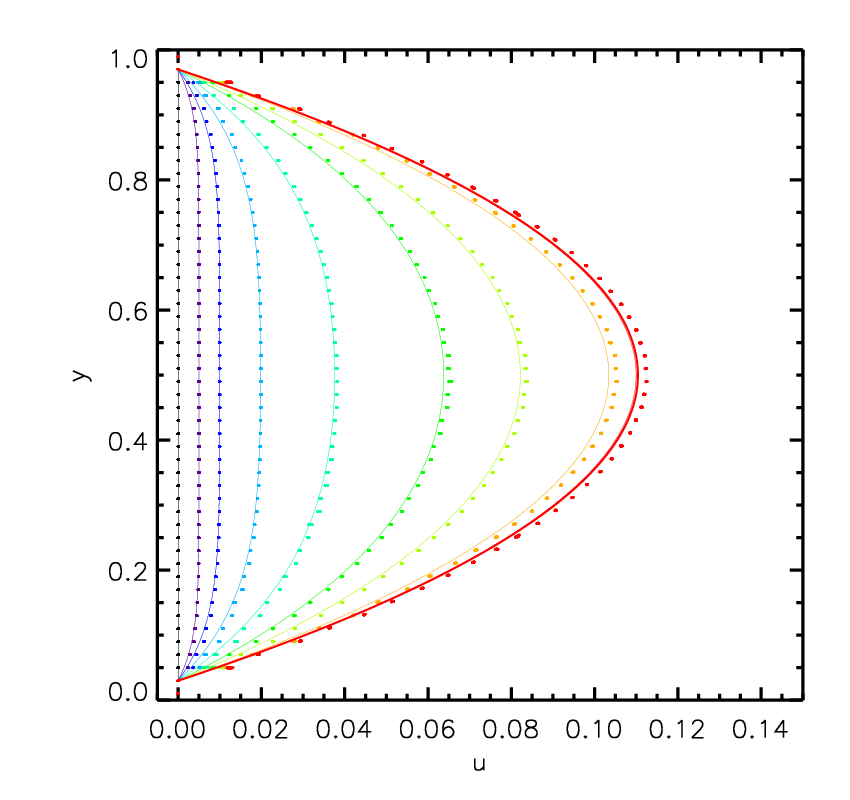}}
\subfigure[Plane Couette
  flow]{\label{fig:plane_couette}\includegraphics[width=8.1cm]{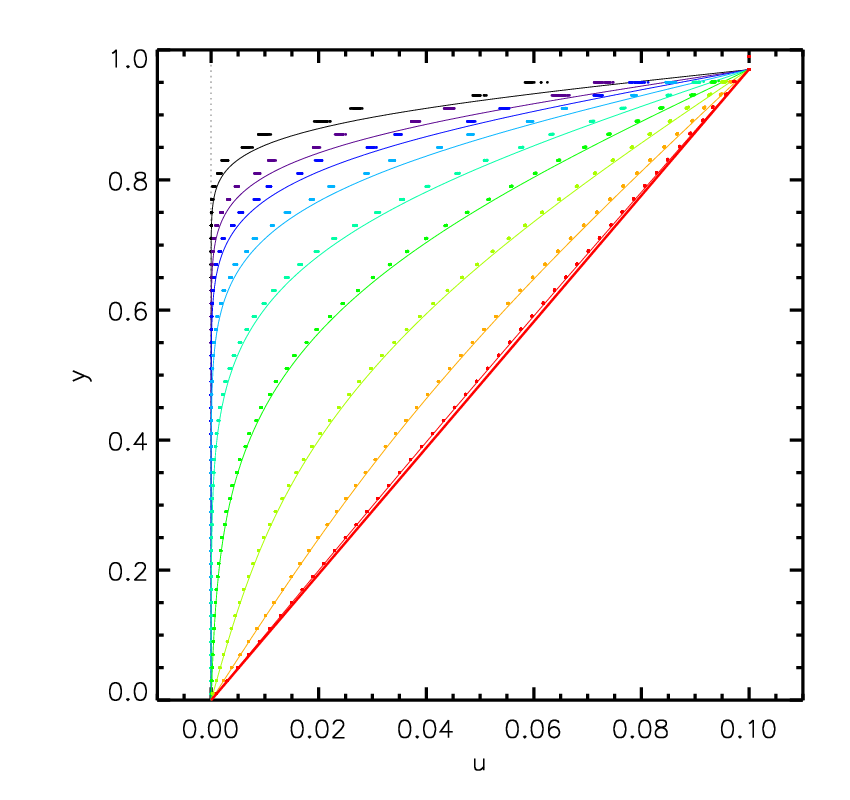}}
\caption{Impulsively started plane Poiseuille and Couette flows as a
  function of time. a) Time evolution of the horizontal velocity
  profile versus vertical distance. Solid curves represent the
  analytic solutions of Eqs.~(\ref{eq:pipe_flow1}) to
  (\ref{eq:pipe_flow3}) for $U=0$ and ${\rm d}p/{\rm d}x=-0.05$, at
  ten different times (time increasing from black to red). The data
  points correspond to all the cell-centered values of velocity along
  $x$ for a $50\times50$ simulation started from rest. b) Time
  evolution of the horizontal velocity profile versus vertical
  distance. Solid curves represent the analytic solutions in
  Eqs.~(\ref{eq:pipe_flow1}) to (\ref{eq:pipe_flow3}) for $U=0.1$ and
  ${\rm d}p/{\rm d}x=0$ at eight different times (time increasing from
  black to red). The data points correspond to all the cell-centered
  values of velocity along $x$ for a $50\times50$ numerical simulation
  started from rest.\label{fig:flow_pipe}}
\end{figure*}

\begin{figure*}
\centering
\subfigure[$t=0.1$]{\includegraphics[width=7.5cm]{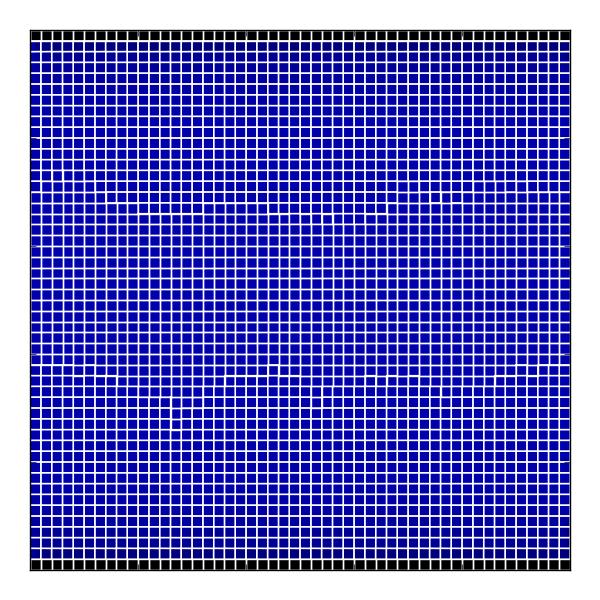}}
\subfigure[$t=0.9$]{\includegraphics[width=7.5cm]{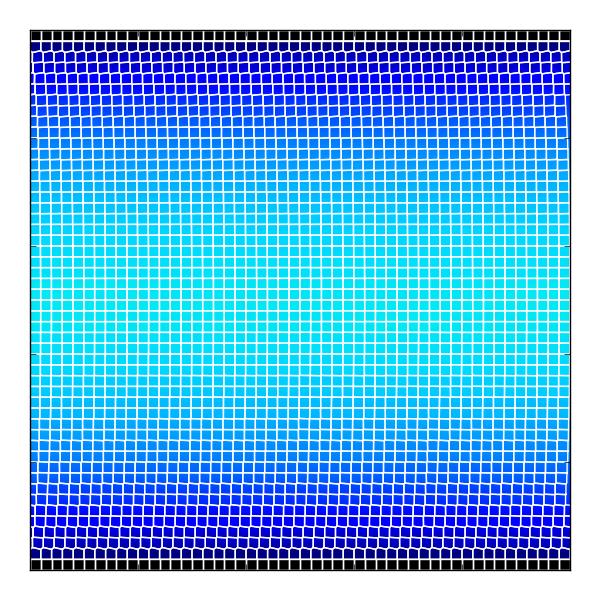}}
\subfigure[$t=2.5$]{\includegraphics[width=7.5cm]{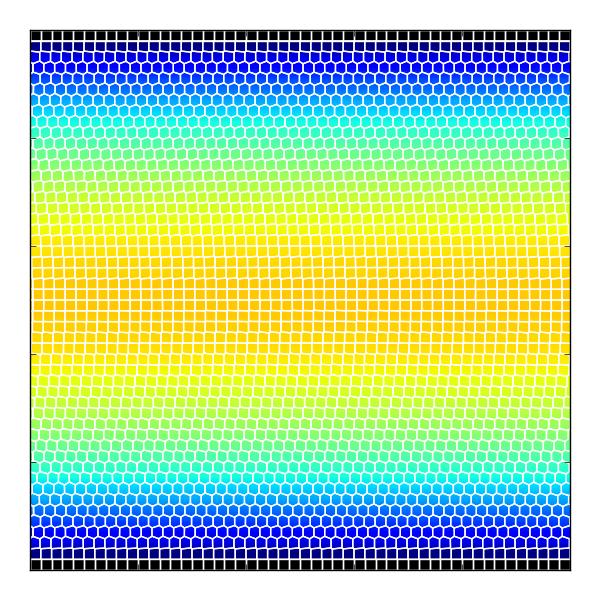}}
\subfigure[$t=5.7$]{\includegraphics[width=7.5cm]{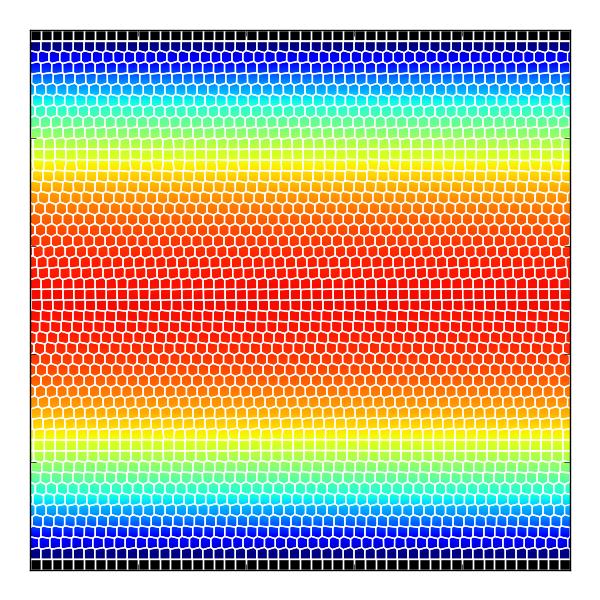}}
\caption{Time evolution of the mesh geometry and the velocity for flow between
  parallel plates.  The horizontal velocity field $u$ for plane
  Poiseuille flow is rendered at four different times. The evolution
  of the velocity field (see Fig~\ref{fig:plane_poiseuille}) is
  accompanied by the evolution of the mesh from an initially Cartesian
  set up (top-left panel) to a fully unstructured grid by the time
  the flow has reached steady state (bottom-right panel). The (linear ) color scale
  ranges from blue ($u=0$) to red ($u=0.12$).\label{fig:plane_couette_mesh}}
\end{figure*}

\subsection{Plane Poiseuille and Couette Flows}

Next, we consider impulsively-started plane Poiseuille and Couette
flows where a fluid between two parallel plates is initially at rest,
and then, suddenly, either pressure gradients or plate motions are
applied.  The time-dependent solution has the form
$\mathbf{v}=(u(y,t),0,0)$, where the horizontal velocity can be
decomposed into steady and time-dependent parts,
$u(y,t)=u_0(y)+\widetilde{u}(y,t)$. In the presence of a pressure
gradient and an upper plate moving at constant speed $U$, the steady
state solution is the well-known expression
\begin{equation}\label{eq:pipe_flow1}
u_0(y)=\frac{yU}{b}-\frac{y}{2\mu}\frac{{\rm d}p}{{\rm d}x}\left(b-y\right),
\end{equation}
for which the special cases $U=0$ and ${\rm d}p/{\rm d}x=0$ are
commonly known as plane Poiseuille flow and plane Couette flow,
respectively.

The time dependent component $\widetilde{u}(y,t)$ is a solution of
Eq.~(\ref{eq:ns_1d}), subject to the initial condition
$\widetilde{u}(y,0)=-u_0(y)$ and the boundary conditions
$\widetilde{u}=0$ at $y=0$ and $y=b$.  By separation of variables, the general solution is
\citep[e.g][]{gra07}
\begin{equation}\label{eq:pipe_flow2}
\widetilde{u}(y,t)=\sum_{n=1}^\infty A_n e^{-n^2\pi^2\nu t/b^2}\sin\frac{n\pi y}{b},
\end{equation}
where the coefficients $A_n$ are determined by the initial condition
\begin{align}\label{eq:pipe_flow3}
A_n&=-\cfrac{\int_0^bu_0(y)\sin\frac{n\pi y}{b}{\rm d}y}{\int_0^b\sin^2\frac{n\pi y}{b}{\rm d}y} \\
&=-\frac{2U(-1)^n}{n\pi}-\frac{2}{b\mu}\frac{{\rm d}p}{{\rm d}x}\left(\frac{b}{n\pi}\right)^3\left[1-(-1)^n\right]~~.
\end{align}

\begin{figure*}
\includegraphics[width=12cm]{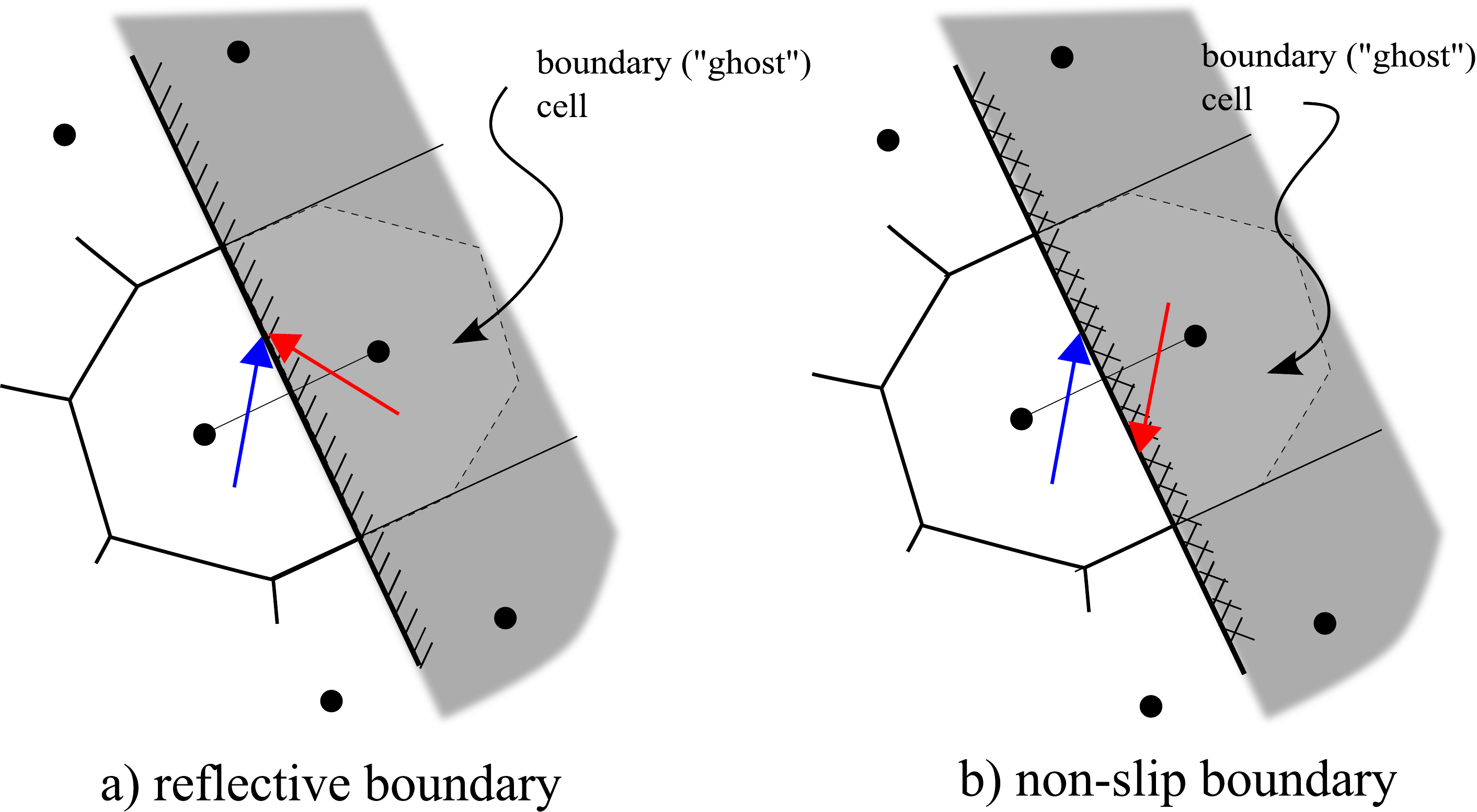}
\caption{Schematic representation of (a) reflective and (b) non-slip boundaries within the computational domain. 
After the spatial and temporal extrapolation steps in the MUSCL-Hancock method (panel e) in Figure~\ref{fig:arepo_scheme}),
the Riemann problem is solved as elsewhere in the domain but with the boundary-side cell mimicking the gas side with either one velocity component
-- the normal one -- reversed  (reflection) or all three (non-slip). \label{fig:boundaries}}
\end{figure*}

The numerical setup for this problem is straightforward. We produce a
Cartesian mesh in the range $[0,1]\times[0,1]$ with a resolution of
$50\times50$. The fluid is originally at rest and its density and
pressure are given by $\rho=P=1$. The equation of state is that of an
ideal gas with adiabatic index $5/3$. To represent the plates, the
uppermost and lowermost rows of cells are replaced by ``solid
boundaries`` at which the no-slip condition is enforced,
i.e.~$v_x=v_y=0$ (see Fig.~\ref{fig:boundaries}). Moving solid
  boundaries are straightforward to implement with a Voronoi
  tessellation mesh. A solid surface can be constructed as a series of
  mesh-generating point pairs, one on each side of the surface, such
  that the common interface -- equidistant to both points -- defines
  the boundary locally (see \citealp{ser01} and \citealp{spr10}).  The
  Voronoi cell on the side of the ``solid" object can regarded as ``a
  ghost cell within the domain". That is, this cell is part of the
  domain discretization process and is tessellated/updated as any
  other normal gas cell. However, when solving the Riemann problem at
  the local interface between a ``solid" cell and a real gas cell,
  boundary conditions are imposed in the same way as boundary
  conditions on the outer box are imposed. For perfectly reflecting
  boundaries, the normal component of the velocity is reflected in the
  ``solid side" or ``outside region" of the interface. For non-slip
  boundaries, the entire velocity vector is reflected, such that the
  velocity at the interface is zero (Figure~\ref{fig:boundaries}).

We run two different test problems. For the first
one, both plates remain at rest and an external gradient of ${\rm
  d}p/{\rm d}x=-0.05$ is imposed. For the second test, the bottom
plate is at rest and the upper plate moves at a constant speed of
$U=0.1$. In both test simulations, the dynamic viscosity coefficient
has been set to $\mu=0.05$. In Figure~\ref{fig:flow_pipe}, we show the time evolution of the
horizontal velocity profile both for the plane Poiseuille and Couette
flows. In both cases, the numerical results match the analytic
expectations very well. In Figures\ref{fig:plane_couette_mesh} we also 
show maps of the velocity
profile and the mesh geometry at different times for the  Poiseuille
case. The grid evolution shows
how the Cartesian structure is progressively lost, but that the
dynamic Voronoi mesh of \arepo\ successfully avoids any mesh-tangling
effects.


\subsection{Time-Dependent Circular Couette Flow}\label{sec:couette}
\begin{figure*}
\includegraphics[width=7.2cm]{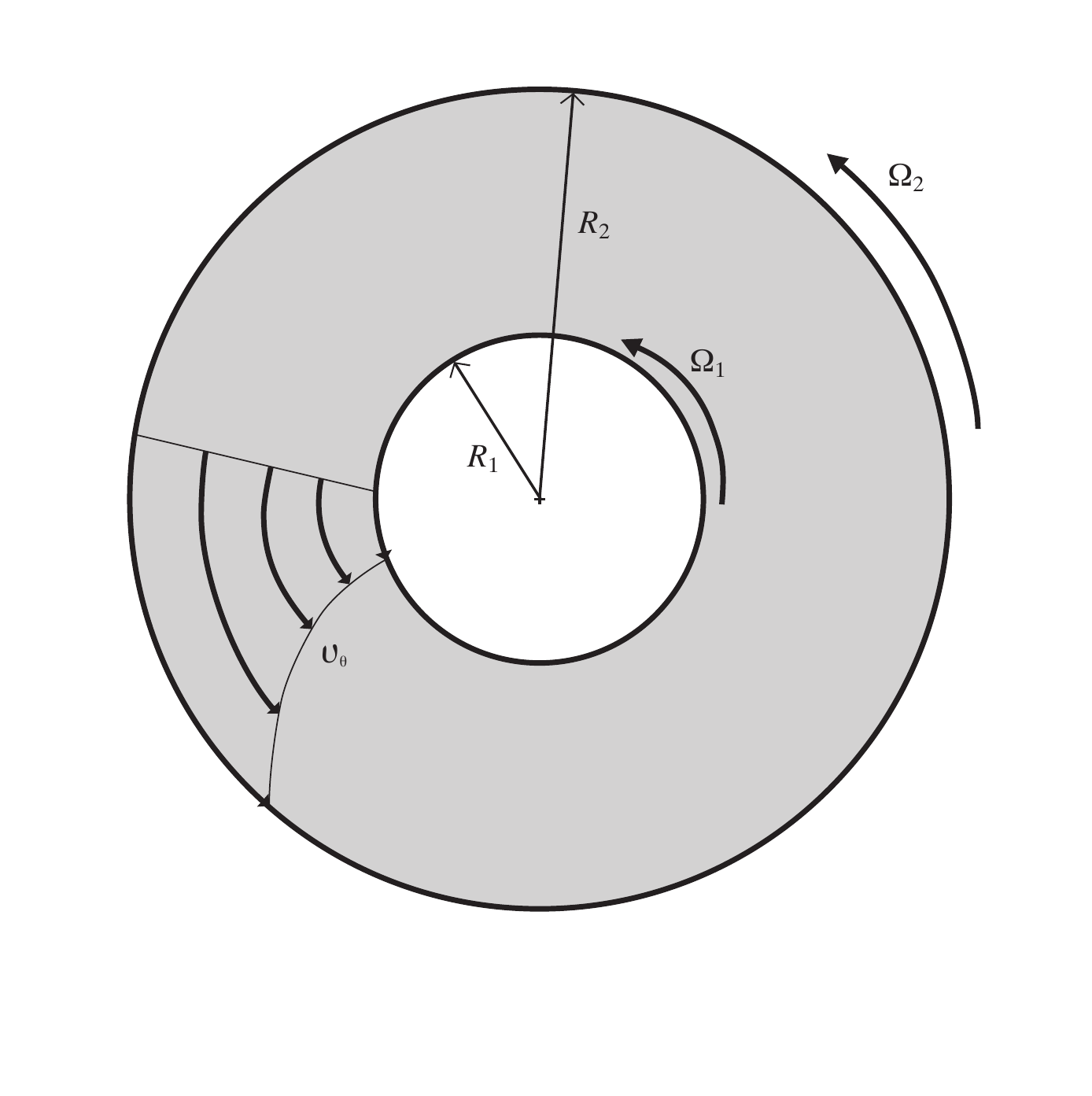}
\includegraphics[width=8cm]{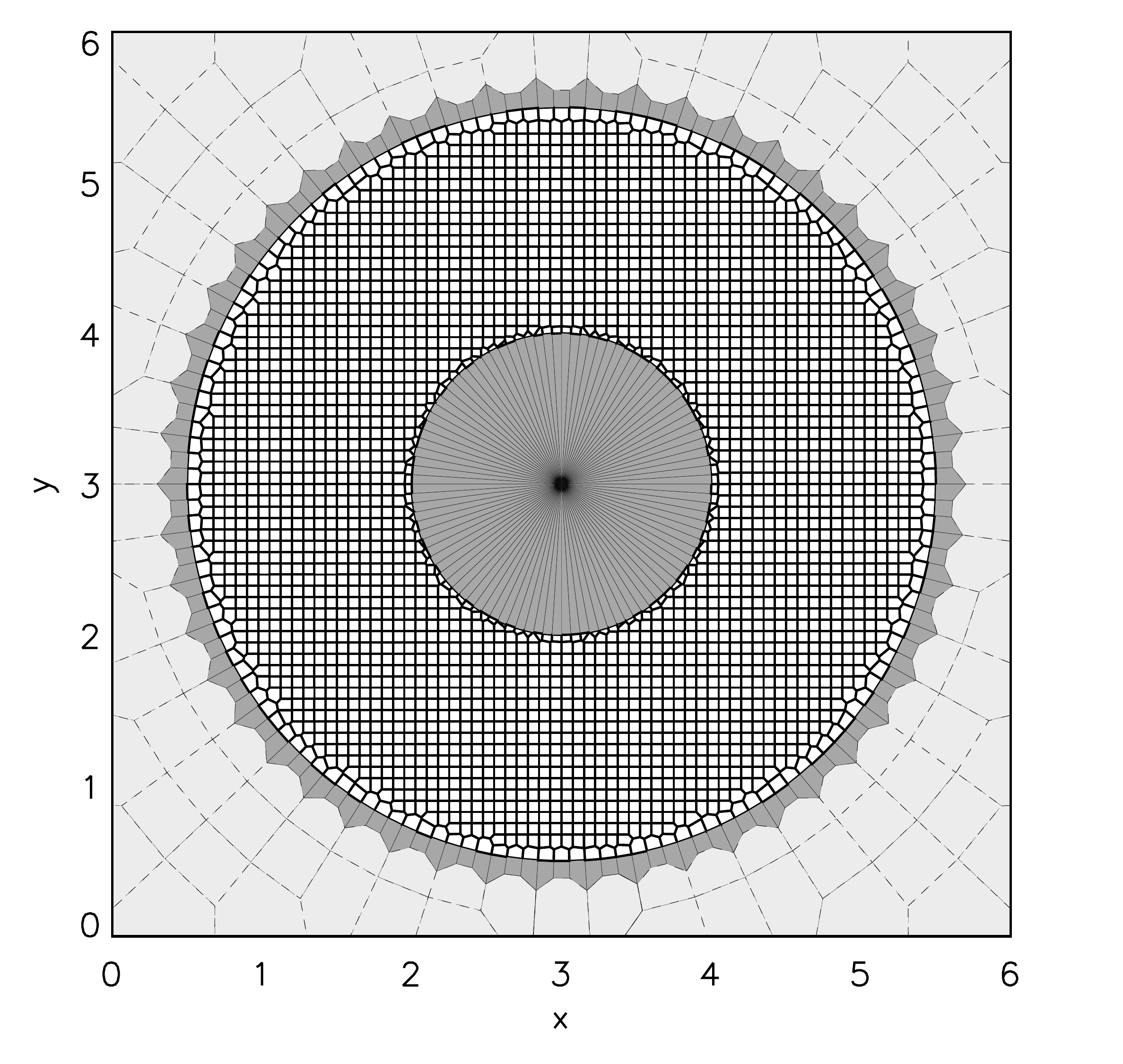}
\caption{Geometry of the circular Couette flow. The left hand panel
shows a schematic view of the two-dimensional problem. The right
hand panel displays the actual initial mesh used in {\small AREPO}
in a setup where we start the problem impulsively from rest. Each
cylindrical boundary (at radii $R_1$ and $R_2$) is generated by
two layers of cells, one side representing the solid cylinder and
the other representing the fluid. These layers of cells are
positioned along circles. The remainder of the fluid cells, originally at
rest, are distributed like a Cartesian grid. The cells outside the
outer cylinder are ``auxiliary cells" and are only included to fill
the computational domain, but do not exert any influence on the
fluid inside the cylinders. The motion of the cylinders is
prescribed to remain constant (with angular velocities $\Omega_1$
and $\Omega_2$), and thus represents a source of kinetic energy. The
motion of the fluid in between the cylinders is induced by means
of the no-slip boundary condition at the contact surface, and the
momentum that is transported in the radial direction through the
shear viscosity.\label{fig:couette_geometry}}
\end{figure*}

We now turn to a more challenging problem, which
highlights the ability of our scheme to deal with geometrically
complex boundary conditions. For purely azimuthal motion, the
NS equations in the radial and tangential directions are
\begin{subequations}
\begin{align}
-\frac{v_\theta^2}{R}&=-\frac{1}{\rho}\frac{{\rm d}P}{{\rm d}R}
\\
\rho\frac{\partial v_\theta}{\partial t}&=\mu\frac{{\rm d}}{{\rm d}R}\left[\frac{1}{R}\frac{{\rm d}}{{\rm d}R}(Rv_\theta)\right]~~.
\end{align}
\end{subequations}
The exact solution of steady flow (i.e.~$\partial v_\theta/\partial t
=0$) between concentric cylinders with boundary conditions
$v_\theta=\Omega_1R_1$ at $R=R_1$, and $v_\theta=\Omega_2R_2$ at
$R=R_2$ is given by \citep[e.g][]{kun08}
\begin{equation}
v_{\theta,0}(R)=\frac{\frac{}{}\left(\Omega_2R_2^2-\Omega_1R_1^2\right)R^2
-\left(\Omega_2-\Omega_1\right)R_2^2R_1^2}{R(R_2^2-R_1^2)}~~,
\end{equation}
where $R_i$  and $\Omega_i$ ($i=1,2$) are the radii and angular velocities of
the respective cylinders.

The impulsively-started version of this problem can be solved analytically 
by separation of variables \citep[see][]{tra68,gra07}. The full solution can thus be
written as $v_{\theta}(R,t)=v_{\theta,0}+\tilde{v}_\theta(R,t)$, where the time-dependent part
has the form
\begin{displaymath}
\tilde{v}_\theta(R,t)=\sum\left\{C_2J_1(nR)+C_2Y_1(nR)\right\}e^{-\nu n^2t},
\end{displaymath}
and $J_1$ and $Y_1$ are Bessel functions of the first and second kind, respectively. This
time-dependent component is subject to the boundary conditions $\tilde{v}_\theta(R,t)=0$ at
$R_1$ and $R_2$, thus allowing us to eliminate  $C_2$:
\begin{displaymath}
\tilde{v}_\theta(R,t)=\sum_{s=1}^\infty \frac{A_s}{Y_1(n_sR_1)}B_1(n_sR)e^{-\nu n_s^2t},
\end{displaymath}
where the $n_s$ are the roots of the equation $B_1(nR)=0$ with
$B_1(nR)\equiv J_1(nR)Y_1(nR_1)-Y_1(nR)J_1(nR_1)$. 

 Finally, the
coefficients $A_s$ are determined by imposing the initial condition
$\tilde{v}_\theta=-v_{\theta,0}$ at $t=0$. To solve for each
coefficient independently, the steady state solution must be written
in terms of a series expansion of $v_{\theta,0}$ in the basis functions $B_1(n_sR)$. After
some algebraic manipulations, we obtain
\begin{displaymath}
\begin{split}
v_{\theta,0}(R)
=\pi\Omega_2R_2\sum_{s=1}^\infty&\frac{J_1(n_s R_2)}{J_1^2(n_sR_1)-J_1^2(n_sR_2)}B_1(n_sR) \\
\times&\left[\cfrac{}{}J_1(n_sR_1)-J_1(n_s R_2)\frac{\Omega_1}{\Omega_2}\frac{R_1}{R_2}\right],
\end{split}
\end{displaymath}
and therefore the time-dependent component is given by
\begin{displaymath}
\begin{split}
\tilde{v}_\theta(R,t)=-\frac{\pi}{R}\sum^\infty_{s=1}
\frac{J_1(n_s R_2)}{J_1^2(n_sR_1)-J_1^2(n_sR_2)}B_1(n_sR)e^{-\nu n_s^2t}\\
\times\left[\frac{}{}\Omega_2R_2J_1(n_sR_1)-J_1(n_s R_2)\Omega_1R_1\right]~~.
\end{split}
\end{displaymath}

 \begin{figure}
\includegraphics[width=7.5cm]{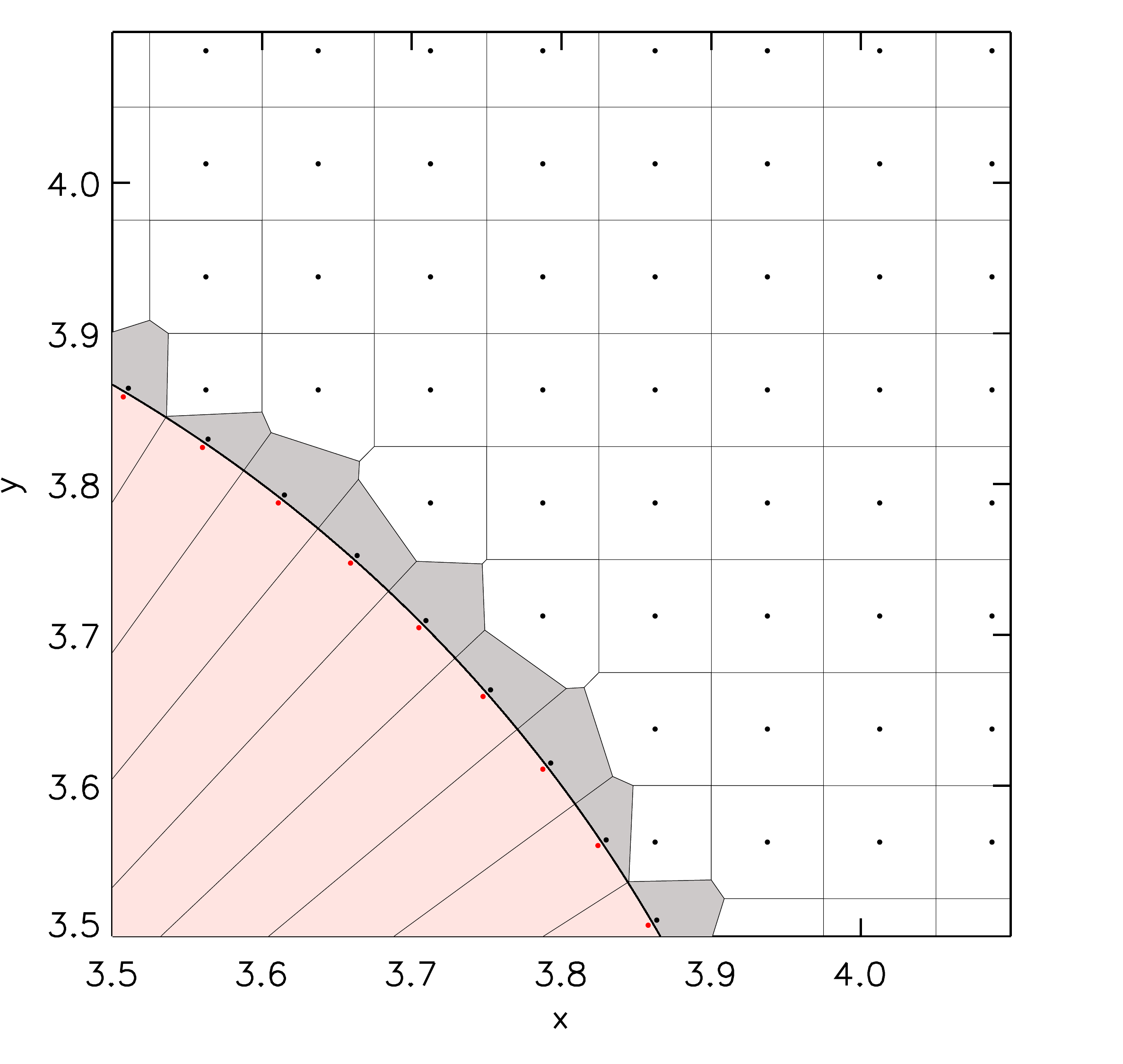}
\includegraphics[width=7.5cm]{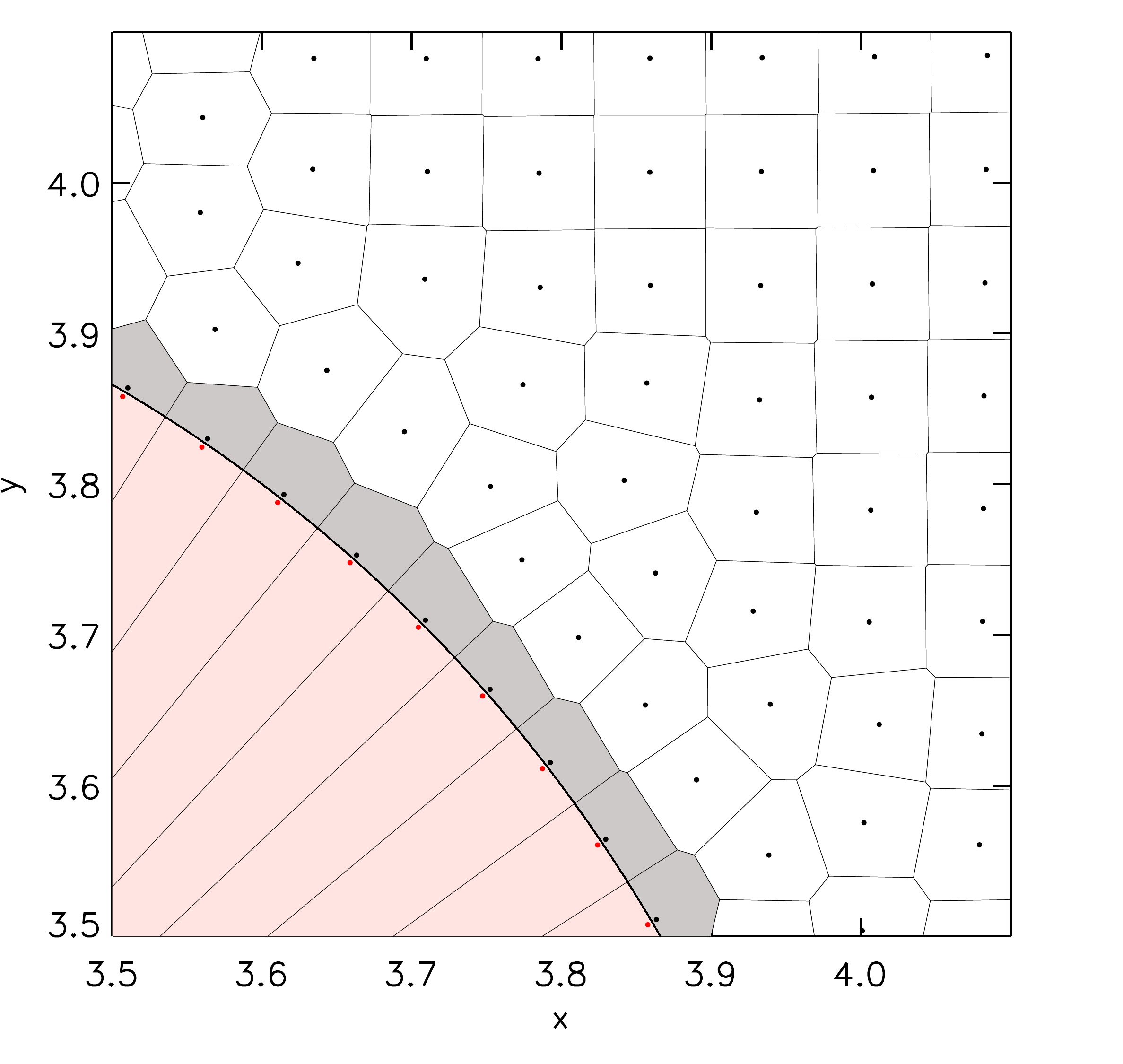}
 \vspace*{-10pt}
\caption{A zoom showing the detailed mesh geometry around the inner
  boundary of the circular Couette flow, at two different times. The
  left hand panel shows a close-up view of the right panel of
  Fig.~\ref{fig:couette_geometry}. The Voronoi faces that make up the
  cylindrical boundary are created by close pairs of points, which
  either lie inside the solid cylinder (red) or on the fluid side
  (black). The gray cells define the contact region of the fluid
  domain with the cylinder; here the no-slip boundary conditions
  are imposed. An analogous geometry applies for the outer cylinder. The
  panel on the right hand side shows the same region of the
  computational domain at a slightly later time, when the mesh filling
  the fluid region has started to react to the motion of the cylinder.\label{fig:couette_mesh_1}}
\end{figure}

Collecting these results, the complete expression for the time-dependent angular velocity 
profile  is 
\begin{equation}\label{eq:couette_analytic}
\begin{split}
\Omega(R,t)=\frac{v_\theta}{R}-\frac{\pi}{R}\sum^\infty_{s=1}
\frac{J_1(n_s R_2)}{J_1^2(n_sR_1)-J_1^2(n_sR_2)}B_1(n_sR)e^{-\nu n_s^2t}\\
\times \left[\frac{}{}\Omega_2R_2J_1(n_sR_1)-J_1(n_s R_2)\Omega_1R_1\right]~~.
\end{split}
\end{equation}

We realize the moving 
boundary conditions in the present case
  through
special Voronoi-cells with prescribed motion and boundary conditions,
as described in \citet{spr10}.  In the present case, we use two sets
of mesh-generating points, each one consisting of a series of
outside-inside pairs located on either side of the boundary and
running parallel to it, so that two circular boundaries of radii $R_1$
and $R_2$ are defined which can be made to rotate at angular
frequencies $\Omega_1$ and $\Omega_2$, respectively. Note that
the only significant technical difference between this
problem and the preceding examples is the way the boundary cells are
prescribed to move; the rest of the numerical scheme remains unaltered.

Figure~\ref{fig:couette_geometry} illustrates the geometry of the
circular Couette flow, and our realization of a suitable mesh in
\arepo.  Since the equations of motion are always solved in the moving
frame of the interfaces, there is no practical difference between
stationary and moving boundaries when they are constructed as a part
of the mesh. Figure~\ref{fig:couette_mesh_1} shows an enlargement of
the mesh at the boundary corresponding to the inner cylinder, which is
represented by a set of Voronoi faces that follow a circular
path. Each one of these Voronoi faces is defined by two mesh
generating points located on either side of the face, one of them
outside the cylinder on the fluid side, the other inside the cylinder
on the side that does not contain fluid.  The right-hand panel of
Fig.~\ref{fig:couette_mesh_1} shows the same region again, but at a
slightly later time. This gives a sense of how the initial Cartesian
mesh between the cylinders reacts to the fluid motion. Since the
latter is azimuthal, the mesh eventually develops an axial geometry,
independent of the initially Cartesian setup.

\begin{figure}
\centering
\includegraphics[width=9cm]{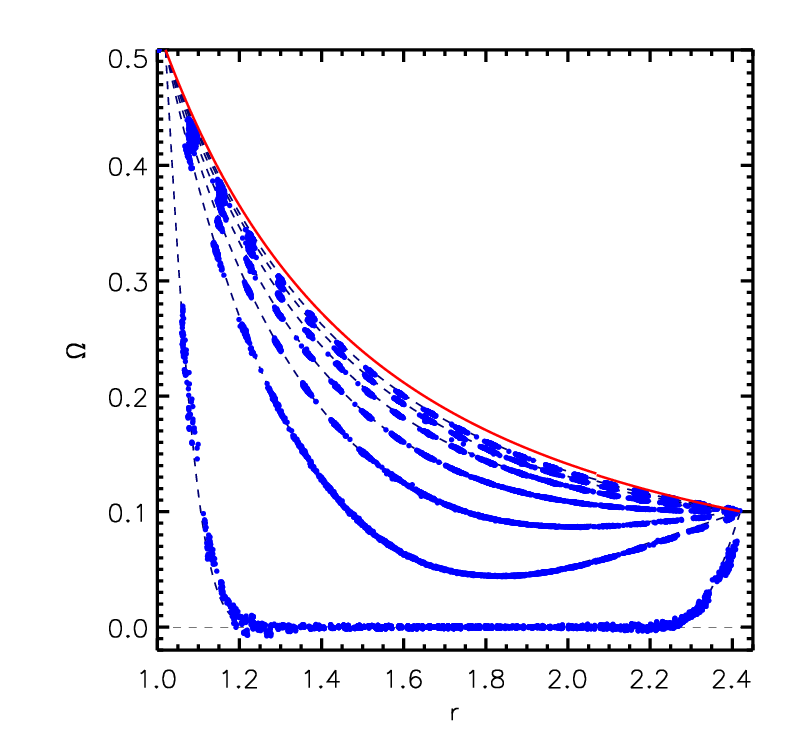}
 \vspace*{-20pt}
\caption{Angular velocity profiles at different times for an
  impulsively started Taylor-Couette flow.  For seven snapshots at
  times $t=0.5$, $20$, $40$, $60$, $80$, $100$, and $120$ we show the
  cell-centered values of $\Omega$, which are plotted as filled blue
  dots for all fluid cells in the calculation. No binning or averaging
  has been performed. The clustering of cell-center points as the
  system evolves is simply a consequence of the mesh adopting an axial
  symmetry in an adaptive fashion. The dashed lines give the
  time-dependent analytic solution of Eq.~(\ref{eq:couette_analytic})
  at the corresponding times. The numerical results are almost
  indistinguishable from the exact solution. The red curve depicts the
  steady state solution to which the time-dependent solution
  eventually converges.\label{fig:couette_flow}}
\end{figure}

Our numerical experiment for this setup has the following
parameters. The initial mesh as described in
Figs.~\ref{fig:couette_geometry} and~\ref{fig:couette_mesh_1} contains
3,254 points, out of which 2,644 are regular fluid cells, 250 are
boundary fluid cells, 250 are solid boundary cells and 110 are unused
auxiliary cells that are only put in to fill up the total mesh area to
an enclosing rectangular shape, as presently required by \arepo. The
radial distance between the cylinders is spanned by 20 cells. The
physical parameters of the Couette flow are $R_1=1$, $R_2=2.5$,
$\Omega_1=0.5$, and $\Omega_2=0.1$, with a dynamic viscosity
coefficient set to $\mu=0.005$. In addition, since the flow is started
from rest, the pressure and density are taken to be uniform with
values $\rho=P=1$. Figure~\ref{fig:couette_flow} shows the time
evolution of the angular velocity profile as it asymptotically
converges to the steady state solution. The agreement of the numerical
data points with the exact analytic solution
(Eq.~\ref{eq:couette_analytic}) is exceptional at all times.

\begin{figure}
\centering
\includegraphics[width=9cm]{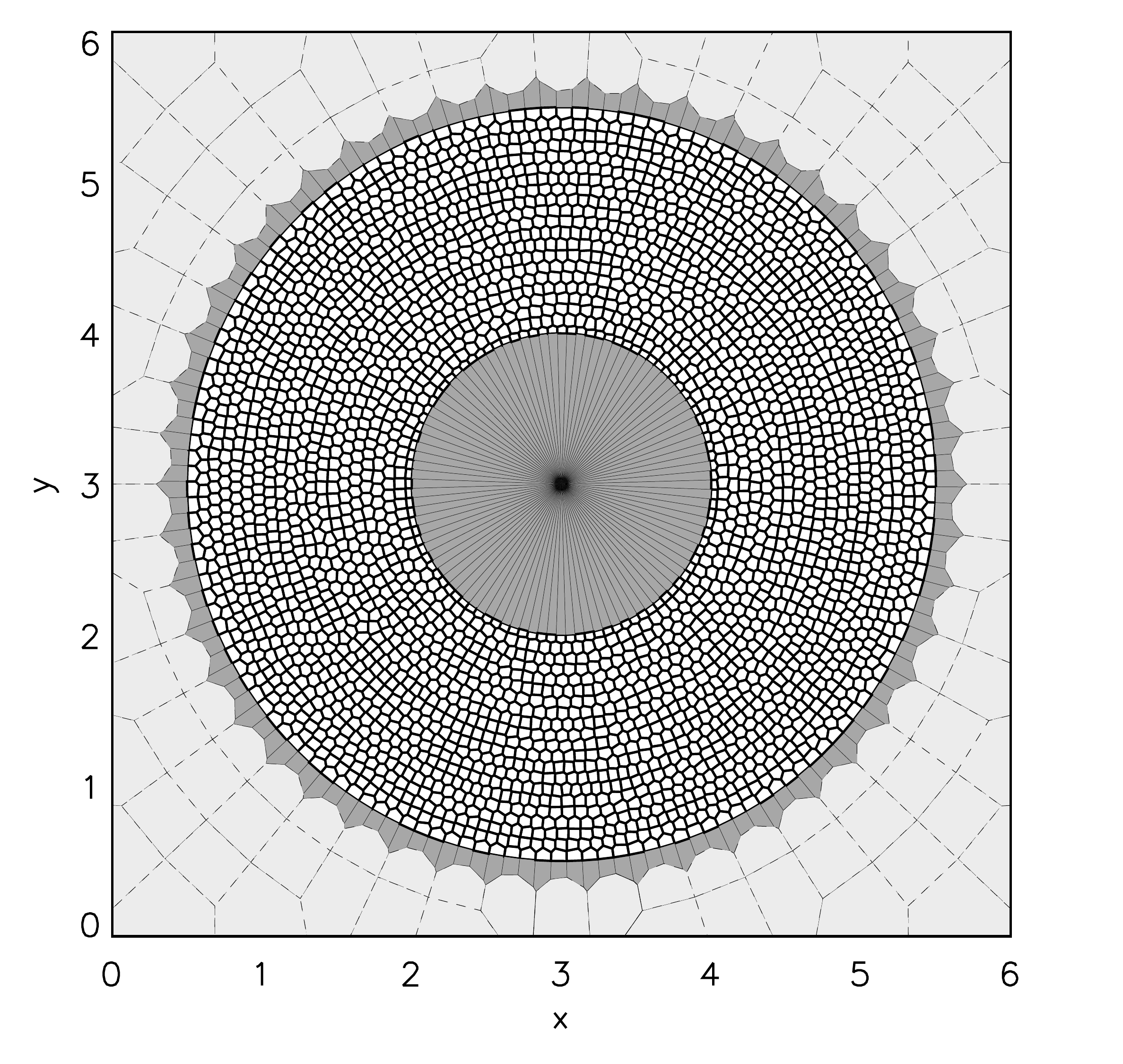}
 \vspace*{-10pt}
\caption{Mesh geometry for the circular Couette flow towards the end of
  the numerical integration. Even though we have started the
  calculation with an initially Cartesian mesh, this structure is
  quickly lost in favor of an on average azimuthal mesh geometry. \label{fig:couette_mesh_2}}
\end{figure}

Finally, we show in Fig.~\ref{fig:couette_mesh_2} the mesh geometry at
the end of the calculation.  Even though we have started the
calculation with an initially Cartesian mesh, the memory of this
geometry is lost during the calculation, and the mesh dynamically
adapts to the azimuthal flow structure present in this problem.  The
transition from a Cartesian grid towards a cylindrical-like mesh can
also be seen in the output sequence of the simulation shown in
Fig.~\ref{fig:couette_flow}, where the values of the radial position
of the cells start to segregate into a set of radial ``bins''.  The
number of these radial clusters corresponds to the average number of
cells along the radial direction.

   It is interesting to comment on the scatter of points --
  especially at the beginning of the simulation -- as seen in the angular
  velocity profile of Figure~\ref{fig:couette_flow}.  This is a
  reflection of the
  challenging initial mesh geometry. Although high-order
  schemes -- fifth or sixth order -- are not sensitive to the
  compliance of the mesh geometry with the flow, second order schemes
  are.  In this particular case, an axially symmetric mesh geometry would
  be more suitable due to the characteristics of the flow. However,
  the main point of this test is to show how the mesh responds to the
  evolution of the problem, achieving rough axisymmetry despite the
  unfavorable initial setup.

  As discussed by \citet{spr10}, our moving Voronoi mesh
  technique needs a ``quality control" to keep cells sufficiently
  regular in order to avoid large errors in the spatial reconstruction.
  However, this modification of the mesh motion comes at a price:
  imagine a very strong compression along one direction (e.g.~due to a
  very strong shock), then the mesh cells will acquire locally a high
  aspect ratio, which our mesh-quality control motions will try to
  eliminate, if needed by breaking the mesh symmetry (cell shapes are
  made ``round'' through small transverse motions). This is what
  happens when we start the Couette flow impulsively on a non-suitable
  mesh.  The introduction of asymmetries in the mesh can influence the
  flow, in particular in situations where fluid instabilities develop
  \citep[see the Kelvin Helmholtz instability test in][]{spr10}, where
  such asymmetric discretization errors can source growing
  perturbations. We note however that also on regular Cartesian meshes
  similar ``grid-sourcing'' errors exists. It appears unlikely
  that the poorer ability of the dynamic Voronoi mesh to maintain
  strict mesh symmetry is particularly detrimental for physical
  applications. 

\subsection{Flow Past a Circular Cylinder}
\begin{figure*}
\centering
\includegraphics[width=16cm]{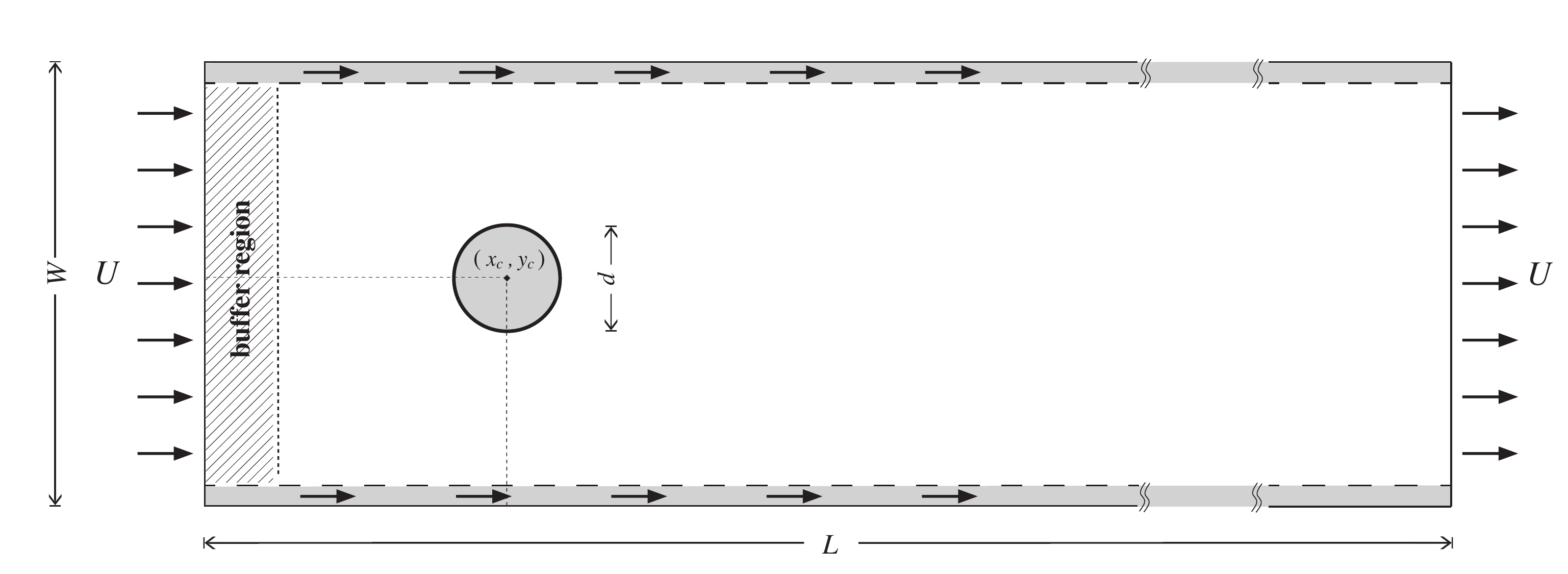}
\caption{Geometry of our wind tunnel set-up with a circular obstacle. \label{fig:wind_tunnel}}
\end{figure*}
We next consider the flow over a circular cylinder immersed in a wind
tunnel. The geometric setup of the problem is shown in
Fig.~\ref{fig:wind_tunnel}. The flow comes from the left at a fixed
horizontal velocity $U$. The upper and lower boundaries are also kept
at constant velocity $U$. Far from the cylinder, at the right end of
the computational domain, we impose again an exit velocity $U$. The
injection and exit regions are forced to have the prescribed inflow
and outflow mass fluxes at all times, something that we numerically
impose through small ``buffer'' regions as labeled in
Fig.~\ref{fig:wind_tunnel}. For static Cartesian grids, this buffer
region does not need to extend more than one cell in the
$x$-direction. However, moving grids require not only the injection of
momentum from the left, but also the injection of new mesh-generating
points, since the wind tunnel will otherwise produce a depletion of
cells at the left end as the mesh generating points drift to the right
in the direction of the flow. We address this issue by letting the mesh
automatically generate new cells through cell splitting, as already
implemented in \arepo\ \citep[see examples in][]{spr10}.  In doing this,
some attention must however be paid to guarantee that the new cells
reproduce the externally imposed inflow boundary conditions, which is
most easily achieved with a sufficiently broad buffer region on the
left end of the wind tunnel that covers the region where new cells are
injected.  Similarly, we employ the ability of \arepo\ to automatically
remove mesh cells to prevent them from piling up on the right end of the
wind tunnel. Altogether, we have created a wind tunnel that
is filled with a mesh that blows with constant velocity from left to
right, in a quasi-stationary state.

\begin{figure}
\centering
\includegraphics[width=8.0cm]{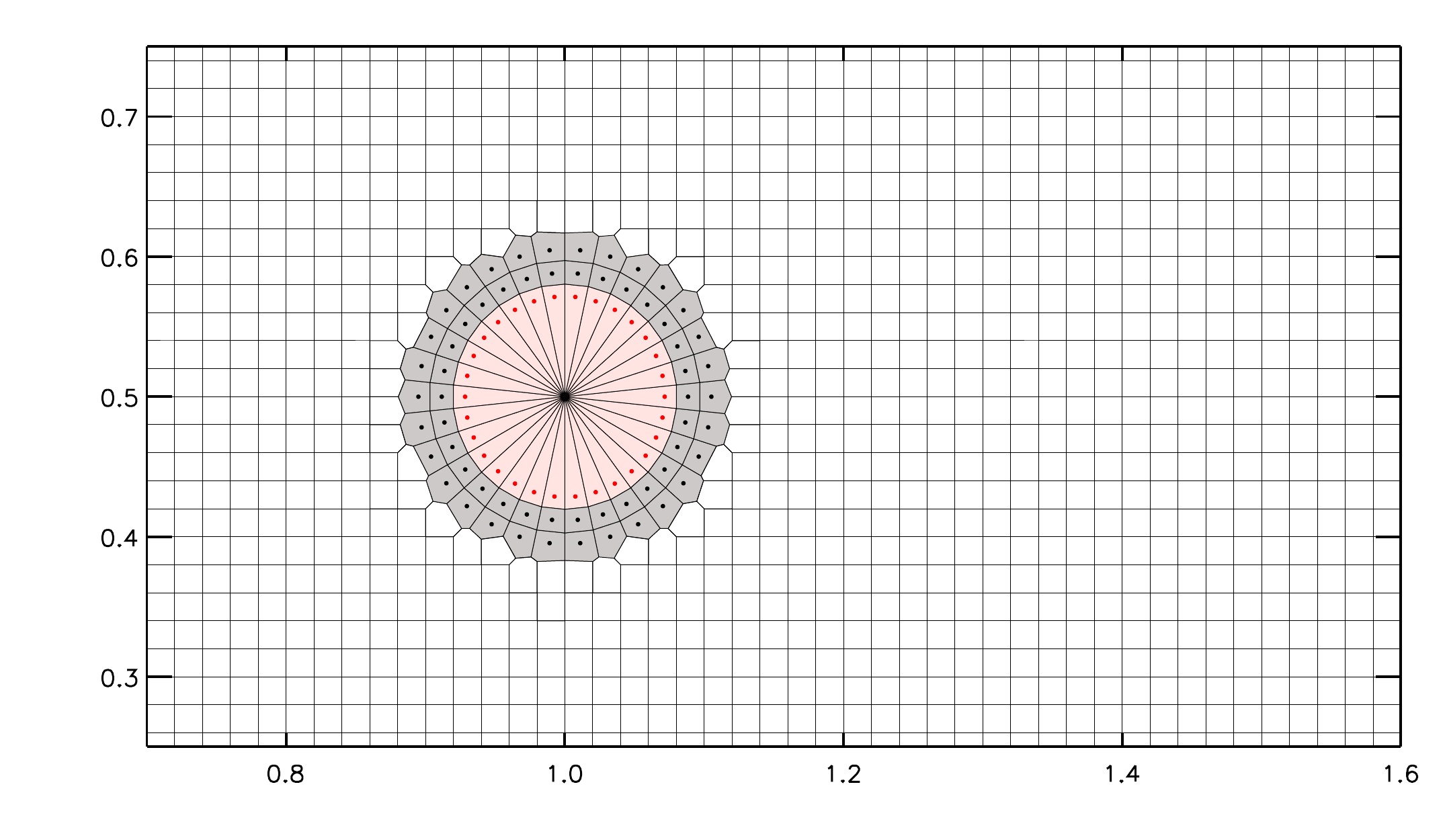}
\includegraphics[width=8.0cm]{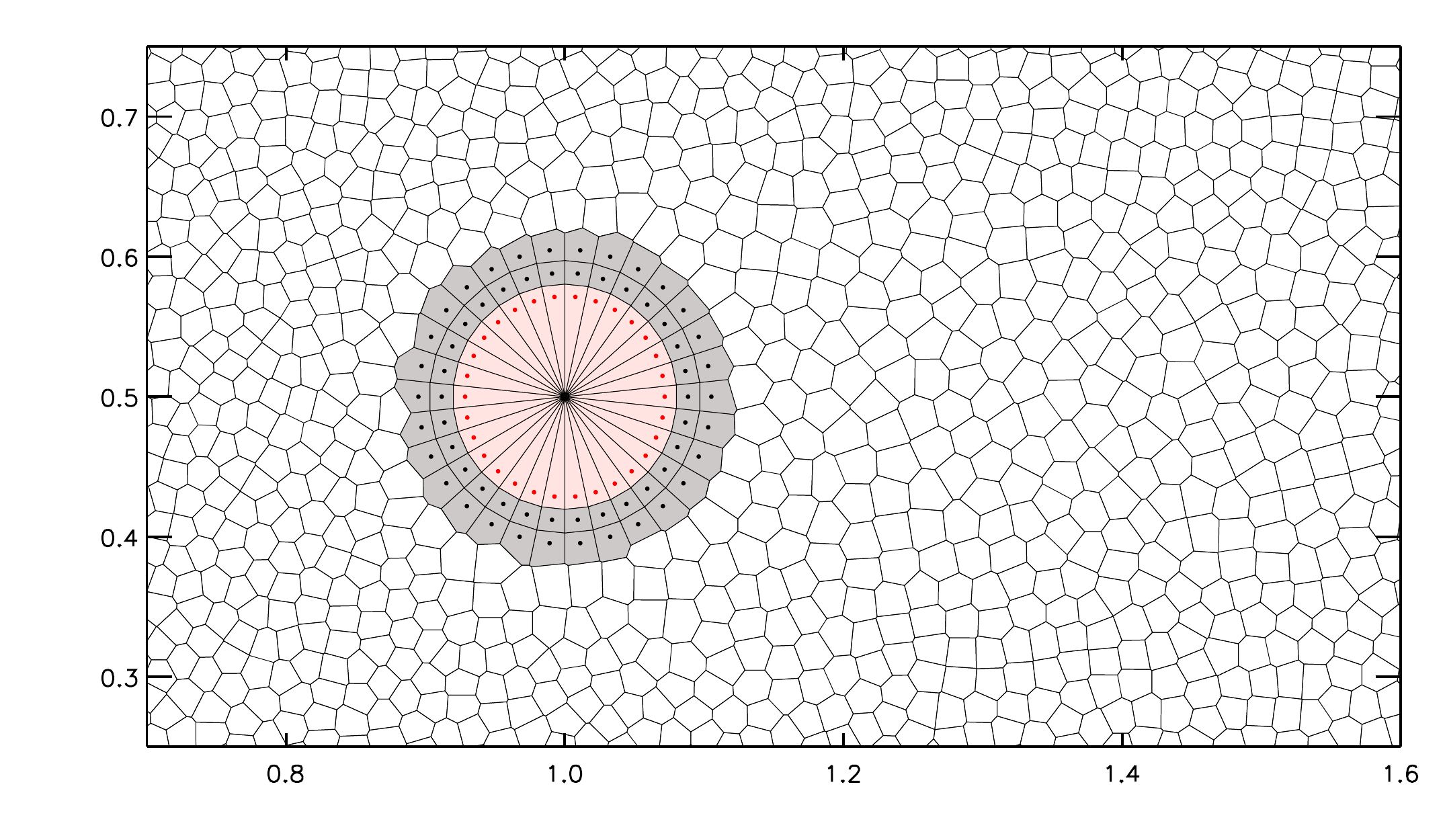}
\caption{Mesh near a circular cylinder inside a wind tunnel. The mesh
  contains both stationary mesh-generating points (defining the solid
  cylinder and two layers of cells used to create the cylindrical
  solid surface) and moving mesh-generating points (the remainder of the
  grid). The upper panel shows the initial setup, which highlights the
  cells representing the solid cylinder, and the two layers of fluid
  cells for which the equations of hydrodynamics are solved as in a
  standard stationary mesh. The total number of cells in the wind
  tunnel is $12,478$ (roughly $250\times50$). The perimeter of the
  cylinder is outlined by 30 cells, and its diameter is equivalent to
  eight cells across. The Voronoi faces in between the red and grey
  cells define the boundary at which the no-slip condition is
  imposed. The lower panel shows the same region at a later
  time. Whereas one layer of mesh-generating points surrounding the
  cylinder has remained stationary, the rest of the background mesh
  has moved downstream and transformed to a generic unstructured
  Voronoi mesh as it moves along with the fluid. \label{fig:cylinder_mesh}}
\end{figure}

The other geometric parameters of the test problem we simulate here
are the diameter $d$ of the cylindrical obstacle, the width $W$ of
the tunnel and its length $L$. We have chosen $W=6.25\,d$ and
$L=5\,W=31.25d\,$, and have scaled all length units such that
$W=1.0$. The flexibility of the Voronoi mesh allows us to easily embed
a cylindrical obstacle within the initially Cartesian background grid
that fills the tunnel.  Fig.~\ref{fig:cylinder_mesh} shows how we can
tailor the mesh construction to reproduce the curved surface of the
cylinder, using techniques similar to those that we used for the
circular Couette flow problem.

The physical properties of the problem are primarily determined by the
external velocity of the flow, $U$, and the dynamic viscosity of the
fluid $\mu$. In our numerical experiments we set the external flow
velocity to $U=0.5$, and combine this with constant initial pressure
and density ($\rho=P=1$). We take the fluid to be described by an
ideal gas equation of state with adiabatic index $\gamma=5/3$. The
characteristic Reynolds number of the problem can then be defined by
\begin{equation}
Re=\frac{U\,d}{\nu}=\frac{U\,d\,\rho}{\mu}
\end{equation}
where $\rho$ might however vary in time and space since the flow is
fully compressible.

We have performed several numerical experiments of this problem using
the viscous module added to \arepo. In each of these simulations, the
Reynolds number is the only relevant quantity being changed. This is
accomplished by changing $\mu$ exclusively, while keeping the other
parameters fixed.  Fig.~\ref{fig:cylinder_mesh} (upper panel) shows
the initial setup for all the runs, which consist of a circular
cylinder plus a Cartesian background grid of $250\times50$ mesh
generating points. The dynamic viscosity coefficient $\mu$ takes five
different values: $2.5\times10^{-2}$, $5\times10^{-3}$,
$2.5\times10^{-3}$, $1.25\times10^{-3}$ and $8.3\times10^{-4}$. These
values correspond to Reynolds numbers of 2, 10, 20, 40 and 60.

\begin{figure*}
\subfigure[$Re=2$]{\includegraphics[width=12.5cm]{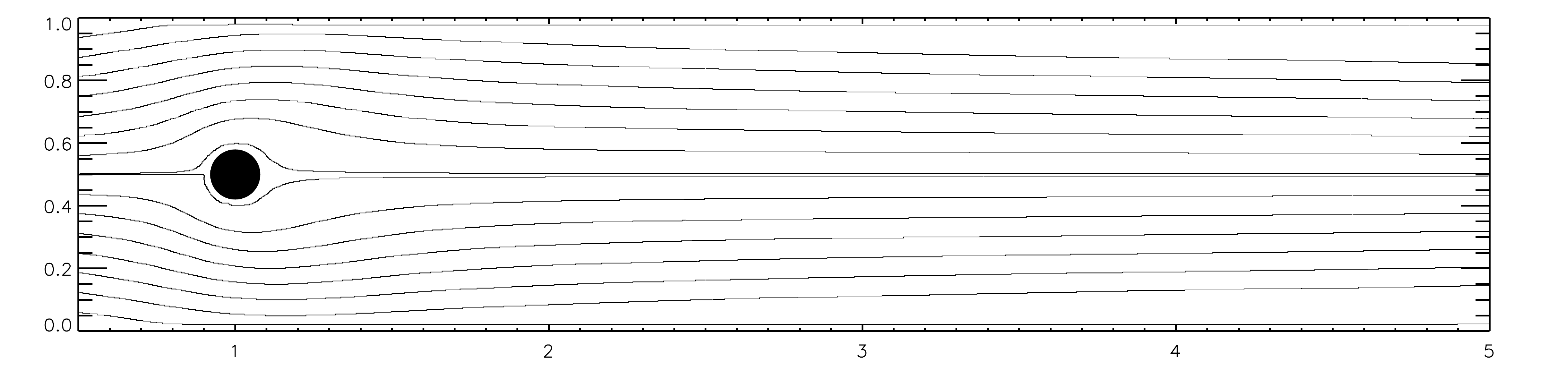}}
\subfigure[$Re=10$]{\includegraphics[width=12.5cm]{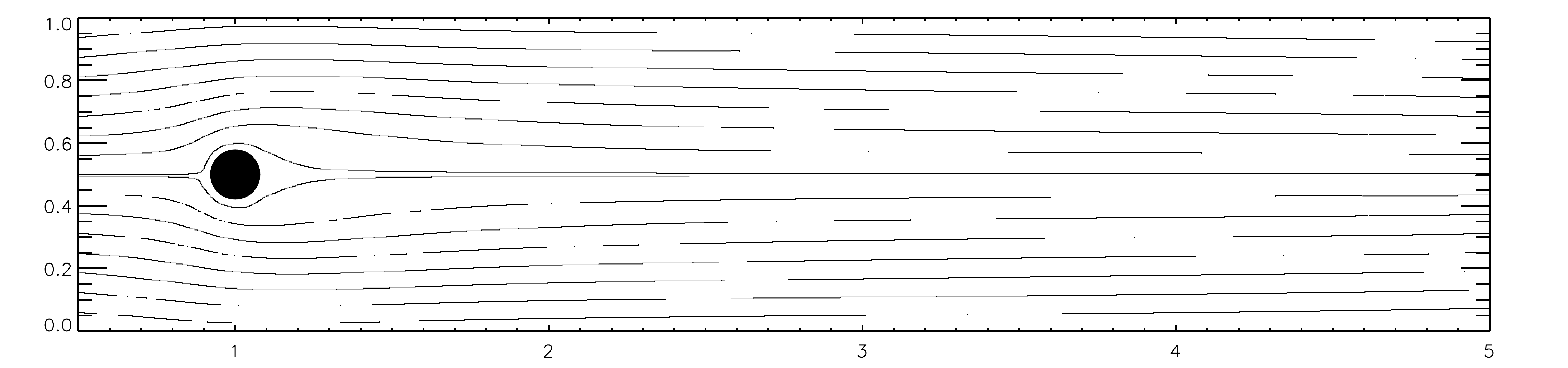}}
\subfigure[$Re=20$]{\includegraphics[width=12.5cm]{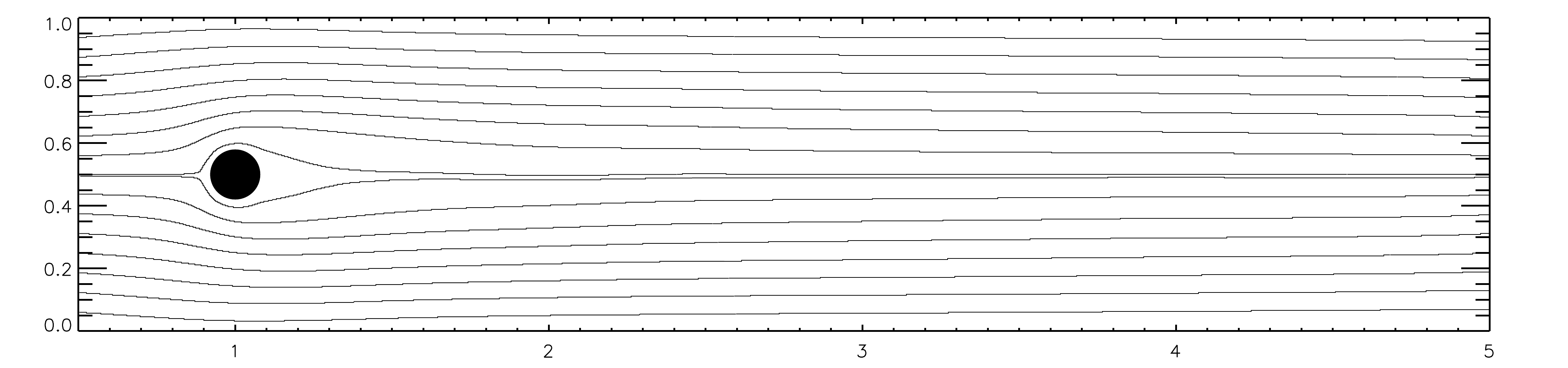}}
\subfigure[$Re=40$]{\includegraphics[width=12.5cm]{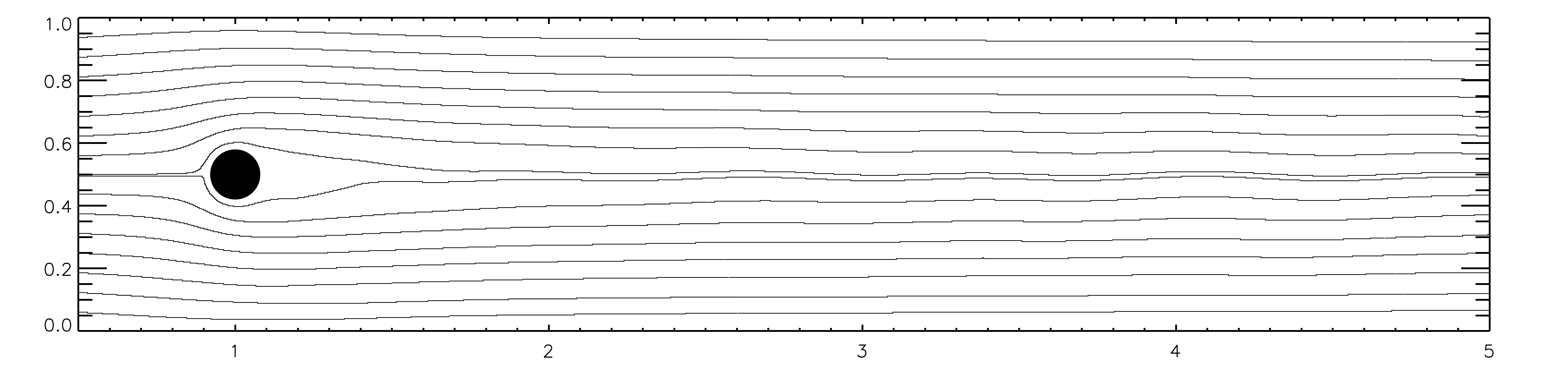}}
\subfigure[$Re=60$]{\includegraphics[width=12.5cm]{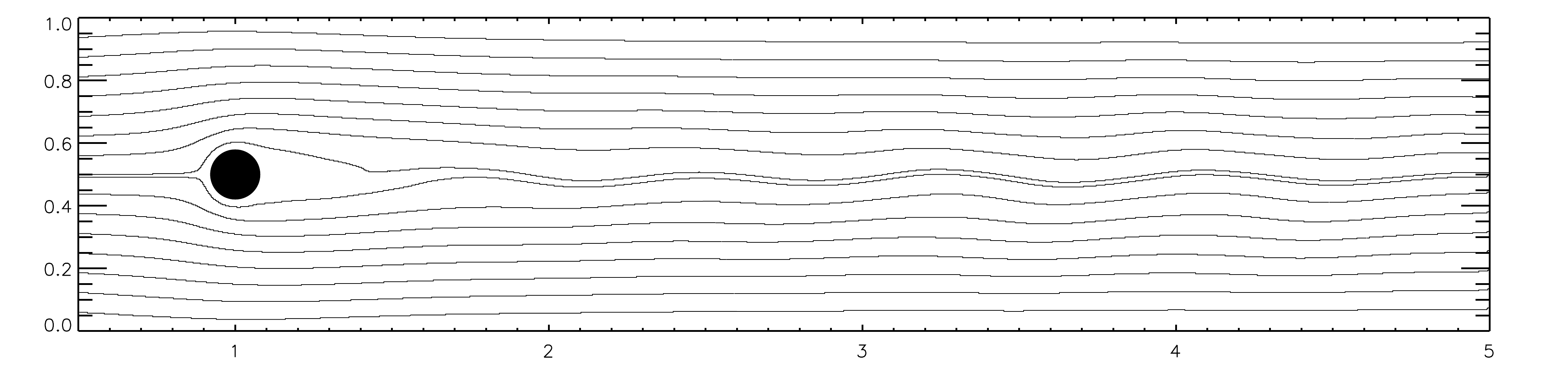}}
\caption{Streamlines for compressible gas flow around a cylinder at
  five different Reynolds numbers, as labeled. \label{fig:cylinder_streamlines1}}
\end{figure*}

For each one of the tests, we show the resulting streamlines at time
$t=9.9$ (or an equivalent dimensionless time of $\bar{t}=tU/d\approx
31.0$) in Fig.~\ref{fig:cylinder_streamlines1}.  Below $Re\sim40$, the
flow is steady and symmetric above and below the cylinder.  As the
Reynolds number increases, the size of the wake behind the cylinder
grows. Although in this example the structure of the wake is poorly
resolved, the increase in $Re$ is accompanied by an increase of
vorticity confined within the wake.


\begin{figure*}
\includegraphics[width=12cm]{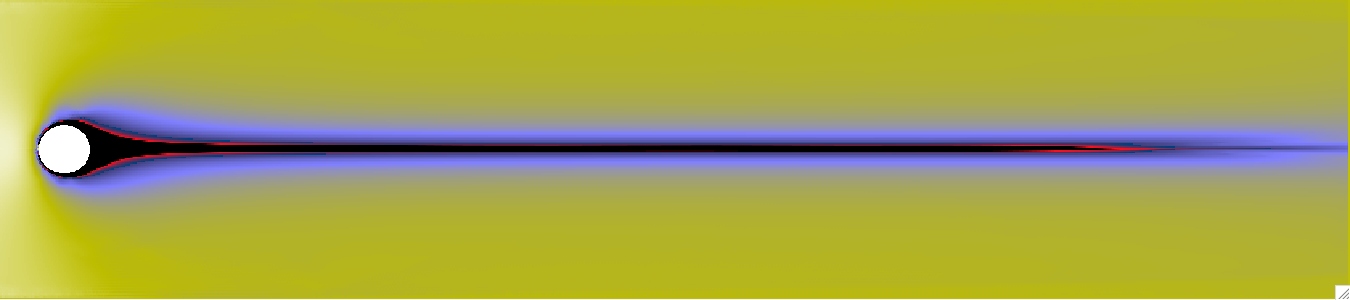}
\includegraphics[width=12cm]{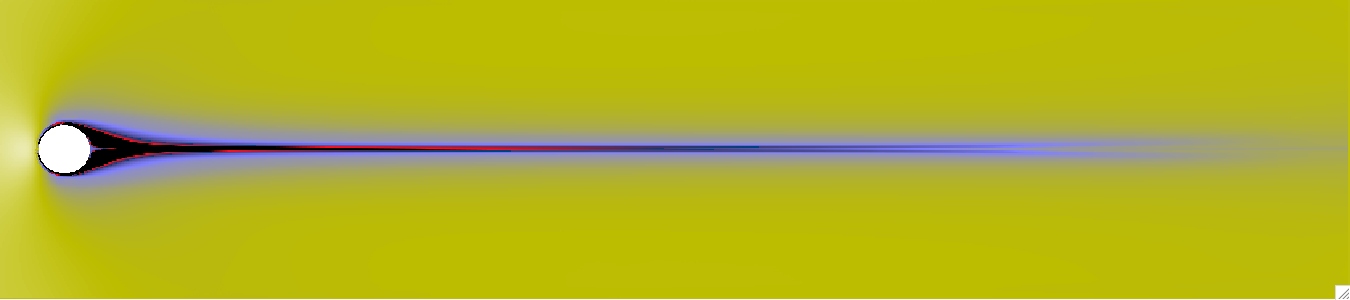}
\includegraphics[width=12cm]{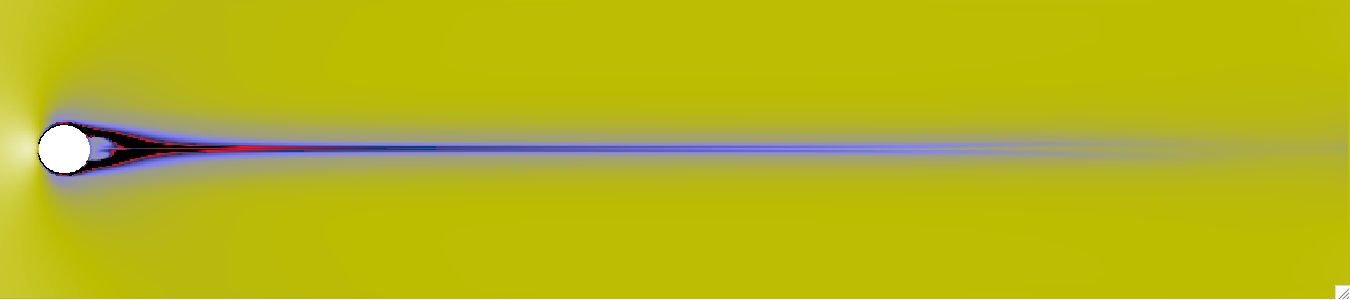}
\includegraphics[width=12cm]{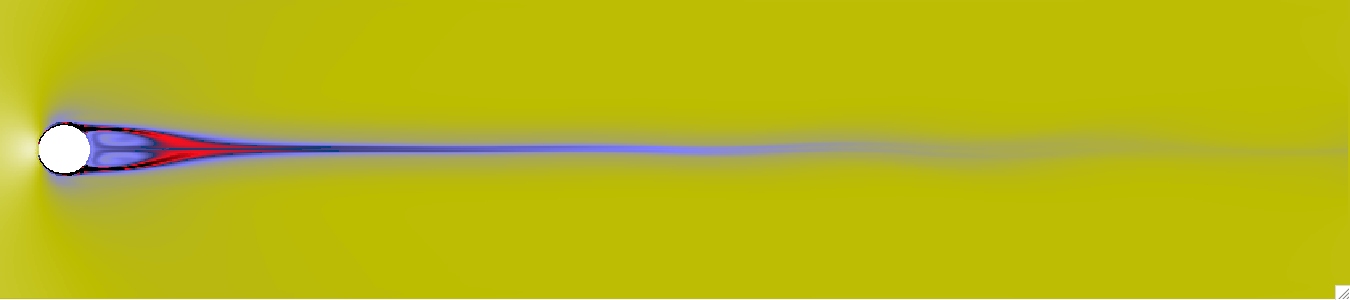}
\includegraphics[width=12cm]{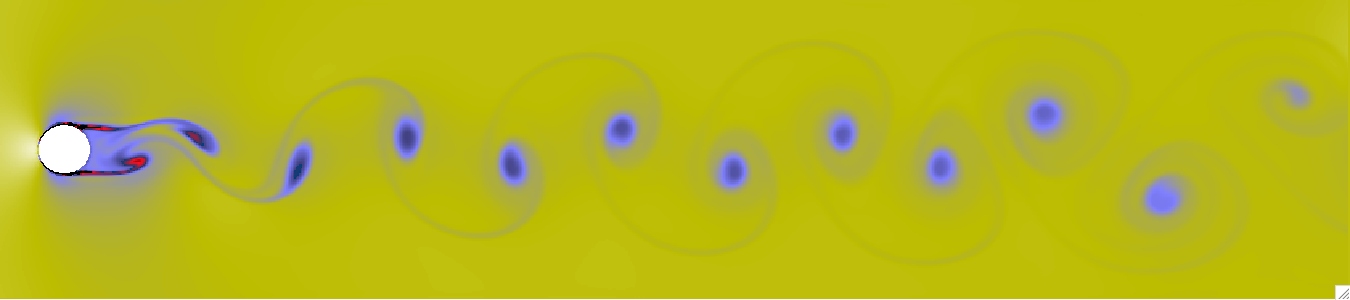}
\caption{Density contrast of compressible flow past a cylinder at five
  different Reynolds numbers, corresponding to $Re=2$, $10$, $20$,
  $40$, $100$, from top to bottom. All five numerical experiments were
  computed with a static Cartesian mesh at moderately high resolution
  ($1000\times250$), where the cell size is 1/32 of the cylinder's
  diameter. \label{fig:cylinder_density}}
\end{figure*}

Above $Re\sim40$, the wake behind the cylinder starts to become
unstable. This can be clearly seen in the streamline pattern of the
$Re=60$ panel. As the wake becomes unstable, the symmetry between the
upper and lower portions of the domain is broken, at which point
the flow becomes unsteady, such that the streamlines are no longer a
valid representation of the Lagrangian trajectories of fluid
parcels. This marks the onset of the {\it von Karman vortex street},
and the eventual transition to fully developed turbulence.

To further illustrate the flexibility of the mesh construction in
\arepo, we can repeat this experiment with the mesh generating points
set to remain static, thus recovering an Eulerian grid code. In
addition, we increase the resolution by a factor of four to better
resolve the wake behind the cylinder. In
Fig.~\ref{fig:cylinder_density}, we show the density contrast for five
different Reynolds numbers. For stationary flow, the density
distribution traces the streamline topology. At $Re=100$, we can
appreciate how the fully developed von Karman vortex street looks for
a compressible gas.

\subsection{Three Dimensions: Taylor-Couette Flow}

Circular Couette flow is a stable, special case of the more complex
and richer three-dimensional Taylor-Couette flow \citep{tay23}. Taylor
found that when the angular velocity of the inner cylinder is
increased above a certain threshold, Couette flow becomes
unstable. After this transition, different states have been
identified, the most famous of which is the Taylor vortex flow,
characterized by axisymmetric toroidal vortices. The diversity of
states for Taylor-Couette flow has been explored in the past, most
notably by \citet{col65} and \citet{and86}. The latter work lists up
to 18 different flow regimes observed in flow between independently
rotating cylinders. Its ``Andereck diagram", which explores the
stability of the Taylor-Couette problem for a variety of Reynolds numbers,
has become the standard benchmark for computational experiments of
flow between rotating cylinders.

\begin{figure}
\centering
\includegraphics[width=7.4cm]{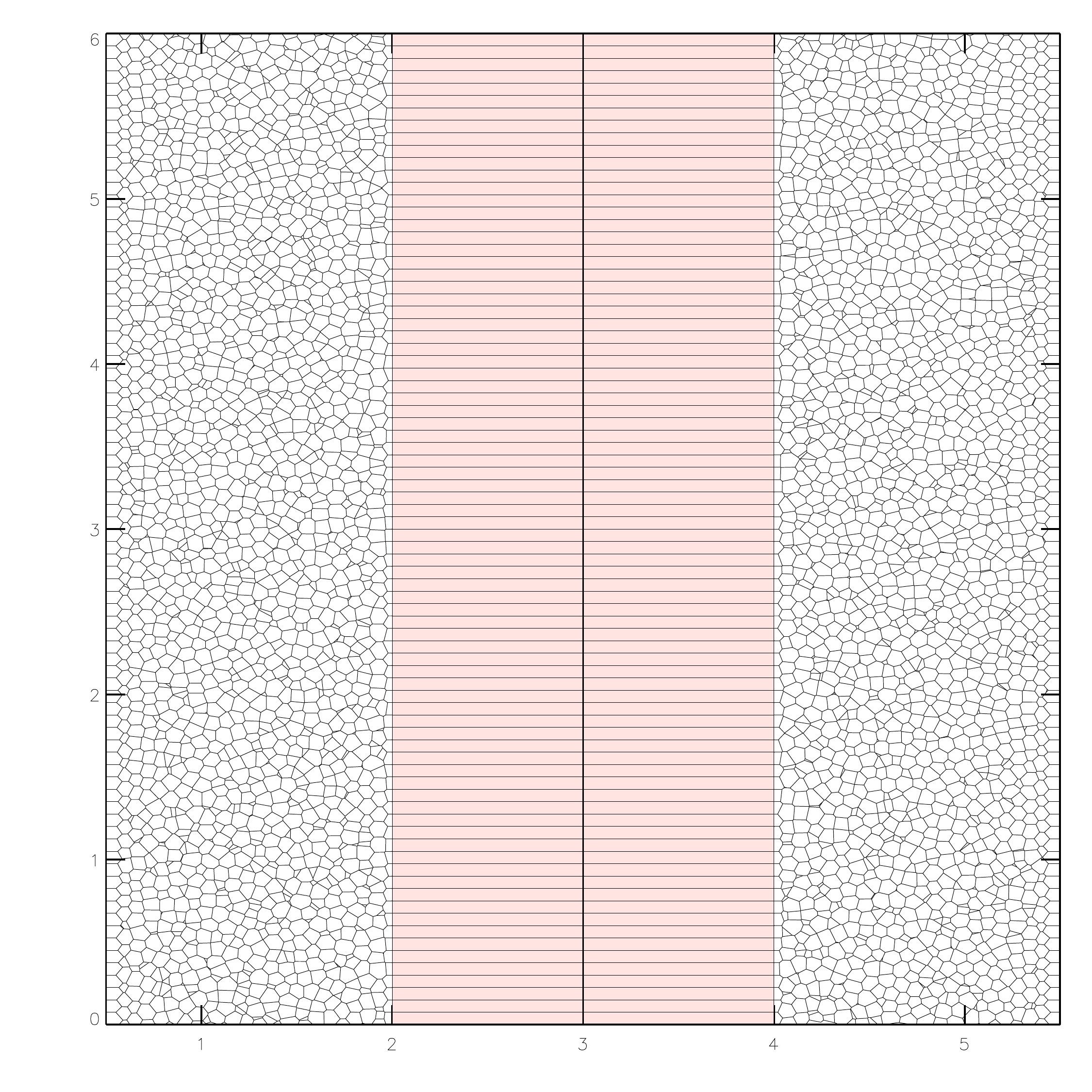}
 \vspace*{-15pt}
\caption{Vertical slice of the three-dimensional Voronoi tessellation
  in Taylor-Couette flow at the time Taylor vortices have
  developed. This same slice is used when visualizing the $v_x$,$v_y$
  and $v_z$ fields (Fig.~\ref{fig:taylor_couette_geometry}b and
  Fig.~\ref{fig:taylor_couette_velocity}).\label{fig:taylor_couette_mesh}}
   \vspace*{-10pt}
\end{figure}

  Although the computational and experimental study of three-dimensional Couette
  flow peaked during the 1980's with the classical works of
  \citet{and86} and \citet{mar84a,mar84b}, in recent years it has
  regained popularity \citep[e.g.][]{don07,avi08,mes09a,mes09b} mainly
  driven by the experimental studies of magnetized and unmagnetized
  rotating flows of \citet{ji01,ji06} and \citet{sis04}, which have
  resulted in significant progress on the characterization of the
  magnetorotaional instability \citep[MRI;][]{bal98} in the
  laboratory.

  In this section, we briefly explore the evolution of
  Taylor-Couette flow on a moving Voronoi mesh.  Although the \arepo\
  code is not specifically designed for problems with symmetric
  geometries where static cylindrical meshes have proven to be more
  suitable, we have included this test to emphasize that our method
  works in three dimensions in an analogous way to the two-dimensional
  examples shown above. It is straightforward to extend the
  two-dimensional Couette flow shown above to three dimensions using
  \arepo.  Since the mesh is obtained from a distribution of
  mesh-generating points, all that is needed is to replicate the
  initial conditions shown in Figure~\ref{fig:couette_flow} in the
  vertical direction (about 80 times) to fill up a cubic box.

   A standard validation for a Taylor-Couette simulation with
  azimuthal and axial periodicity may include, for example (Avila,
  private communication): obtaining perfect axial symmetry at low
  Reynolds number (i.e.~circular Couette flow ), followed by obtaining
  the first bifurcation to axially symmetric Taylor vortices, and by
  reaching the second bifurcation to wavy vortices. These transitions
  occur sequentially as the angular velocity of the inner cylinder is
  increased while keeping the outer cylinder stationary (see the phase
  diagram of \citealp{and86}). However, it is not the purpose of this
  section to explore these transitions exhaustively; we only want to
  show that the third dimension works with our technique. We thus have
  focused on a particular configuration: counter-rotating
  Taylor-Couette flow, for which it is easy to obtain axially
  symmetric Taylor vortices \citep[although these might relax back to
  Couette flow after several rotation periods; e.g.][]{lia99}. The
  geometry described in Fig.~\ref{fig:couette_geometry} is replicated
  in the vertical direction such that the computational domain is now
  a cube of dimensions $6\times6\times6$, with periodic boundary
  conditions in the $z$-direction. The initially Cartesian mesh will
  eventually relax in all directions as the flow evolves
  (Fig.~\ref{fig:taylor_couette_mesh}). The cylinder is effectively
  infinite, like in the two-dimensional case, except that this time
  there is no imposed symmetry along the $z$-direction. We choose the
  cylinder radii to be $R_1=1.0$ and $R_2=2.5$, just like in the 2D
  example, and the respective angular velocities are $\Omega_1=0.8$
  $\Omega_2=-0.5$ (counterrotating). The dynamic viscosity is
  $\mu=0.005$ and the fluid is started from rest with $\rho=P=1$. The
  inner and outer Reynolds numbers \citep[$Re_{i}\equiv
  R_i\Omega_i(R_2-R_1)\rho/\mu$; e.g.][]{lia99} are $R_1=240$ and
  $R_2=-375$, respectively.

\begin{figure*}
\centering
\subfigure[]{\includegraphics[width=6.5cm]{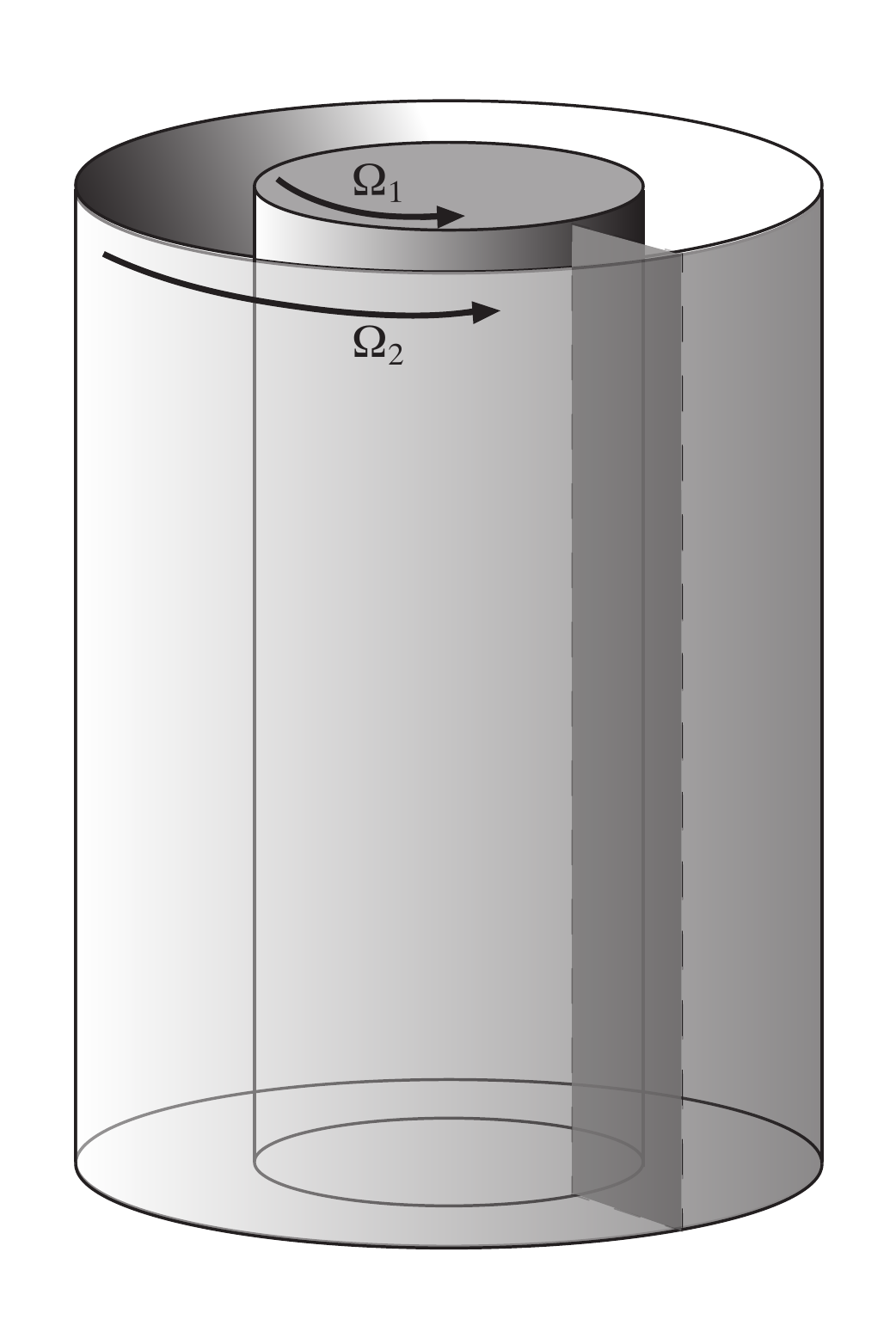}}
\subfigure[]{\includegraphics[width=9.5cm]{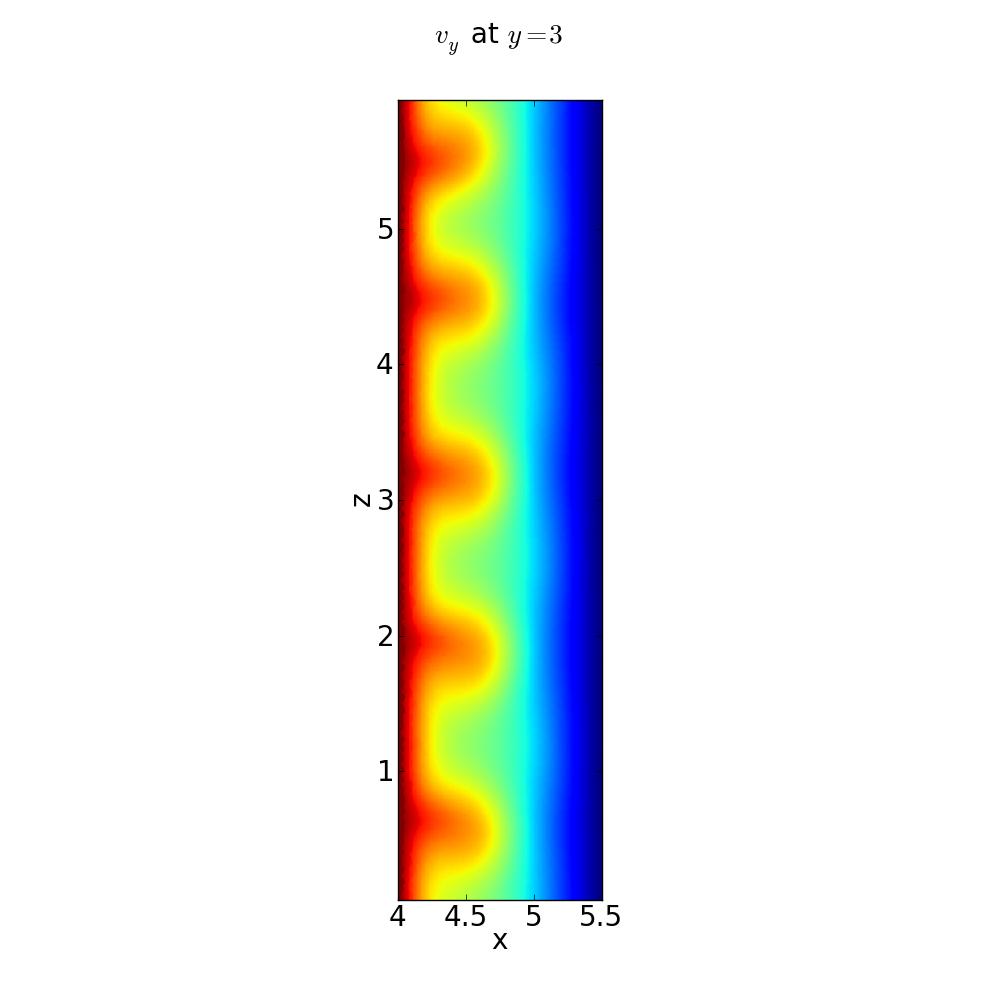}}
\caption{a) Illustration of the three-dimensional flow between two
  independently rotating cylinders. The figure shows a plane along the
  radial direction where the local velocity field is evaluated. b)
  Velocity field in the $y$-direction between the cylinders for a
  slice defined by $y=3$ (i.e. along the diameter of both
  cylinders). For this particular plane, $v_y$ is equivalent to the
  azimuthal velocity $v_\theta$. The color scale goes from
  $v_y=\Omega_2R_2=-1.25$ (blue) to $v_y=\Omega_1R_1=0.8$ (red). This
  example shows that $v_\theta$ is no longer independent of $z$. Thus
  the two-dimensional solution of Eq.~(\ref{eq:couette_analytic}) is no
  longer valid. \label{fig:taylor_couette_geometry}}
\end{figure*}

\begin{figure*}
\centering
\subfigure[]{\includegraphics[width=8.5cm]{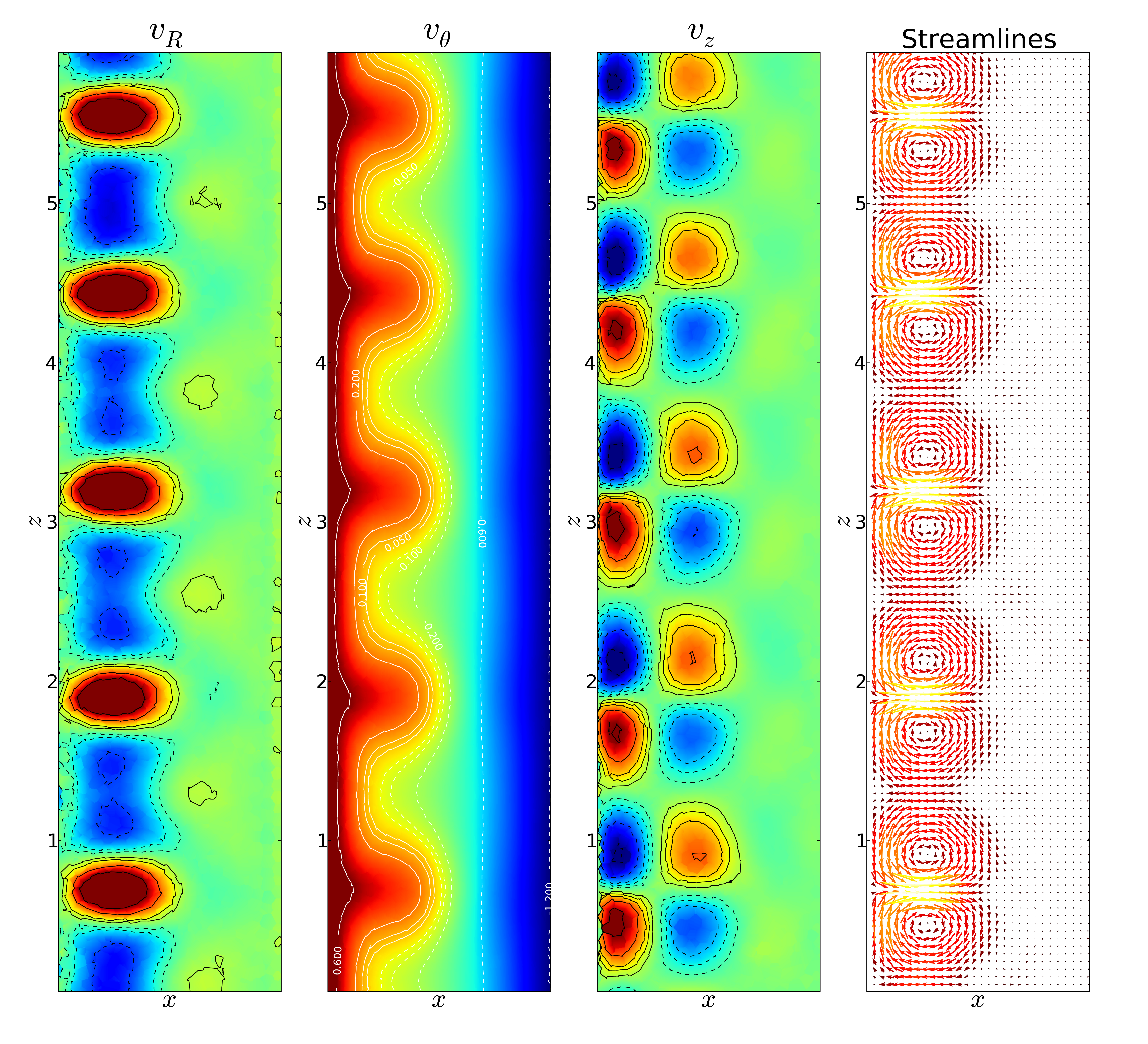}}
\subfigure[]{\includegraphics[width=8.5cm]{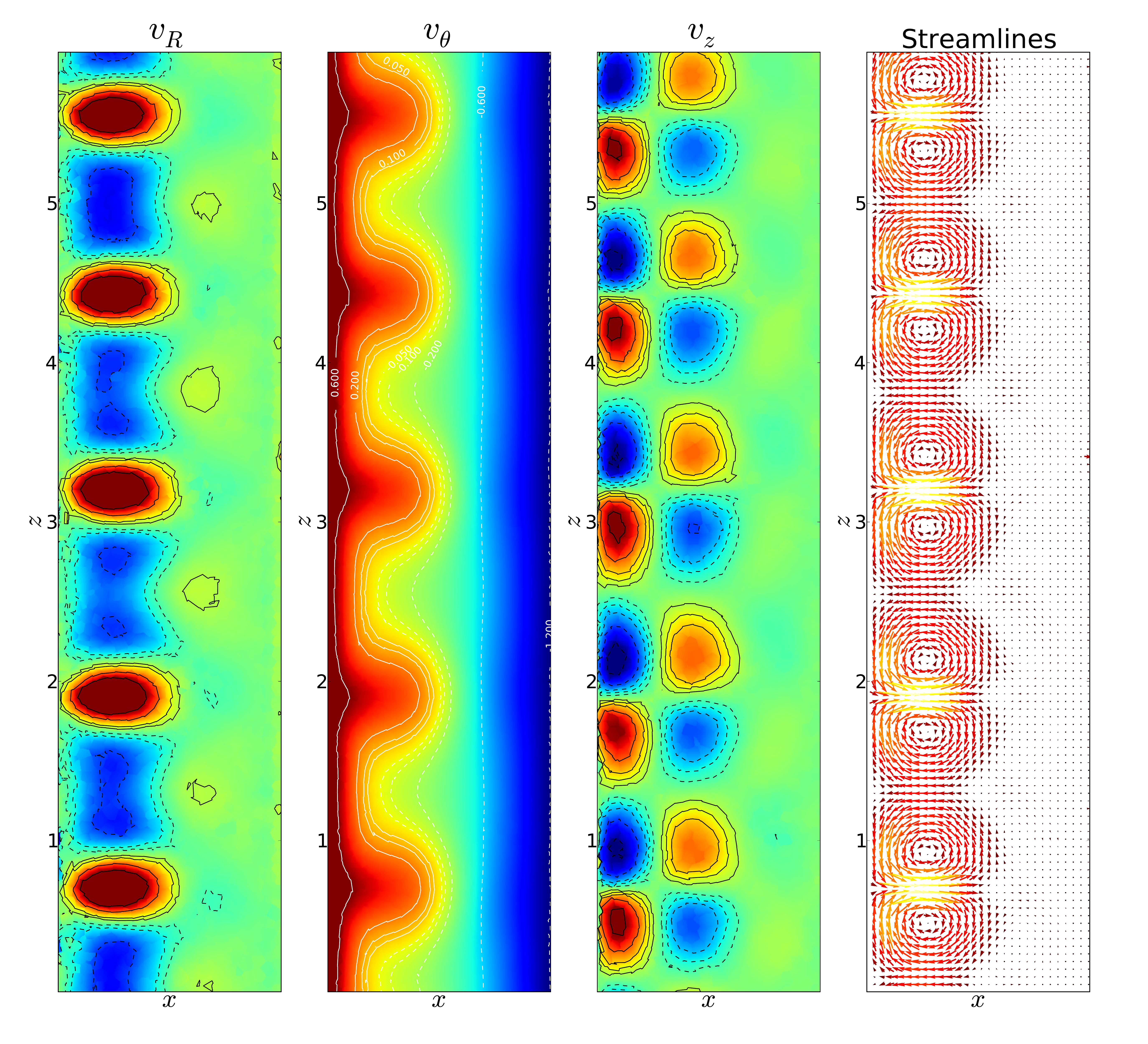}}
\vspace*{-5pt}
\caption{ Velocity structure of axisymmetric Taylor vortex flow at time $t=119$ at two
different meridian planes separated by $90^\circ$. The poloidal velocity field ($v_R,v_z$) is color mapped
in the linear range  [$-0.08$ (blue), $+0.08$ (red)], while the azimuthal field ($v_\theta$) is color mapped
in the linear range [$-1.2$ (blue), $0.6$(red)]. The streamlines illustrate the vector field in the poloidal plane,
showing with clarity the nature of Taylor vortices.}
\label{fig:taylor_couette_velocity}
\vspace*{-5pt}
\end{figure*}

The geometry of the problem is shown in
Fig.~\ref{fig:taylor_couette_geometry}a. A vertical slice is taken at
a time when the Taylor vortices have developed (the corresponding
sliced mesh is shown in Fig.~\ref{fig:taylor_couette_mesh}).  The
azimuthal velocity on that slice shows deviations from the symmetry in
$z$ present in the circular Couette regime
(Fig.~\ref{fig:taylor_couette_geometry}b). Looking at the poloidal
velocity field on that same slice ($v_x$ and $v_z$ in
Fig.~\ref{fig:taylor_couette_velocity}) one can appreciate, near the
inner cylinder, the circular vertical motion characteristic of the
Taylor vortices.

  In Figure~\ref{fig:taylor_couette_velocity}, we show the velocity
  field of this Taylor-Couette experiment at time $t=128$ for two
  different slices of the volume: (a) along the $x$-axis, and (b)
  along the $y$-axis (i.e. at $90^\circ$ from the first slice).
  Except for the numerical noise, the two solutions are nearly
  indistinguishable, evidence of a global axially symmetric Taylor
  vortex flow (for a very similar configuration, see Fig.~3 in
  \citealp{lia99}). This flow starts to develop at time $t\sim60$ and
  remains essentially unaltered for several rotation periods. At much
  longer time scales, the flow would presumably decay back to a
  two-dimensional Couette flow as seen in the roughly similar test
  carried out by \citet{lia99}.

\section{Concluding Remarks}

We have presented a new numerical approach for solving the
three-dimensional, compressible NS equations on a dynamic
mesh using the new astrophysical code {\small AREPO}. This novel
approach, an extension of the finite volume method, defines the
computational mesh as a Voronoi tessellation moving with the local
flow. The advantages of using a dynamic Voronoi mesh for transient and
stationary flows under diverse boundary conditions has been
addressed. The implicit adaptivity of the quasi-Lagrangian mesh
elements, in addition to the well-behaved topological properties of
the Voronoi tessellation, ensure both geometric flexibility and low
numerical diffusivity. In addition, the shock capturing,
second-order-accurate finite-volume scheme implemented in the
rest-frame of each moving cell provides high accuracy.

We have described in detail the algorithm used to estimate the viscous
diffusion of momentum across inter-cell boundaries. Our scheme
produces smoothly varying estimates of the viscous terms, resulting in
accurate and stable solutions. The method extends previously known
finite-volume formulations of the NS equations with the
introduction of a new reconstruction scheme that represents a
compromise between the use of piece-wise constant gradients and fully
consistent quadratic-reconstruction schemes.
 
  For pure hydrodynamic flow, the CPU time consumption of our code
  per timestep is typically quite a bit higher than for structured
  mesh codes or SPH codes, for the same number of resolution
  elements. In three dimensions, the factor is close to $2$ relative
  to SPH (if 64 smoothing neighbors are used), and up to $\sim 3$
  relative to a Cartesian mesh codes.  The additional computational
  time goes mostly into the Voronoi mesh construction overhead, which
  is simply not needed by a structured mesh code, and also into an
  enlarged computational cost for the flux computations. The latter
  comes about because of a larger number of faces per cell (in 3D,
  there are 6 sides for a cubical cell, but for a 3D Voronoi mesh, we
  have of order $\sim 12$ sides for each polyhedral cell). It is
  however important to note that other, problem-dependent factors
  should be taken into account when assessing the performance in
  practice. For example, if large bulk velocities are present, our
  method can take considerably larger timesteps than a corresponding
  fixed mesh code. Also, because the advection errors are reduced in
  our scheme, fewer cells are required to reach a given accuracy, so
  that our code can then end up being computationally more
  efficient. We also note that once self-gravity is included (as in
  many of our primary target applications in astrophysics), the
  relative speed difference in the hydrodynamic part between the
  structured fixed mesh and our moving Voronoi mesh becomes much less
  of an issue, because the cost of calculating self-gravity
  sufficiently accurately for arbitrary geometries substantially
  reduces the relative importance of the hydrodynamical cost.

As part of our study, we have verified the reliability of our new
method through a series of example calculations that range from simple
flows with known analytic solutions to traditional experiments of
well-known quantitative behavior. The demonstrated ability of the
scheme to reproduce exact solutions as a function of time, even if the
flow is started impulsively from rest, is reassuring. These examples
also show the flexibility of the scheme in the presence of different
solid surfaces moving in diverse ways. In all of these examples, the
overall structure of the numerical scheme is identical, and the
boundary conditions are set solely by the prescribed motion of the
surfaces, which consist of controlled collections of Voronoi cells.

  Although we have tested the performance of \arepo\ in
  configurations possessing a high degree of symmetry, it is in
  complex asymmetric problems where the moving-mesh approach would
  show all its power. The flexibility of the Lagrangian nature of the
  mesh will allow us to simulate, for example, complex astrophysical
  objects where viscosity is presumed to play a significant role. One
  such problem is the simulation of accretion disks around young
  stars. Although angular momentum transport in accretion disks is
  attributed to turbulence (most likely of magneto-hydrodynamic
  nature), this process is usually modeled both analytically
  \citep[e.g.][]{sha73,lyn74,pri81,lin87} as well numerically
  \citep[e.g.][just to name a few]{kle92,
    mas00,ang02,val06,paa06,mud09} by laminar flow in the presence of
  turbulent viscosity (Boussinesq approximation to eddy viscosity),
  due to the computational cost of global models of
  magneto-hydrodynamic disks. Another application of viscous flow is
  the plasma viscosity at galaxy cluster scales
  \citep[e.g.][]{sij06}. However, it is likely that in such systems
  viscosity, as well as thermal conduction, is anisotropic
  (\citealp{bra65}; see \citealp{don09} for an example). In such a
  case, the viscous stress tensor in Eq.~(\ref{eq:viscous_tensorA}) can
  be easily generalized to include the up to seven independent
  viscosity coefficients \citep{lifshitz}. It will be particularly
  exciting to couple the local anisotropy directly to the magnetic
  field topology, with the latter calculated self-consistently using
  a recent magnetohydrodynamics implemention in {\small AREPO} \citep{pak11}.

Its powerful flexibility will make {\small AREPO} an interesting code
both for astrophysical simulations of viscous flow, but
potentially also in engineering applications where the ability to cope
with curved and moving boundaries is particularly attractive.

\section*{acknowledgements}
{The simulations in this paper were run on the Odyssey cluster
 supported by the FAS Science Division Research Computing Group 
 at Harvard University. We are thankful to Joseph Barranco,  Paul Duffel,
 Patrik Jonsson,  Andrew MacFadyen and Debora Sijacki for helpful discussions.}  



 \appendix

\section{Gradient Extrapolation Coefficients}

The extrapolation of the velocity gradients
(e.g. Eq.~\ref{eq:velocity_extrapolation}) requires a numerical
estimate of the gradient matrix as well as an estimate for the time
derivative of the gradient. For the latter, the tensors
$A_{\alpha\beta b}$ and $B_{\alpha\beta b a}$ are needed
(Eq.~\ref{eq:euler_derivatives}). Both tensors depend on the
cell-centered scalar quantities as well as their gradients. The values
of $A_{\alpha\beta b}$ are \citep[e.g.][]{tor09}

\begin{equation}
A_{\alpha \beta x}=A_{\alpha\beta1}=\begin{pmatrix}
v_x &\rho&0&0&0\\
0 &v_x&0&0&1/\rho\\
0 &0&v_x&0&0\\
0 &0&0&v_x&0\\
0 &\gamma P&0&0&v_x\\
\end{pmatrix},
\end{equation}
\begin{equation}
A_{\alpha\beta y}=A_{\alpha\beta2}=\begin{pmatrix}
v_y &0&\rho&0&0\\
0 &v_y&0&0&0\\
0 &0&v_y&0&1/\rho\\
0 &0&0&v_y&0\\
0 &0&\gamma P&0&v_y\\
\end{pmatrix},
\end{equation}
\begin{equation}
A_{\alpha\beta z}=A_{\alpha\beta3}=\begin{pmatrix}
v_z &0&0&\rho&0\\
0 &v_z&0&0&0\\
0 &0&v_z&0&0\\
0 &0&0&v_z&1/\rho\\
0 &0&0&\gamma P&v_z\\
\end{pmatrix}.
\end{equation}
The tensor $B_{\alpha\beta ba}\equiv\partial_a A_{\alpha \beta b}=A_{\alpha\beta b,a}$  (with  $a,b=x,y,z\;\text{or }1,2,3$ and $\alpha,\beta=0,1,2,3,4$) has components:
\begin{align}
B_{\alpha0xa}&=
\begin{pmatrix}
\partial_xv_x &\partial_yv_x&\partial_zv_x\\
0 &0&0\\
0 &0&0\\
0 &0&0\\
0 &0&0\\
\end{pmatrix}~~,
\end{align}
\begin{align}
B_{\alpha0ya}&=
\begin{pmatrix}
\partial_xv_y &\partial_yv_y&\partial_zv_y\\
0 &0&0\\
0 &0&0\\
0 &0&0\\
0 &0&0\\
\end{pmatrix}~~,
\end{align}
\begin{align}
B_{\alpha0za}&=
\begin{pmatrix}
\partial_xv_z &\partial_yv_z&\partial_zv_z\\
0 &0&0\\
0 &0&0\\
0 &0&0\\
0 &0&0\\
\end{pmatrix}~~,
\end{align}
\begin{align}
B_{\alpha1xa}&=
\begin{pmatrix}
\partial_x\rho &\partial_y\rho&\partial_z\rho\\
\partial_xv_x &\partial_yv_x&\partial_zv_x\\
0 &0&0\\
0 &0&0\\
\gamma\partial_xP&\gamma\partial_yP&\gamma\partial_zP\\
\end{pmatrix}~~,
\end{align}
\begin{align}
\nonumber\\
B_{\alpha1ya}&=
\begin{pmatrix}
0 &0&0\\
\partial_xv_y &\partial_yv_y&\partial_zv_y\\
0 &0&0\\
0 &0&0\\
0 &0&0\\
\end{pmatrix}~~,
\end{align}
\begin{align}
B_{\alpha1za}&=
\begin{pmatrix}
0 &0&0\\
\partial_xv_z &\partial_yv_z&\partial_zv_z\\
0 &0&0\\
0 &0&0\\
0 &0&0\\
\end{pmatrix}~~,
\end{align}
\begin{align}
B_{\alpha2xa}&=
\begin{pmatrix}
0 &0&0\\
0 &0&0\\
\partial_xv_x &\partial_yv_x&\partial_zv_x\\
0 &0&0\\
0 &0&0\\
\end{pmatrix}~~,
\end{align}
\begin{align}
B_{\alpha2ya}&=
\begin{pmatrix}
\partial_x\rho &\partial_y\rho&\partial_z\rho\\
0 &0&0\\
\partial_xv_y &\partial_yv_y&\partial_zv_y\\
0 &0&0\\
\gamma\partial_xP&\gamma\partial_yP&\gamma\partial_zP\\
\end{pmatrix}~~,
\end{align}
\begin{align}
B_{\alpha2za}&=
\begin{pmatrix}
0 &0&0\\
0 &0&0\\
\partial_xv_z &\partial_yv_z&\partial_zv_z\\
0 &0&0\\
0 &0&0\\
\end{pmatrix}~~,
\end{align}
\begin{align}
B_{\alpha3xa}&=
\begin{pmatrix}
0 &0&0\\
0 &0&0\\
0 &0&0\\
\partial_xv_x &\partial_yv_x&\partial_zv_x\\
0 &0&0\\
\end{pmatrix}~~,
\end{align}
\begin{align}
B_{\alpha3ya}&=
\begin{pmatrix}
0 &0&0\\
0 &0&0\\
0 &0&0\\
\partial_xv_y &\partial_yv_y&\partial_zv_y\\
0 &0&0\\
\end{pmatrix}~~,
\end{align}
\begin{align}
B_{\alpha3za}&=
\begin{pmatrix}
\partial_x\rho &\partial_y\rho&\partial_z\rho\\
0 &0&0\\
0 &0&0\\
\partial_xv_z &\partial_yv_z&\partial_zv_z\\
\gamma\partial_xP&\gamma\partial_yP&\gamma\partial_zP\\
\end{pmatrix}~~,
\end{align}
\begin{align}
B_{\alpha4xa}&=
\begin{pmatrix}
0 &0&0\\
-\cfrac{\partial_x\rho}{\rho^2} &-\cfrac{\partial_y\rho}{\rho^2}&-\cfrac{\partial_z\rho}{\rho^2}\\
0 &0&0\\
0 &0&0\\
\partial_xv_x &\partial_yv_x&\partial_zv_x\\
\end{pmatrix}~~,
\end{align}
\begin{align}
B_{\alpha4ya}&=
\begin{pmatrix}
0 &0&0\\
0 &0&0\\
-\cfrac{\partial_x\rho}{\rho^2} &-\cfrac{\partial_y\rho}{\rho^2}&-\cfrac{\partial_z\rho}{\rho^2}\\
0 &0&0\\
\partial_xv_y &\partial_yv_y&\partial_zv_y\\
\end{pmatrix}~~,
\end{align}
\begin{align}
B_{\alpha4za}&=
\begin{pmatrix}
0 &0&0\\
0 &0&0\\
0 &0&0\\
-\cfrac{\partial_x\rho}{\rho^2} &-\cfrac{\partial_y\rho}{\rho^2}&-\cfrac{\partial_z\rho}{\rho^2}\\
\partial_xv_z &\partial_yv_z&\partial_zv_z\\
\end{pmatrix}~~.
\end{align}

\end{document}